\documentclass[journal,12pt,onecolumn,draftclsnofoot,a4paper]{IEEEtran}

\usepackage{cite}
\usepackage{subcaption}
\usepackage{comment}
\usepackage[hidelinks]{hyperref}
\usepackage{amsthm}

\theoremstyle{definition}
\newtheorem{definition}{Definition}

\theoremstyle{plain}
\newtheorem{lemma}{Lemma}

\usepackage{amsmath,amssymb,amsfonts}
\usepackage[acronym,shortcuts]{glossaries}
\usepackage{graphicx}
\usepackage{textcomp}
\usepackage{xcolor}
\def\BibTeX{{\rm B\kern-.05em{\sc i\kern-.025em b}\kern-.08em
    T\kern-.1667em\lower.7ex\hbox{E}\kern-.125emX}}

\usepackage{tikz}
\usetikzlibrary{shapes, arrows, positioning}
\usepackage{pgfplots}
\usetikzlibrary{arrows,automata, positioning}
\usetikzlibrary{calc, chains,
                fit,
                positioning,
                shapes}
\usetikzlibrary{patterns}
\usepackage{multirow}

\newcommand{\6}{\mathbf }
\newacronym{AR4JA}{AR4JA}{accumulate, repeat-by-4, and jagged accumulate}
\newacronym{AWGN}{AWGN}{additive white Gaussian noise}
\newacronym{BER}{BER}{bit error rate}
\newacronym{BP}{BP}{belief propagation}
\newacronym{BPSK}{BPSK}{binary phase shift keying}
\newacronym{BW}{BW}{bandwidth}
\newacronym{BCH}{BCH}{Bose--Chaudhuri--Hocquenghem}
\newacronym{BICM-ID}{BICM-ID}{bit-interleaved coded modulation with
iterative decoding}
\newacronym{CCSDS}{CCSDS}{Consultative Committee for Space Data Systems}
\newacronym{CW}{CW}{continuous-wave}
\newacronym{DS-SS}{DS-SS}{direct-sequence spread spectrum}
\newacronym{FER}{FER}{frame error rate}
\newacronym{JSI}{JSI}{jamming state information}
\newacronym{LDPC}{LDPC}{low-density parity-check}
\newacronym{LLR}{LLR}{log-likelihood ratio}
\newacronym{MI}{MI}{mutual information}
\newacronym{PEXIT}{PEXIT}{protograph-based extrinsic information transfer}
\newacronym{PN}{PN}{pseudonoise}
\newacronym{PSD}{PSD}{power spectral density}
\newacronym{SC-LDPC}{SC-LDPC}{spatially coupled low-density parity-check}
\newacronym{SIR}{SIR}{signal-to-interference ratio}
\newacronym{SPA}{SPA}{sum product algorithm}
\newacronym{SWD}{SWD}{sliding window decoding}

\begin{document}

\markboth{BATTAGLIONI ET AL.}{SC-LDPC CODES FOR PULSED-JAMMING-RESILIENT SPACE TELEMETRY}

\title{Analysis and Design of Spatially Coupled LDPC Codes for Pulsed-Jamming-Resilient Space Telemetry}

\author{Massimo~Battaglioni,~\IEEEmembership{Senior Member,~IEEE},
Mauro~M.~M.~Costantino,~\IEEEmembership{Student Member,~IEEE},
Rebecca~Giuliani,~\IEEEmembership{Student Member,~IEEE},
Franco~Chiaraluce,~\IEEEmembership{Senior Member,~IEEE},
and~Marco~Baldi,~\IEEEmembership{Senior Member,~IEEE}%
\thanks{Massimo Battaglioni, Mauro M. M. Costantino, Rebecca Giuliani, Franco Chiaraluce, and Marco Baldi are with the Department of Information Engineering, Università Politecnica delle Marche, 60131 Ancona, Italy. Franco Chiaraluce is also with the Consorzio Nazionale Interuniversitario per le Telecomunicazioni, 43124 Parma, Italy. E-mail: m.battaglioni@univpm.it; m.m.m.costantino@pm.univpm.it; r.giuliani@pm.univpm.it; f.chiaraluce@univpm.it; m.baldi@univpm.it.}%
\thanks{Corresponding author: M. Battaglioni.}%
\thanks{This publication was produced whilst Rebecca Giuliani was enrolled on the PhD programme in Information Engineering at the Polytechnic University of the Marche, 39th cycle, with the support of a scholarship funded by Ministerial Decree No. 118 of 2 March 2023, under the PNRR, funded by the European Union, NextGenerationEU, Mission 4 ``Education and Research'', Component 1 ``Strengthening the provision of educational services: from nurseries to universities'', Investment 4.1 ``Expansion of the number of PhD programmes and innovative PhDs for public administration and cultural heritage''. The work of Mauro Michele Maria Costantino is supported by the Italian National Cybersecurity Agency (ACN) under the programme for promotion of XLII cycle PhD research in cybersecurity (CUP I32B24001750005). The views expressed are those of the authors and do not represent the funding institutions.}%
\thanks{Preliminary, simulation-only results related to this work were presented in part at the IEEE International Conference on Communications (ICC) 2026~\cite{giuliani2026icc} and at the International Conference on Software, Telecommunications and Computer Networks (SoftCOM) 2026~\cite{GiulianiSoftCOM2026}. The analysis, the code design, and most of the numerical results reported here are new.}%
}
\maketitle

\begin{abstract}
Jamming is a major threat to space telemetry downlinks, which operate under severe power and bandwidth constraints. This paper develops a jamming-aware analysis and design framework for protograph-based spatially coupled low-density parity-check (SC-LDPC) codes. For pulsed jamming with jamming state information (JSI) at the receiver, we derive a two-state protograph-based extrinsic information transfer (PEXIT) recursion that tracks the messages originating from clean and jammed symbols separately and merges them at the check nodes through a mutual-information-matched Gaussian approximation. For non-interleaved pulsed jamming, we map each pulse realization, through the serialization of the lifted codeword, to a profile of jammed fractions over the variable-node types, and we prove that, under random lifting, this profile determines the local channel-state statistics. The resulting position-dependent thresholds are the objective of a constrained differential-evolution search over protographs, coupling memories, and edge spreadings. At different code rates, the designed ensembles are compared with the Consultative Committee for Space Data Systems (CCSDS) accumulate, repeat-by-4, and jagged accumulate (AR4JA) LDPC codes whose block length matches the SC-LDPC constraint length, with longer AR4JA codes whose block length is comparable with the size of the sliding window decoder, and with non-optimized SC-LDPC codes. Under pulsed jamming with random pulse positions and with JSI, the ensembles designed over random pulse realizations, decoded through sliding window decoding, reach a bit error rate of $10^{-4}$ at lower signal-to-noise ratios than all the benchmarks.
\end{abstract}


\section{Introduction}

Reliable telemetry is a fundamental requirement of space missions, since the downlink carries spacecraft status information, payload data, and scientific measurements to ground stations. Space telemetry links are typically power-limited and may operate at very low received signal levels, particularly in deep-space missions. Their design therefore relies heavily on forward error correction and carefully optimized synchronization and receiver-processing architectures~\cite{tutorial}.

The \ac{CCSDS} telemetry standard includes convolutional, Reed--Solomon, turbo, and \ac{LDPC} coding options~\cite{TMst}. Among them, the protograph-based \ac{AR4JA} family provides an attractive trade-off among decoding performance, implementation complexity, and spectral efficiency~\cite{DivsalarAR4JA,CCSDS130G3}. These codes have been extensively characterized over \ac{AWGN} channels and have also been considered in end-to-end studies of space links affected by propagation impairments such as solar scintillation~\cite{Baldisolar,Fino2017,baldi2020solar}. Their robustness to intentional or unintentional jamming in long-block telemetry applications, however, has received considerably less attention.

Jamming may severely reduce the reliability of a space link, particularly when interference is injected directly into the receiver bandwidth. Depending on its time and frequency occupancy, the jammer may generate continuous, pulsed, partial-band, tone, or noise-like interference. Conventional countermeasures include spread-spectrum signaling, frequency hopping, interleaving, interference detection and suppression, adaptive receiver processing, and channel coding. Coding-based countermeasures have been investigated in terrestrial and military systems, including block, convolutional, and concatenated codes under partial-band, multitone, and pulsed-noise jamming~\cite{Ma1984,Wang1989,Frank1996,Chia2002}, turbo codes and \ac{BICM-ID} for partial-band-jammed and frequency-hopping links~\cite{jordan1998,Dai2010}, and \ac{LDPC} codes designed against follower jamming~\cite{Kim2016}. Channels in which blocks of symbols are hit by interference were analyzed from an information-theoretic viewpoint in~\cite{McEliece1984}, and the benefit of jamming state side information in coded systems is well established~\cite{SimonSS}. These studies demonstrate the potential of channel coding as a jamming countermeasure, but they do not directly address the long-block standardized codes and decoding architectures used in space telemetry.

Space-specific investigations have mainly focused on the telecommand uplink. The robustness of satellite telecommand links against different jamming strategies was studied in~\cite{Martinelli2012}, while advanced coding schemes, including binary and nonbinary \ac{LDPC} codes, turbo codes, and \ac{BCH} codes, were evaluated in~\cite{BALDIMILCOM,Baldi2013ttc,BaldiIJ}. These works also examined the effects of \ac{DS-SS}, interleaving, receiver knowledge of the jamming state, and different jammer waveforms. Telecommand systems, however, are characterized by short messages, stringent command-detection requirements, and latency and synchronization constraints that differ substantially from those of telemetry downlinks. Therefore, the findings of those works cannot be transferred directly to the long-block \ac{LDPC} codes used for high-rate telemetry.

In this paper, we focus specifically on space-telemetry links affected by jamming and study a family of codes that, to the best of our knowledge, has not yet been considered in the context of jamming-affected transmissions (apart from our preliminary works~\cite{giuliani2026icc,GiulianiSoftCOM2026}); we also propose a technique for designing codes optimized for use in this context.
The family of codes we consider are protograph-based \ac{SC-LDPC} codes. These codes were originally introduced as \ac{LDPC} convolutional codes in~\cite{Felstrom1999}. Their main asymptotic advantage over block codes is due to the so-called threshold saturation, through which belief-propagation decoding approaches the maximum-a-posteriori threshold of the underlying uncoupled ensemble over binary memoryless symmetric channels~\cite{Kudekar2013}. Protograph-based constructions provide a structured framework for code design~\cite{Mitchell2015}, while windowed decoding bounds the latency and memory requirements by processing only a limited portion of the coupled chain at each decoding step~\cite{Iyengar2012}. Spatial coupling also improves robustness when the channel varies along the coupled chain, as shown for block-fading channels~\cite{ulHassan2014}, and windowed decoding has been analyzed for channels with erasure bursts~\cite{Iyengar2012}. These properties make \ac{SC-LDPC} codes suitable for continuous telemetry transmissions and provide the structural flexibility required to optimize their behavior under nonuniform interference.

The use of \ac{SC-LDPC} codes under jamming requires an analytical tool that quantifies how the decoder reacts to nonuniform symbol reliabilities, and a design method that exploits it.
Starting from these considerations, the main contributions of this paper are as follows.

\begin{itemize}
\item We extend the \ac{PEXIT} recursion of~\cite{liva_pexit} to the two channel states induced by pulsed jamming with \ac{JSI}. The recursion tracks the messages generated by clean and jammed symbols separately and merges them at the check nodes into a Gaussian message with the same \ac{MI} as the two-state mixture. Under ideal interleaving, this yields a homogeneous two-class recursion.
\item For non-interleaved pulsed jamming, we map each time-domain pulse realization, through the serialization of the lifted codeword, into a type-dependent profile of jammed fractions that accounts for partial overlaps between pulses and variable-node types. We prove that, under random lifting, this profile determines the channel-state statistics of every fixed-depth computation neighborhood, which yields a position-dependent \ac{PEXIT} recursion for the terminated coupled chain, whose thresholds are within $0.03$~dB of those obtained through Monte Carlo density evolution. Type-dependent channels have already been used in \ac{PEXIT} analysis, e.g., for block-fading channels~\cite{Pulini2012}, but not for jamming channels, for which we introduce the mapping from non-interleaved pulse realizations to type-dependent state mixtures and its use as a design objective.
\item We design new \ac{SC-LDPC} code ensembles at rates $1/2$, $2/3$, and $4/5$ by optimizing the uncoupled protograph, the coupling memory, and the edge spreading through a constrained differential-evolution search that minimizes a tail average of position-dependent thresholds over random pulse realizations. We also consider a fixed pulse pattern, used both in the design and in the simulations, to test the predictive value of the thresholds.
\item We compare the designed codes, decoded through \ac{SWD}, with three benchmarks: the \ac{CCSDS} \ac{AR4JA} codes with $k=1024$, whose block length is approximately equal to the constraint length of the \ac{SC-LDPC} codes; the \ac{CCSDS} \ac{AR4JA} codes with $k=4096$, whose block length is comparable with the size of the sliding window decoder; and non-optimized terminated \ac{SC-LDPC} codes, which separate the gain due to spatial coupling from the gain due to the proposed design. The comparison is carried out with and without \ac{JSI} at all rates and, at rate $1/2$ and without \ac{JSI}, also for different jammer duty cycles and with ideal interleaving. We also compare \ac{SWD} and block decoding in terms of the number of coded symbols to be received per decoding step and of decoding latency.
\end{itemize}
The numerical results show that, at rates $1/2$ and $2/3$, spatial coupling alone already outperforms the \ac{AR4JA} code with $k=1024$ under pulsed jamming, and that the ensembles designed for random pulse realizations provide further gains, especially in the waterfall region at rates $1/2$ and $4/5$.

This paper substantially extends our conference papers~\cite{giuliani2026icc,GiulianiSoftCOM2026}, which relied only on Monte Carlo simulations of \ac{SC-LDPC} codes that were not designed for the jammed channel and were decoded over the whole frame. More precisely, in~\cite{giuliani2026icc}, two regular $(3,6)$ \ac{SC-LDPC} codes were compared with the rate-$1/2$ \ac{CCSDS} \ac{AR4JA} code for a single duty cycle. In~\cite{GiulianiSoftCOM2026}, \ac{SC-LDPC} codes with randomly generated protographs were compared with the \ac{AR4JA} codes at rates $1/2$, $2/3$, and $4/5$, without \ac{JSI}. Compared to those works, this paper makes the following key contributions:
\begin{itemize}
\item It develops a two-class \ac{PEXIT} analysis for interleaved pulsed jamming and a position-dependent analysis for the non-interleaved case, checked through Monte Carlo density evolution.
\item It uses the resulting profile-dependent thresholds to design the protographs, coupling memories, and edge-spreading matrices of new \ac{SC-LDPC} code ensembles; the ensembles of~\cite{GiulianiSoftCOM2026} are used here only as a non-optimized reference.
\item It decodes the \ac{SC-LDPC} codes through \ac{SWD}, compares their symbol-reception requirements and latency with those of block decoding,  and extends the numerical assessment to the availability of \ac{JSI} at all rates and to symbol interleaving.
\end{itemize}

The remainder of the paper is organized as follows. Section~\ref{sec:preli} introduces the system model, the \ac{SC-LDPC} code ensembles, the pulsed-jamming model, and the interleaving model. Section~\ref{sec:pexitjam} develops the two-state Gaussian-approximation \ac{PEXIT} recursion and applies it to interleaved pulsed jamming. Section~\ref{sec:pexit_noninterleaved} develops the position-dependent recursion for non-interleaved pulsed jamming. Section~\ref{sec:design} formulates the design of \ac{SC-LDPC} code ensembles for a generic set of jamming profiles. Section~\ref{sec:designed} applies it to a single deterministic pulse pattern and to random pulse realizations, and presents the selected ensembles. Section~\ref{sec:decoder} characterizes the symbol reception and decoding latency of \ac{SWD} and block decoding. Section~\ref{sec:results} presents the finite-length simulation results. Finally, Section~\ref{sec:concl} concludes the paper.

\section{Preliminaries and system model}\label{sec:preli}

In this section we introduce the necessary background regarding the communication channel with jamming and the family of \ac{SC-LDPC} codes we consider.

\begin{figure*}[th!] \centering \resizebox{0.98\textwidth}{!}{    \begin{tikzpicture}[auto, node distance=2.5cm, >=latex]
    
        \tikzstyle{block} = [draw, rectangle, minimum height=10mm, minimum width=20mm, align=center]
        \tikzstyle{sum} = [draw, circle, inner sep=1pt]
        \tikzstyle{signal} = [coordinate]
    
        \node[signal] (txdata) {Tx data};
        \node[block, right of=txdata] (enc) {Encoding};
        \node[block, right of=enc, node distance=4.5cm] (spread){Modulation};
 
        \node[sum, right of=spread, node distance=3.4cm] (sum1) {$+$};

        \node[sum, right of=sum1, node distance=2cm] (sum2) {$+$};
    
        \node[signal, below of=sum1, node distance=1.2cm] (noise) {};
        \node[signal, below of=sum2, node distance=1.2cm] (jam) {};
        
        \node[block, right of=sum2, node distance=3.7cm] (despread) {Demodulation};
        \node[block, right of=despread, node distance=4cm] (dec) {Decoding};
        \node[signal, right of=dec, node distance=2.5cm] (rxdata) {Rx data};

        \node[signal, below of=despread, node distance=1.2cm] (jsi) {};

        \draw[->] (txdata) -- node[above] {Tx data} (enc);
        \draw[->] (enc) -- node[above] {codewords} (spread);

        \draw[->] (spread) -- node[above] {Tx signal} (sum1);
        \draw[->] (noise) -- node[right] {noise} (sum1);
        
        \draw[->] (sum1) -- (sum2); 
        \draw[->] (jam) -- node[right] {jamming} (sum2); 
        \draw[->] (sum2) -- node[above] {Rx signal} (despread); 
        
        \draw[->] (despread) -- (dec);
        \draw[->, dashed] (jsi) -- node[right] {jamming state information (optional)} (despread);
        \draw[->] (dec) -- node[above] {Rx data} (rxdata);
    
    \end{tikzpicture}} \caption{System model of the jammed telemetry link} \label{fig:systemmodel} \end{figure*}
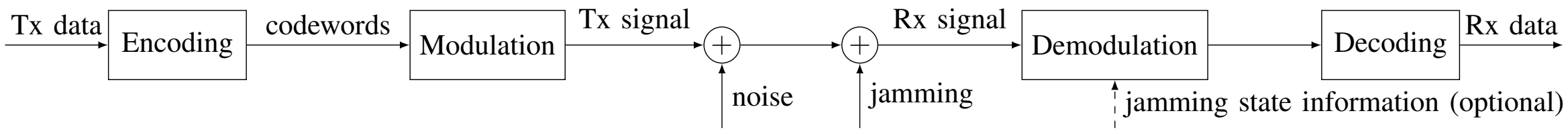

The communication scenario analyzed in this work is schematically illustrated in Fig.~\ref{fig:systemmodel}. Message bits are partitioned into blocks, encoded with the selected error-correcting code, and transmitted with \ac{BPSK} modulation.

During propagation, the transmitted signal is affected by \ac{AWGN} with two-sided \ac{PSD} $N_0/2$ and by jamming, the latter taking the form of some interference occupying the same frequency band. The received signal is demodulated with or without knowledge of the interference state, i.e., with or without \ac{JSI}, and the resulting soft decisions are passed to the decoder.
Let $T_b=1/R_b$ denote the information-bit duration. With a code of length $n$ and design rate $R$ (see Section~\ref{sec:preli}-\ref{subsec:SC-LDPC}), the coded-symbol duration and energy are $T_e=RT_b$ and $E_e=RE_b$, and we index the coded symbols by $\eta=0,\ldots,n-1$. The receiver obtains the decision variable of the $\eta$-th coded symbol by projecting the received signal onto the unit-energy basis function
\begin{equation}
    g_{\eta}(t)=\sqrt{\frac{2}{T_e}}\,P_{T_e}(t-\eta T_e)\cos(\omega_0 t),
    \label{eq:fj_coded}
\end{equation}
where $P_{T_e}(t)$ equals one for $0\leq t<T_e$ and zero otherwise, and $\omega_0=2\pi f_0$ is the angular carrier frequency. Let $W_{\rm s}$ denote the bandwidth occupied by the transmitted signal.\footnote{With \ac{DS-SS}, $g_\eta(t)$ is multiplied by the spreading waveform and $W_{\rm s}$ is the spread bandwidth. Since the spreading waveform has unit magnitude, the projection of a jammer with flat \acs{PSD} over $W_{\rm s}$ has the same variance as without spreading, and the processing gain is entirely captured by $J_0=P_J/W_{\rm s}$, as defined  in~\eqref{eq:average_jamming_psd}. All results are therefore expressed in terms of $E_b/J_0$ and hold with or without \ac{DS-SS}.}

\subsection{Terminated SC-LDPC codes}\label{subsec:SC-LDPC}

A binary linear block code of length $n$ and dimension $k$ is the null space of a parity-check matrix $\mathbf{H}$, i.e., the set of vectors $\6c$ such that $\6c\6H^\top =\60$, where $\60$ is the all-zero vector and $\top$ denotes transposition.

A bipartite graph $\mathcal{G}(\mathbf{H})$, referred to as a Tanner graph~\cite{Tanner1981}, can be constructed from the parity-check matrix $\mathbf{H}$. A protograph-based \ac{SC-LDPC} code is obtained by coupling a sequence of copies of a small base Tanner graph, called \emph{protograph}~\cite{Thorpe2003}, so that the edges of each copy connect nodes in $m_s+1$ consecutive spatial positions, where $m_s$ is the syndrome former memory order (simply referred to as coupling memory hereafter) of the code. The base matrix \( \mathbf{B} \) of a protograph is a small \( r_0 \times n_0 \) matrix with integer entries denoting the number of edges between the corresponding check and variable nodes. To form the spatially coupled structure, the edges of \( \mathbf{B} \) are distributed across \( m_s + 1 \) component base matrices \( \{ \mathbf{B}_0, \mathbf{B}_1, \ldots, \mathbf{B}_{m_s} \} \) such that $\mathbf{B}=\sum_{\ell=0}^{m_s}\mathbf{B}_\ell$.
These component matrices can then be arranged diagonally to form a terminated convolutional base matrix as follows:
\begin{equation}
\mathbf{B}_{\mathrm{SC}} =
\begin{bmatrix}
\mathbf{B}_0      &              &              &              \\
\mathbf{B}_1      & \mathbf{B}_0 &              &              \\
\vdots            & \mathbf{B}_1 & \ddots       &              \\
\mathbf{B}_{m_s}  & \vdots       & \ddots       & \mathbf{B}_0 \\
                  & \mathbf{B}_{m_s} & \ddots   & \mathbf{B}_1 \\
                  &                & \ddots     & \vdots       \\
                  &                &             & \mathbf{B}_{m_s}
\end{bmatrix}.
\label{eq:BSC}
\end{equation}
Outside the diagonal band formed by the component matrices \(\mathbf{B}_0, \mathbf{B}_1, \ldots, \mathbf{B}_{m_s}\), all the remaining entries of \(\mathbf{B}_{\mathrm{SC}}\) are zeros. The matrix \(\mathbf{B}_{\mathrm{SC}}\) can then be \emph{lifted} by a factor \(M\), i.e., each nonzero entry is replaced by a sum of one or more \(M \times M\) permutation matrices; zero entries are instead replaced by all-zero matrices. This produces the full binary parity-check matrix \(\6H\) of the terminated \ac{SC-LDPC} code~\cite{Mitchell2015}, describing a code with frame length \(LMn_0\), where \(L\) is the number of spatial positions (i.e., replicas of the base matrices
\(\mathbf{B}_0, \mathbf{B}_1, \ldots, \mathbf{B}_{m_s}\)) in \(\mathbf{B}_{\mathrm{SC}}\). After lifting by a factor \( M \), the parity-check matrix will be of size \((L + m_s)r_0M \times Ln_0M\). The (nominal) design rate of the resulting \ac{SC-LDPC} code is $R=R_{\infty}-m_sr_0/(n_0L)$, 
where \(R_{\infty} =1- \frac{r_0}{n_0}\) denotes the asymptotic code rate.  The actual rate of the lifted code is $R_{\rm actual}=1-\operatorname{rank}_{\mathbb{F}_2}(\mathbf{H})/(Ln_0M)\geq R$, since in terminated \ac{SC-LDPC} codes the parity-check matrix does not necessarily have full row rank. Unless otherwise stated, energy normalizations use $R$. Finally, the code's syndrome former constraint length is defined as $\nu_s=Mn_0(m_s+1)$. The collection of codes that can be obtained by lifting a given base matrix is called an \emph{ensemble}.

\subsection{Pulsed jamming model}\label{subsec:pulsed}

Let us describe the \textit{pulsed} jamming model we consider~\cite{SimonSS,Hall2005}.
The jamming signal $j(t)$ is modeled as band-limited Gaussian noise that is intermittently transmitted. The time axis is divided into intervals of duration $T$, and within each interval the jammer is active for a duration $D$. The active interval may start at an arbitrary position within each interval of duration $T$, and this position may change from one interval to the next. The jammer duty cycle is therefore $\rho=D/T$, with $0<\rho\leq1$. 
With some abuse of terminology, we refer to $T$ as the jamming \emph{period}. Continuous jamming corresponds to the limiting case $\rho=1$.
Without information on the jamming state, pulsed jamming can be more harmful than continuous jamming with the same average power: for uncoded \ac{BPSK} with negligible thermal noise, the worst-case duty cycle is inversely proportional to $E_b/J_0$~\cite{SimonSS}, so that the jammer benefits from short pulses when $E_b/J_0$ is large. With ideal \ac{JSI} and soft-decision decoding, instead, spreading the jammer power over more symbols is more harmful, as shown at the end of this subsection.

Let $P_J$ denote the average received jamming power over an interval of duration $T$. Since the jammer is active only for a fraction $\rho$ of the time, the jamming power received during an active interval is $P'_J=P_J/\rho$, 
whereas no jamming power is received outside the active interval. Consequently, for fixed average jamming power, reducing the duty cycle concentrates the jamming power over a smaller fraction of the transmitted symbols. We denote by
\begin{equation}
    J_0=\frac{P_J}{W_{\rm s}}
    \label{eq:average_jamming_psd}
\end{equation}
the equivalent average one-sided jamming \ac{PSD} over the bandwidth $W_{\rm s}$. During an active interval, the corresponding one-sided jamming \ac{PSD} is $J'_0=P'_J/W_{\rm s}=J_0/\rho$, 
and is assumed constant over the considered bandwidth.
The corresponding ratio between the energy per information bit and the average jamming \ac{PSD} is $E_b/J_0$.

The jamming contribution to the decision variable associated with the $\eta$-th coded symbol is obtained by projecting $j(t)$ onto the basis function $g_\eta(t)$ in~\eqref{eq:fj_coded}, namely
$u_\eta=\int_{\eta T_e}^{(\eta+1)T_e}j(t)g_\eta(t)\,\mathrm{d}t$. 
For a coded symbol entirely covered by an active jamming interval, $u_\eta$ is a zero-mean Gaussian random variable with variance
\begin{equation}
    \sigma_{\mathrm{pulse}}^2
    =
    \frac{J'_0}{2}
    =
    \frac{J_0}{2\rho}.
    \label{eq:pulsed_jamming_variance}
\end{equation}
For a coded symbol outside the active intervals, the jamming contribution is zero.

\begin{figure*}[tb!]
    \centering
    \begin{subfigure}[t]{0.48\textwidth}
        \centering
        \resizebox{\linewidth}{!}{\definecolor{razzmatazz}{HTML}{F61067}

\begin{tikzpicture}
\def\Tb{2}
\def\H{0.6}

\fill[razzmatazz]  (0,0) rectangle (6,\H);
\fill[cyan!90] (6,0) rectangle (8,\H);

\draw (0,0) rectangle (8,\H);

\foreach \x in {2,4,6}
    \draw[gray] (\x,-0.3) -- (\x,1.2);

\foreach \a/\b in {0/2,2/4,4/6,6/8}{
    \draw[<->,blue!60!black] (\a,1.0) -- (\b,1.0);
    \node at ({(\a+\b)/2},1.25) {$T_e$};
}

\draw[<->,blue!60!black] (0,-0.6) -- (6,-0.6);
\node at (3,-0.85) {$D$};

\draw[<->,blue!60!black] (0,-1.2) -- (8,-1.2);
\node at (4,-1.45) {$T$};

\foreach \x in {0,6,8}
    \draw[gray] (\x,0) -- (\x,-1.3);

\end{tikzpicture}}
        \caption{Example of a jamming period with $L_{\rm p}=4$ and $L_{\rm b}=3$.}
        \label{fig:PulsedCase3}
    \end{subfigure}
    \hfill
    \begin{subfigure}[t]{0.48\textwidth}
        \centering
        \resizebox{\linewidth}{!}{\definecolor{grapesoda}{HTML}{8D3B72}
\definecolor{pacificblue}{HTML}{58A4B0}
\definecolor{razzmatazz}{HTML}{F61067}
\definecolor{shamrock}{HTML}{499F68}
\definecolor{carrotcolor}{HTML}{ea9010}

\begin{tikzpicture}

\def\Tb{2}
\def\H{0.6}

\fill[shamrock] (0,0) rectangle (8,\H);
\fill[carrotcolor]  (0.4,0) rectangle (0.9,\H);
\fill[carrotcolor]  (7.3,0) rectangle (7.8,\H);

\draw (0,0) rectangle (8,\H);

\foreach \a/\b in {0/8}{
    \draw[<->,blue!60!black] (\a,1.0) -- (\b,1.0);
    \node at ({(\a+\b)/2},1.25) {$T_{\text{code}}$};
}

\draw[{<[scale=0.6]}-{>[scale=0.6]},blue!60!black] (0.4,-0.1) -- (0.9,-0.1);
\node at (0.65,-0.4) {$D$};

\draw[{<[scale=0.6]}-{>[scale=0.6]},blue!60!black] (7.3,-0.1) -- (7.8,-0.1);
\node at (7.5,-0.4) {$D$};

\foreach \a/\b in {0/4,4/8}{
    \draw[<->,blue!60!black] (\a,-1.0) -- (\b,-1.0);
    \node at ({(\a+\b)/2},-1.45) {$T$};
}

\foreach \x in {0,8}
    \draw[gray] (\x,1.2) -- (\x,-1.3);

\foreach \x in {4}
    \draw[gray] (\x,0) -- (\x,-1.3);
    
\foreach \x in {0.4,0.9,7.3,7.8}
    \draw[gray] (\x,\H) -- (\x,-0.3);

\end{tikzpicture}}
        \caption{Example of a codeword spanning two jamming periods and, consequently, affected by two jamming bursts, which start at different positions within their periods.}
        \label{fig:T_code}
    \end{subfigure}
    \caption{Examples of the considered pulsed-jamming model.}
    \label{fig:pulsed_examples}
\end{figure*}
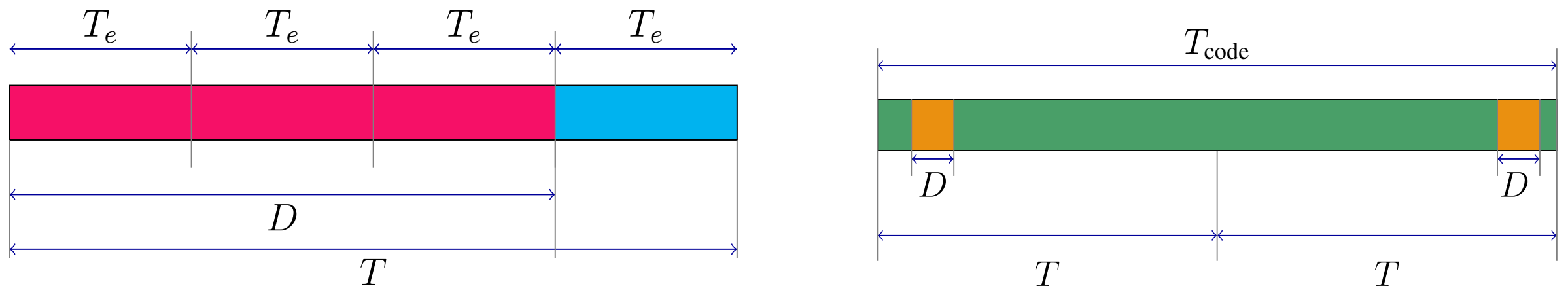

Strictly speaking, a jamming signal confined to the bandwidth $W_{\rm s}$ is not white with respect to the rectangular basis functions in~\eqref{eq:fj_coded}, whose spectrum is not band-limited, so that its projections onto different symbol intervals are slightly correlated. As common, we neglect this effect and treat the jamming signal during an active interval as white Gaussian noise with two-sided \ac{PSD} $J'_0/2$. Since the basis functions $g_\eta(t)$ are orthonormal, this yields the variance in~\eqref{eq:pulsed_jamming_variance} and makes the projected disturbances $u_\eta$ of different coded symbols independent.

The temporal effect of pulsed jamming also depends on the relationship between $D$, $T$, and the coded-symbol duration $T_e$. In the following, we consider $D\geq T_e$ and, for simplicity but without affecting the significance of the results, assume that both $D$ and $T$ are integer multiples of $T_e$. We additionally assume that the transitions between active and inactive intervals are aligned with coded-symbol boundaries, so that no coded symbol is only partially affected by a pulse. Let $L_{\rm p}\triangleq T/T_e$ and $L_{\rm b}\triangleq D/T_e$ denote the number of coded symbols in one jamming period and in one active pulse, respectively (see Fig.~\ref{fig:PulsedCase3}), so that $\rho=L_{\rm b}/L_{\rm p}$. Each period therefore contains $L_{\rm b}$ consecutive jammed symbols and $L_{\rm p}-L_{\rm b}$ unjammed symbols. The total disturbance variance at the coded-symbol decision variable is $N_0/2$ for the unjammed symbols and
\begin{equation}
    \sigma_{\mathrm{tot,p}}^2
    =
    \frac{N_0}{2}
    +
    \frac{J'_0}{2}
    =
    \frac{N_0}{2}
    +
    \frac{J_0}{2\rho}
    \label{eq:pulsed_total_variance}
\end{equation}
for the jammed symbols.

Let $T_{\mathrm{code}}$ denote the duration of a codeword. We focus on the case $T\leq T_{\mathrm{code}}$, so that a single codeword may be affected by multiple jamming pulses.
This is the most relevant regime for the considered application since telemetry codewords are long, and the coupled chains of \ac{SC-LDPC} codes are much longer: at a coded-symbol rate of $1$~Msymbol/s, a codeword of $2048$ symbols lasts about $2$~ms, i.e., longer than the period of any pulsed jammer or interferer with a pulse repetition frequency above $500$~Hz.
The ratio $T_{\mathrm{code}}/T$ is not necessarily an integer. For simplicity, we assume that the beginning of the first codeword coincides with the beginning of a jamming period, whereas the active interval may start at an arbitrary position within each period. Under this assumption, the codeword contains $\lfloor T_{\mathrm{code}}/T\rfloor$ complete jamming periods, each including one active interval of duration $D$. If $T_{\mathrm{code}}/T$ is not an integer, the remaining part of the codeword intersects one additional partial jamming period. Depending on the relative position of the active interval, the corresponding pulse may be entirely contained in the codeword, may not intersect it, or may be truncated by the codeword boundary. Fig.~\ref{fig:T_code} illustrates the simpler case in which $T$ divides $T_{\mathrm{code}}$. Numbering the jamming periods from $q=0$ at the beginning of the codeword, we denote by $o_q\in\{0,\ldots,L_{\rm p}-L_{\rm b}\}$ the offset, in coded-symbol intervals, of the pulse within the $q$-th period, so that this pulse occupies the coded symbols $qL_{\rm p}+o_q,\ldots,qL_{\rm p}+o_q+L_{\rm b}-1$. In general, the offsets of different periods are not equal, as in Fig.~\ref{fig:T_code}.

Besides the number and positions of the pulses within a codeword, the impact of jamming also depends on the extent to which the jammed symbols are corrupted. During an active interval, the ratio between the coded-symbol energy and the jamming \ac{PSD} is
\begin{equation}
    \frac{E_e}{J'_0}=\rho R\,\frac{E_b}{J_0}.
    \label{eq:active_sir}
\end{equation}
Unless otherwise stated, the numerical results of this paper are obtained with $E_b/J_0=5$~dB and $\rho=0.1$, the latter obtained with $L_{\rm p}=1280$ and $L_{\rm b}=128$. For a coded-symbol rate of $1$~Msymbol/s, these values of $L_{\rm p}$ and $L_{\rm b}$ correspond to $T=1.28$~ms and $D=128~\mu\mathrm{s}$, and both durations scale inversely with the symbol rate. We fix the jammer parameters because the design method of Section~\ref{sec:design} optimizes the code for a given jammer, and we choose these values for the following reasons.

With $E_b/J_0=5$~dB and $\rho=0.1$, \eqref{eq:active_sir} gives $E_e/J'_0$ between about $-8$ and $-6$~dB for the code rates we consider, i.e., $1/2$, $2/3$, and $4/5$. Hence, the jammed symbols carry little information, and the decoder must recover them mainly through the code constraints. At the same time, the average jamming power is moderate: as shown in Section~\ref{sec:designed}-\ref{subsec:iteration_budget}, homogeneous jamming increases the decoding threshold of the considered rate-$1/2$ ensembles by about $0.5$~dB with respect to the jamming-free case, and, for the designed rate-$1/2$ ensemble, the positions of the bursts along the coupled chain increase it by a further $0.07$--$0.14$~dB. This is the regime that motivates the analysis developed in Sections~\ref{sec:pexitjam} and~\ref{sec:pexit_noninterleaved}, which tracks the messages originating from jammed and unjammed symbols separately and, without interleaving, accounts for the positions of the bursts along the coupled chain.

The duty cycle $\rho=0.1$, instead, is not the most harmful choice for a jammer with a given average power when the receiver has ideal \ac{JSI}. Let $C(x)$ denote the capacity of the \ac{BPSK}-input \ac{AWGN} channel with signal-to-noise ratio $x$. With ideal \ac{JSI}, the capacity of the pulsed-jamming channel is $(1-\rho)\,C(E_e/N_0)+\rho\,C\bigl(E_e/(N_0+J'_0)\bigr)$, and the smallest $E_b/N_0$ for which it reaches the code rate increases with $\rho$. For instance, with $E_b/J_0=5$~dB and rate $1/2$, it is $0.34$, $0.68$, $1.11$, $1.52$, and $1.93$~dB for $\rho=0.025$, $0.1$, $0.25$, $0.5$, and $1$, respectively, against $0.19$~dB without jamming, and it also increases with $\rho$ at rates $2/3$ and $4/5$. Therefore, $\rho=0.1$ models pulsed interference whose duty cycle is not chosen against a receiver with \ac{JSI}, such as a jammer designed against conventional receivers without \ac{JSI}, for which, with uncoded \ac{BPSK} and negligible thermal noise, the most harmful duty cycle is about $0.709/(E_b/J_0)$, i.e., about $0.22$, $0.07$, and $0.02$ for $E_b/J_0=5$, $10$, and $15$~dB~\cite{SimonSS}, or unintentional pulsed interference, e.g., from radar systems. Other values of $\rho$ are considered in Section~\ref{sec:results}-\ref{subsec:rho_results}. For $E_b/J_0=10$ and $15$~dB, we have verified that the ranking of the considered codes is preserved; the corresponding results are not reported for space reasons.

\subsection{Interleaving}
\label{subsec:interleaving}

Unless otherwise stated, no symbol interleaving is used, i.e., the coded symbols are transmitted in the order in which they appear in the codeword, so that each pulse hits $L_{\rm b}$ consecutive coded symbols. This is the main setting of this paper because, for \ac{SC-LDPC} codes decoded through \ac{SWD}, an interleaver would alter the correspondence between the temporal order of the received symbols and the spatial positions processed by the decoding window: an interleaver spanning several spatial positions would delay the decoding of each window, and an interleaver spanning the whole frame would remove the latency advantage of \ac{SWD}. As a reference, we also consider ideal interleaving. We do not model it through a specific permutation, but through an independent state allocation, in which each coded symbol is jammed with probability $\rho$, independently of the other symbols. An interleaver approaches this allocation only if it spans many jamming periods, i.e., many multiples of $L_{\rm p}$ coded symbols. With $L_{\rm p}=1280$, this is longer than the \ac{AR4JA} codewords with $k=1024$, which contain at most $2048$ coded symbols, and than one spatial position of the designed \ac{SC-LDPC} codes, which contains at most $1028$ coded symbols (see Table~\ref{tab:reception}). Hence, ideal interleaving would require interleaving across several codewords or spatial positions, at the cost of additional latency. Moreover, unlike a permutation of a fixed pulse pattern within one codeword, the independent allocation does not preserve the number of jammed symbols in each codeword. For these reasons, ideal interleaving is used only as a reference, and it is not shown in Fig.~\ref{fig:systemmodel}.

\section{PEXIT analysis for interleaved pulsed jamming}
\label{sec:pexitjam}

In this section, we develop a Gaussian-approximation \ac{PEXIT} analysis for the pulsed-jamming channel of Section~\ref{sec:preli}-\ref{subsec:pulsed} under the ideal interleaving of Section~\ref{sec:preli}-\ref{subsec:interleaving}, in which all variable-node types see the same asymptotic fractions of unjammed and jammed symbols. The resulting two-class recursion is written for generic type-dependent state probabilities, so that the same update rules also cover the \ac{AWGN} channel and the non-interleaved case of Section~\ref{sec:pexit_noninterleaved}.

\subsection{Gaussian-approximation PEXIT preliminaries}
\label{subsec:pexit_prelim}

Let us first recall the notation of the \ac{PEXIT} framework of Liva and Chiani~\cite{liva_pexit}, which we borrow for our analysis.
In a protograph, each node represents a \emph{node type}. When the protograph is lifted by a factor \(M\), every protograph node gives rise to \(M\) copies in the resulting Tanner graph, and all such copies are said to belong to the same type. Let
\[
    \mathbf{B}
    =
    [b_{i,j}]
    \in
    \mathbb{Z}_{\geq 0}^{N_{\rm c}\times N_{\rm v}}
\]
denote the base matrix used in the analysis, where \(b_{i,j}\) is the number of edges between check-node type \(i\) and variable-node type \(j\), and $N_{\rm c}$ and $N_{\rm v}$ denote the number of check-node types and the number of variable-node types, respectively. Whenever \(b_{i,j}>0\), the pair \((i,j)\) identifies an \emph{edge type} of the protograph. If \(b_{i,j}>1\), this edge type contains parallel edges.   For an uncoupled protograph, we have $N_{\rm c}=r_0$ and $N_{\rm v}=n_0$. For an \ac{SC-LDPC} ensemble, \(\mathbf{B}\) is the terminated coupled base matrix \(\mathbf{B}_{\mathrm{SC}}\) introduced in Section~\ref{sec:preli}, with $N_{\rm c}=(L+m_s)r_0$ and $N_{\rm v}=Ln_0$.

Following the original \ac{PEXIT} analysis~\cite{liva_pexit}, decoder messages are characterized through their \ac{MI}
under a Gaussian approximation. More precisely, each scalar message is represented by an equivalent Gaussian \ac{LLR}, and its reliability is tracked through the corresponding \ac{MI}. The function
\begin{equation}
\begin{split}
J(\sigma)
=
1-
\int_{-\infty}^{+\infty}
&
\frac{
\exp\!\left[
-\dfrac{(\xi-\sigma^2/2)^2}{2\sigma^2}
\right]
}{
\sqrt{2\pi\sigma^2}
}
\\
&\times
\log_2\!\left(1+e^{-\xi}\right)
\,\mathrm{d}\xi
\end{split}
\label{eq:J_function}
\end{equation}
gives the \ac{MI} associated with a Gaussian \ac{LLR} whose conditional distribution, given the transmitted symbol \(x=+1\), has mean \(\sigma^2/2\) and variance \(\sigma^2\) (the same holds for $x=-1$, with opposite sign of the mean and same variance). We also define
\begin{equation*}
    \psi(I)
    \triangleq
    \left[J^{-1}(I)\right]^2,
    \label{eq:psi_function}
\end{equation*}
which is the variance of the equivalent Gaussian \ac{LLR} having mutual information \(I\). Therefore, the \ac{PEXIT} recursion does not track the full message densities, but propagates \ac{MI} values along the edge types of the protograph.

Let \(I_{{\rm E}v}(i,j)\) and \(I_{{\rm E}c}(i,j)\) denote, respectively, the \ac{MI} of a variable-to-check and a check-to-variable message associated with edge type \((i,j)\).  For an \ac{AWGN} channel, all variable-node types receive the channel \ac{MI} $I_{\rm ch}=J(\sigma_{\rm ch})$, with $\sigma_{\rm ch}^{2}=8R\gamma$, where $R$ is the design rate of the ensemble and $\gamma=E_b/N_0$.\footnote{For a punctured variable-node type $j$, the node remains in the protograph recursion, but its channel mutual information is set to zero (in both channel states, in the two-state recursion of Section~\ref{sec:pexitjam}-\ref{subsec:pexit_pulsed}). The rate entering the channel normalization is the effective rate per transmitted symbol.} The conventional recursion~\cite{liva_pexit} starts from $I_{{\rm E}c}(i,j)=0$, alternates variable-node and check-node updates of $I_{{\rm E}v}(i,j)$ and $I_{{\rm E}c}(i,j)$ over the edge types, and declares convergence when the a posteriori \acp{MI} of all variable-node types approach~$1$. Its update rules are reported in Section~\ref{sec:pexitjam}-\ref{subsec:pexit_pulsed} in a two-state form, which reduces to the conventional recursion when all the coded symbols are in the same channel state. The threshold definition and its numerical computation are also given there.

\subsection{Two-state PEXIT recursion and homogeneous pulsed jamming}
\label{subsec:pexit_pulsed}

We first consider the pulsed-jamming channel under ideal \ac{JSI} at the receiver, assuming that the receiver knows which transmitted symbols are affected by jamming and the jamming variance in each state. Such state information can be obtained in practice from receiver-side energy or received-power measurements, following classical energy-detection principles for unknown signals in noise~\cite{urkowitz1967energy}, as also considered in measurement-based jamming detection approaches for wireless networks~\cite{xu2005feasibility}. The ideal-\ac{JSI} model is used as a benchmark for assessing the intrinsic robustness of the code and of the associated decoding metric, independently of the potential inaccuracies of practical \ac{JSI} estimation from the received signal, such as false alarms and missed detections.

Recall that, according to the pulsed-jamming model described in Section~\ref{sec:preli}-\ref{subsec:pulsed}, each transmitted coded symbol is either entirely unaffected by the jammer or entirely affected by it. Our main interest is the non-interleaved case, in which each pulse hits contiguous transmitted symbols and which is analyzed in Section~\ref{sec:pexit_noninterleaved}. In this section, we first consider the homogeneous case, in which every variable-node type has the same probability of being jammed, for two reasons. First, it describes the ideal-interleaving reference of Section~\ref{sec:preli}-\ref{subsec:interleaving}. Second, the two-state recursion derived for it is reused without changes in Section~\ref{sec:pexit_noninterleaved}, where the common state probabilities are replaced by type-dependent ones. In the homogeneous case, we do not preserve the deterministic time positions of the active intervals within the codeword. Instead, we retain only the fraction of jammed symbols and assume that the corresponding state labels are randomly mixed across the lifted codeword positions, as in the independent state allocation of Section~\ref{sec:preli}-\ref{subsec:interleaving}.

Let \(\zeta_{j,m}\in\{0,1\}\) denote the jamming state label assigned to the \(m\)-th lifted copy of variable-node type \(j\), with \(j=1,\ldots,N_{\rm v}\) and \(m=1,\ldots,M\). The value \(\zeta_{j,m}=0\) identifies an unjammed coded symbol, whereas \(\zeta_{j,m}=1\) identifies a jammed coded symbol. Under the homogeneous allocation, these labels are modeled as independent Bernoulli random variables with parameter \(\rho\), namely
\[
    \Pr\{\zeta_{j,m}=0\}=1-\rho,
    \qquad
    \Pr\{\zeta_{j,m}=1\}=\rho .
\]
If $N_{j,M}^{(1)}=\sum_{m=1}^{M}\zeta_{j,m}$ denotes the number of jammed lifted copies of variable-node type $j$, then, by the law of large numbers, $N_{j,M}^{(1)}/M\to\rho$ in probability as $M\to\infty$. Thus, asymptotically, every variable-node type contains a fraction \(1-\rho\) of unjammed copies and a fraction \(\rho\) of jammed copies.

Moreover, since the state labels are assigned independently of the random lift, the states of any fixed number of distinct variable-node copies appearing in a fixed-depth computation neighborhood are independent Bernoulli-\(\rho\) random variables. This justifies the homogeneous two-class description used in the PEXIT recursion described next, where all variable-node types experience the same asymptotic mixture of unjammed and jammed coded symbols.

We first compute the \ac{MI} that initializes the recursion. Conditioned on the state label $\zeta_{j,m}=s$, the projected observation normalized by $\sqrt{E_e}$ is $y_\eta=x_\eta+z_\eta^{(s)}$, with $x_\eta\in\{-1,+1\}$ and $z_\eta^{(s)}$ zero-mean Gaussian with variance $\sigma_{z,0}^2=N_0/(2E_e)$ for unjammed symbols and $\sigma_{z,1}^2=(N_0+J'_0)/(2E_e)$ for jammed symbols. With ideal \ac{JSI}, the receiver uses the state-matched \ac{LLR} $L_{\rm ch}^{(s)}=2y_\eta/\sigma_{z,s}^2$, which, conditioned on $x_\eta=+1$, is Gaussian with mean $2/\sigma_{z,s}^2$ and variance $4/\sigma_{z,s}^2$. By the convention in~\eqref{eq:J_function}, the channel \ac{MI} of the symbols in state $s$ is
\begin{align}
    I_{\rm ch}^{(s)}&=J\big(\sigma_{\rm ch}^{(s)}\big),\qquad s\in\{0,1\},
    \label{eq:pulsed_state_channel_mi}
    \end{align}
    with
    \begin{align}
    \big(\sigma_{\rm ch}^{(0)}\big)^2&=\frac{8E_e}{N_0}=8R\gamma,\qquad
    \big(\sigma_{\rm ch}^{(1)}\big)^2=\frac{8E_e}{N_0+J'_0}.
    \label{eq:pulsed_state_sigma}
\end{align}

Throughout the paper, the jammer strength is fixed through $\gamma_J\triangleq E_b/J_0$, while thresholds are computed over $\gamma$. With $J'_0=J_0/\rho$, \eqref{eq:pulsed_state_sigma} gives
\begin{equation}
    \left(\sigma_{\rm ch}^{(1)}\right)^2
    =
    \frac{8R\gamma}{1+\gamma/(\rho \gamma_J)},
    \qquad
    \lim_{\gamma\to\infty}\left(\sigma_{\rm ch}^{(1)}\right)^2
    =
    8R\rho \gamma_J ,
    \label{eq:pulsed_sigma_fixed_jammer}
\end{equation}
where the limit corresponds to vanishing thermal noise, i.e., $N_0\to0$ with $E_b$ and $J_0$ fixed. Hence, increasing $E_b/N_0$ at fixed $E_b/J_0$ does not remove the jammer: the reliability of the jammed symbols saturates, and some ensembles or profiles may never satisfy the convergence test in~\eqref{eq:pulsed_convergence}, in which case their threshold is $+\infty$.

We are now ready to present the two-state \ac{PEXIT} recursion. To reuse the same update rules in Section~\ref{sec:pexit_noninterleaved}, we let the probabilities of the two channel states depend on the variable-node type: $p_{j,0}$ and $p_{j,1}=1-p_{j,0}$ denote the fractions of unjammed and jammed copies of variable-node type $j$. Under the homogeneous allocation considered in this section,
\begin{equation}
    p_{j,0}=1-\rho,
    \qquad
    p_{j,1}=\rho,
    \qquad
    j=1,\ldots,N_{\rm v}.
    \label{eq:hom_state_probabilities}
\end{equation}

\begin{itemize}
    \item \emph{Initialization.}
    The state-dependent channel \acp{MI} are initialized as
    \[
        I_{\rm ch}^{(s)}
        =
        J\!\left(\sigma_{\rm ch}^{(s)}\right),
        \qquad
        s\in\{0,1\},
    \]
    where \(\sigma_{\rm ch}^{(0)}\) and \(\sigma_{\rm ch}^{(1)}\) are obtained by \eqref{eq:pulsed_state_sigma}. The iterative messages are initialized by setting
    \[
        I_{{\rm E}c}(i,j)=0
    \]
    for all edge types satisfying \(b_{i,j}>0\).

    \item \emph{Variable-to-check update.}
    Let \(I_{{\rm E}v}^{(s)}(i,j)\) denote the \ac{MI} of a variable-to-check message associated with edge type \((i,j)\), conditioned on the originating variable-node copy being in state
    \(s\). For every \(s\in\{0,1\}\) and every edge type \((i,j)\) satisfying \(b_{i,j}>0\), compute
   \begin{equation}
    \begin{split}
    I_{{\rm E}v}^{(s)}(i,j)
    =
    J\Bigg(
    \Bigg[
        &
        \sum_{i'=1}^{N_{\rm c}}
        b_{i',j}
        \psi\!\left(
            I_{{\rm E}c}(i',j)
        \right)
        \\
        &-
        \psi\!\left(
            I_{{\rm E}c}(i,j)
        \right)
        +
        \psi\!\left(
            I_{\rm ch}^{(s)}
        \right)
    \Bigg]^{1/2}
    \Bigg).
    \end{split}
    \label{eq:pulsed_vn_update}
\end{equation}
  \item \emph{State-mixture approximation.}
    The messages entering a check node along edge type $(i,j)$ originate from variable-node copies in either state, so their population is a two-component mixture with weights $p_{j,0}$ and $p_{j,1}$. With ideal \ac{JSI}, each component is the density of a true \ac{LLR}, i.e., a symmetric density $a$, satisfying $a(-\xi)=e^{-\xi}a(\xi)$, and, since the jamming state does not depend on the transmitted symbol, the mixture is also symmetric. For a symmetric density $a$, the \ac{MI} between the transmitted symbol and the message is $1-\int_{-\infty}^{+\infty}a(\xi)\log_2\left(1+e^{-\xi}\right)\mathrm{d}\xi$~\cite{RichardsonUrbanke2008}, which~\eqref{eq:J_function} evaluates for a Gaussian density, and it is therefore linear in $a$. Hence, the \ac{MI} of the mixture is
    \begin{equation}
        \overline{I}_{{\rm E}v}(i,j)=p_{j,0}I_{{\rm E}v}^{(0)}(i,j)+p_{j,1}I_{{\rm E}v}^{(1)}(i,j).
        \label{eq:pulsed_mi_average}
    \end{equation}
    To retain a scalar recursion, the mixture is represented by a Gaussian message with the same \ac{MI}, i.e., with parameter \(J^{-1}\!\left(\overline{I}_{{\rm E}v}(i,j)\right)\).

    \item \emph{Check-to-variable update.}
    For every edge type \((i,j)\) satisfying \(b_{i,j}>0\), compute
    \begin{equation}
    \begin{split}
    I_{{\rm E}c}(i,j)
    =
    1
    -
    J\Bigg(
    \Bigg[
        &
        \sum_{j'=1}^{N_{\rm v}}
        b_{i,j'}
        \psi\!\left(
            1-\overline{I}_{{\rm E}v}(i,j')
        \right)
        \\
        &-
        \psi\!\left(
            1-\overline{I}_{{\rm E}v}(i,j)
        \right)
    \Bigg]^{1/2}
    \Bigg).
    \end{split}
    \label{eq:pulsed_cn_update}
    \end{equation}

    \item \emph{APP mutual-information evaluation.}
    After each check-to-variable update, the state-conditioned a posteriori \ac{MI} associated with variable-node type \(j\) is computed as
    \begin{equation}
    \begin{split}
    I_{\rm APP}^{(s)}(j)
    =
    J\Bigg(
    \Bigg[
        &
        \sum_{i=1}^{N_{\rm c}}
        b_{i,j}
        \psi\!\left(
            I_{{\rm E}c}(i,j)
        \right)
        \\ & +
        \psi\!\left(
            I_{\rm ch}^{(s)}
        \right)
    \Bigg]^{1/2}
    \Bigg),
    \end{split}
    \label{eq:pulsed_app}
    \end{equation}
    for \(s\in\{0,1\}\).

    \item \emph{Convergence test.}
    Only state classes that occur with nonzero probability for variable-node type $j$, namely those in $\mathcal{A}_j\triangleq\{s\in\{0,1\}:p_{j,s}>0\}$, are included in the test. Successful convergence is declared when
    \begin{equation}
        \min_j
        \min_{s\in\mathcal{A}_j}
        I_{\rm APP}^{(s)}(j)
        \geq
        1-\epsilon
        \label{eq:pulsed_convergence}
    \end{equation}
    within at most \(I_{\max}\) iterations, where \(\epsilon>0\) is a numerical tolerance.
\end{itemize}

Notice that the check-to-variable message in \eqref{eq:pulsed_cn_update} is not conditioned on the state of the destination variable-node copy. Indeed, the extrinsic message sent along an edge is computed from the messages entering the other edges of the check node. Therefore, the same \(I_{{\rm E}c}(i,j)\) is supplied to both the unjammed and jammed copies of variable-node type \(j\).

Equations~\eqref{eq:pulsed_vn_update}--\eqref{eq:pulsed_app} define the two-state \ac{PEXIT} recursion. Besides the standard Gaussian approximation of each message population, its only additional approximation is the Gaussian representation of the state mixture in~\eqref{eq:pulsed_mi_average}. Their combined effect on the thresholds is assessed through Monte Carlo density evolution in Section~\ref{sec:designed}-\ref{subsec:single_profile_design}. With $p_{j,1}=0$ for all $j$, $\mathcal{A}_j=\{0\}$, and the recursion reduces to the conventional single-class recursion~\cite{liva_pexit} for the \ac{AWGN} channel with $I_{\rm ch}=I_{\rm ch}^{(0)}$. With $p_{j,1}=1$ for all $j$, i.e., $\rho=1$, it reduces to the single-class recursion for an \ac{AWGN} channel with one-sided \ac{PSD} $N_0+J_0$. With~\eqref{eq:hom_state_probabilities}, it is the homogeneous two-class recursion for interleaved pulsed jamming.

For fixed jammer parameters, a fixed ensemble, and fixed state probabilities $p_{j,s}$, the \emph{\ac{PEXIT} threshold}\footnote{The reason for using the infimum instead of the minimum is two-fold. First, with the convention $\inf\emptyset=+\infty$, the definition also covers the ensembles and profiles for which~\eqref{eq:pulsed_convergence} is never satisfied. Furthermore, the set of values of $\gamma$ for which~\eqref{eq:pulsed_convergence} is satisfied need not contain its lower endpoint, as for the conventional iterative-decoding threshold. Since thresholds are computed by bisection over $\gamma$, the distinction does not matter in the numerical results. The same applies to~\eqref{eq:nonint_profile_threshold} in Section~\ref{sec:pexit_noninterleaved}.} is
\begin{equation}
    \gamma_{\rm th}
    \triangleq
    \inf
    \left\{
        \gamma:
        \text{\eqref{eq:pulsed_convergence} is satisfied}
    \right\},
    \label{eq:generic_pexit_threshold}
\end{equation}
and we denote it by $\gamma_{\rm th}^{\rm hom}$ under~\eqref{eq:hom_state_probabilities}.

In our numerical evaluations, thresholds are computed by bisection over $\gamma$, with the jammer parameters fixed. Bisection applies because the success test is monotone in $\gamma$: the channel \acp{MI} are nondecreasing in $\gamma$, and all the \ac{PEXIT} updates preserve this ordering. Since~\eqref{eq:pulsed_convergence} is tested within $I_{\max}$ iterations, $\gamma_{\rm th}$ is an iteration-constrained threshold. For terminated \ac{SC-LDPC} chains, convergence is governed by the propagation of the decoding wave~\cite{Aref2013}, and the required number of iterations grows with $L$; the maximum number of iterations $I_{\max}=2000$ used in the design captures the coupling gain, as quantified in Section~\ref{sec:designed}-\ref{subsec:iteration_budget}. If the test is never satisfied, the threshold is $+\infty$.

\section{PEXIT analysis for pulsed jamming in the non-interleaved case}
\label{sec:pexit_noninterleaved}

We now remove the ideal-interleaving abstraction of Section~\ref{sec:pexitjam}-\ref{subsec:pexit_pulsed}. In principle, the resulting position-dependent description can be applied to any protograph-based ensemble, once a serialization of the lifted variable-node copies has been fixed. In this work, however, the focus is on terminated \ac{SC-LDPC} code ensembles, because the main contribution of the paper is their analysis and design under non-interleaved pulsed jamming. For most codes, a burst of jammed symbols is more harmful than the same number of jammed symbols scattered at random. With ideal \ac{JSI}, however, for codewords spanning many jamming periods, the capacity of the channel depends only on the fraction of jammed symbols and not on their positions, so that this penalty is due to the code and to the decoder, and can be reduced by the code design. In a coupled code, the bursts also hit specific spatial positions of the chain and therefore interact with the propagation of the decoding wave, which the analysis below accounts for.

Without symbol interleaving, the pulsed jammer affects contiguous portions of the transmitted sequence. Since the lifted copies of the variable-node types occupy specific transmitted-symbol positions, the fraction of jammed copies may vary from one variable-node type to another. For instance, under a fixed serialization, a pulse may overlap only the last symbols assigned to one
variable-node type, cover all the symbols assigned to one or more subsequent types, and then overlap only the first symbols assigned to another type. The goal of this section is therefore to determine a type-dependent jamming profile from the overlap between the pulse pattern and the serialized variable-node copies. The position-dependent \ac{PEXIT} recursion then replaces the global state weights $1-\rho$ and $\rho$ in \eqref{eq:hom_state_probabilities} with the corresponding type-dependent weights. The homogeneous recursion is recovered when every variable-node type has jammed fraction $\rho$.

\subsection{Finite-length and type-dependent state profiles}

Throughout this section, $\mathbf{B}_{\rm SC}=[b_{i,j}]\in\mathbb{Z}_{\geq 0}^{N_{\rm c}\times N_{\rm v}}$ is the terminated coupled base matrix used in the position-dependent \ac{PEXIT} recursion, with $N_{\rm v}=Ln_0$ variable-node types and $N_{\rm c}=(L+m_s)r_0$ check-node types (see Section~\ref{sec:pexitjam}-\ref{subsec:pexit_prelim}). Thus, $j\in\{1,\ldots,N_{\rm v}\}$ is a global variable-node-type index, related to the spatial position $t\in\{0,\ldots,L-1\}$ and the local variable-node type $a\in\{1,\ldots,n_0\}$ through $j=tn_0+a$.

For lifting factor $M$, let $v_{j,m}$ denote the $m$-th copy of variable-node type $j$, with $m\in\{1,\ldots,M\}$, so that the transmitted codeword has length $n=MN_{\rm v}=LMn_0$. A deterministic bijection $\varphi_M:\{1,\ldots,N_{\rm v}\}\times\{1,\ldots,M\}\to\{0,\ldots,n-1\}$ specifies the serialization order, i.e., $\varphi_M(j,m)$ is the transmitted-symbol position assigned to copy $v_{j,m}$. In this work, we use the type-contiguous serialization
\begin{equation}
    \varphi_M(j,m)
    =
    (j-1)M+(m-1),
    \qquad
    j=1,\ldots,N_{\rm v},
    \label{eq:nonint_contiguous_serialization}
\end{equation}
so that the transmitted positions occupied by variable-node type $j$ form the set $\mathcal{P}_{j,M}=\{(j-1)M,\ldots,jM-1\}$. The analysis below holds for any bijection $\varphi_M$; the type-contiguous serialization~\eqref{eq:nonint_contiguous_serialization} is used throughout the paper.

We now describe the non-interleaved pulsed-jamming pattern in the symbol domain, where, consistently with Section~\ref{sec:preli}-\ref{subsec:pulsed}, $L_{\rm p}=T/T_e$ and $L_{\rm b}=D/T_e$ are the number of coded symbols in one jamming period and in one active pulse, respectively.

\begin{definition}
Let $Q_M\triangleq\lceil n/L_{\rm p}\rceil$ be the number of jamming periods intersecting the considered codeword, and let $o_q$, $q\in\{0,\ldots,Q_M-1\}$, be the pulse offsets, expressed in coded-symbol intervals, defined in Section~\ref{sec:preli}-\ref{subsec:pulsed}. The vector $\mathbf{o}=(o_0,\ldots,o_{Q_M-1})$ specifies one realization of the pulse locations over the considered codeword.
\end{definition}

In the \emph{random-offset model}, the entries of $\mathbf{o}$ are random variables.\footnote{A natural choice is to take them independent and uniformly distributed over $\{0,\ldots,L_{\rm p}-L_{\rm b}\}$.}

\begin{definition}
\label{def:nonint_profile}
For a given realization $\mathbf{o}$, let $\chi_M(\eta;\mathbf{o})=1$ if transmitted position $\eta\in\{0,\ldots,n-1\}$ is jammed, i.e., if there exists $q\in\{0,\ldots,Q_M-1\}$ such that $qL_{\rm p}+o_q\leq\eta<qL_{\rm p}+o_q+L_{\rm b}$, and $\chi_M(\eta;\mathbf{o})=0$ otherwise. The corresponding jammed support is $\mathcal{J}_M(\mathbf{o})\triangleq\{\eta:\chi_M(\eta;\mathbf{o})=1\}$, and the state of copy $v_{j,m}$ is $\chi_M(\varphi_M(j,m);\mathbf{o})$. The empirical jammed fraction of variable-node type $j$ is
\begin{equation}
    \rho_{j,M}(\mathbf{o})
    \triangleq
    \frac{
        \left|
            \mathcal{P}_{j,M}
            \cap
            \mathcal{J}_M(\mathbf{o})
        \right|
    }{M}.
    \label{eq:nonint_empirical_fraction}
\end{equation}
\end{definition}

The empirical profile in~\eqref{eq:nonint_empirical_fraction} is evaluated at a design lifting factor, denoted by $M_{\rm d}$. For a pulse-offset realization $\mathbf{o}$, we define
\begin{equation}
    \widehat{\rho}_j(\mathbf{o})
    \triangleq
    \rho_{j,M_{\rm d}}(\mathbf{o}),
    \qquad
    j=1,\ldots,N_{\rm v},
    \label{eq:nonint_design_profile}
\end{equation}
and collect these values in
\begin{equation}
    \widehat{\boldsymbol{\rho}}(\mathbf{o})
    =
    \left(
        \widehat{\rho}_1(\mathbf{o}),
        \ldots,
        \widehat{\rho}_{N_{\rm v}}(\mathbf{o})
    \right).
    \label{eq:nonint_design_profile_vector}
\end{equation}
The vector $\widehat{\boldsymbol{\rho}}(\mathbf{o})$ describes the fractions of jammed copies induced by the physical pulse pattern, the adopted serialization, and the lifting scale considered in the code design.

Once $\widehat{\boldsymbol{\rho}}(\mathbf{o})$ has been obtained, it is kept fixed in the ensemble analysis. Consider a sequence of random protograph lifts of size $\widetilde{M}$ in which variable-node type $j$ contains
\[
    \left\lfloor
        \widetilde{M}
        \widehat{\rho}_j(\mathbf{o})
    \right\rfloor
\]
jammed copies and the remaining copies are unjammed. For each variable-node type, the set of jammed copy indices is fixed independently of the random permutations defining the lift. Owing to the exchangeability of the lifted copy indices, only the cardinality of this set enters the local analysis. As $\widetilde{M}\rightarrow\infty$, a fixed-depth computation neighborhood samples the two channel states with probabilities
\begin{equation}
    p_{j,0}
    =
    1-\widehat{\rho}_j(\mathbf{o}),
    \qquad
    p_{j,1}
    =
    \widehat{\rho}_j(\mathbf{o}).
    \label{eq:nonint_profile_probabilities}
\end{equation}
The meaning of the lifting variable $\widetilde{M}$ used in this ensemble is different from the design lifting factor $M_{\rm d}$ used to obtain the physical overlap profile.

The asymptotic limit thus concerns the random lifting, with the profile fixed at its value for the design lifting factor $M_{\rm d}$. This is the relevant regime for code design: the profile captures the physical overlap between pulses and variable-node types at the lifting factor of the finite-length code, whereas letting the physical lifting factor grow with a fixed jamming period would progressively turn every profile into the homogeneous one.

\subsection{Position-dependent PEXIT recursion}
\label{subsec:nonint_pexit_recursion}

Let us apply the two-state \ac{PEXIT} recursion of Section~\ref{sec:pexitjam}-\ref{subsec:pexit_pulsed} to the type-dependent profile $\widehat{\boldsymbol{\rho}}(\mathbf{o})$ defined in~\eqref{eq:nonint_design_profile_vector}, i.e., with the state probabilities~\eqref{eq:nonint_profile_probabilities}.
The following lemma explains why the deterministic non-interleaved
state sequence can be represented, at the level of a local computation
neighborhood, by type-dependent state probabilities.

\begin{lemma}
\label{lem:nonint_sampling}
Consider the fixed-depth computation neighborhood rooted at a lifted edge chosen uniformly at random among the $\widetilde{M}$ copies of a fixed protograph edge type, and condition on this neighborhood being tree-like. Let $\kappa$ be the number of distinct variable-node copies appearing in the neighborhood, let $j_1,\ldots,j_{\kappa}$ be their variable-node types, and let $S_h^{(\widetilde{M})}$ denote the channel state of the $h$-th copy. For any $s_1,\ldots,s_{\kappa}\in\{0,1\}$,
\begin{equation}
\begin{split}
    &
    \Pr\!\left\{
        S_1^{(\widetilde{M})}=s_1,\ldots,
        S_{\kappa}^{(\widetilde{M})}=s_{\kappa}
        \,\middle|\,
        j_1,\ldots,j_{\kappa},
        \widehat{\boldsymbol{\rho}}(\mathbf{o})
    \right\}
    \\
    &\qquad =
    \prod_{h=1}^{\kappa}
    p_{j_h,s_h}
    +
    O\!\left(
        \frac{\kappa^2}{\widetilde{M}}
    \right).
\end{split}
\label{eq:nonint_state_sampling}
\end{equation}
Hence, for every fixed computation depth, the channel-state labels in a local tree-like neighborhood become asymptotically independent, with probabilities determined by the type-dependent profile.
\end{lemma}

The proof is given in the Appendix.

Lemma~\ref{lem:nonint_sampling} connects the time-domain overlap pattern to the protograph recursion. Although the non-interleaved state sequence is structured in time, the random lift samples it locally according to the type-dependent probabilities $p_{j,0}$ and $p_{j,1}$. Therefore, the same \ac{MI}-matched state-mixture approximation used in Section~\ref{sec:pexitjam}-\ref{subsec:pexit_pulsed} can be applied separately to each variable-node type.

Lemma~\ref{lem:nonint_sampling} requires the set of jammed copies of each type to be independent of the lifting permutations. This holds for the randomized lifting adopted in Section~\ref{sec:design}-\ref{subsec:finite_length_lifting}, in which every lifted block is relabeled by random row and column permutations: a burst covering contiguous copies of one variable-node type is scattered over the copies of the adjacent check-node types, so that the random lift acts as an interleaver within each block of $M$ symbols. The profile $\widehat{\boldsymbol{\rho}}(\mathbf{o})$ therefore captures the spatial localization of the bursts along the coupled chain, i.e., across variable-node types and spatial positions, which is the feature that interacts with the decoding wave. For structured liftings, such as circulant liftings or the permutations of the \ac{CCSDS} \ac{AR4JA} codes, contiguous copies of a variable-node type are connected to contiguous copies of the adjacent check-node types, and the local state statistics also depend on the lifting structure.

The position-dependent \ac{PEXIT} recursion for the non-interleaved pulsed-jamming ensemble with profile $\widehat{\boldsymbol{\rho}}(\mathbf{o})$ is therefore the recursion~\eqref{eq:pulsed_vn_update}--\eqref{eq:pulsed_convergence}, run on $\mathbf{B}_{\rm SC}$ with ideal \ac{JSI} and with the type-dependent state probabilities~\eqref{eq:nonint_profile_probabilities}. These probabilities enter the state averaging~\eqref{eq:pulsed_mi_average} and determine the set $\mathcal{A}_j$ of state classes included in the convergence test~\eqref{eq:pulsed_convergence}: a partially jammed variable-node type, with $0<\widehat{\rho}_j(\mathbf{o})<1$, contributes both state classes, whereas a clean or fully jammed type contributes only one state class. As in the homogeneous case, the check-to-variable message is not conditioned on the state of the destination variable-node copy: by Lemma~\ref{lem:nonint_sampling}, the states of the other variable-node copies connected to the same check node are asymptotically independent of the destination state.

The above recursion analyzes flooding decoding over the entire terminated chain, including both terminated boundaries, and the resulting full-chain thresholds are used as the design objective. The full-chain threshold is the asymptotic target of \ac{SWD}, since window-decoding thresholds approach it as the window size grows~\cite{Iyengar2012}; the finite-length performance under \ac{SWD} with $W=5(m_s+1)$ is assessed in Section~\ref{sec:results}.
For fixed jammer parameters and a fixed type-dependent profile $\widehat{\boldsymbol{\rho}}(\mathbf{o})$, the corresponding \ac{PEXIT} threshold is
\begin{equation}
    \gamma_{\rm th}
    \left(
        \widehat{\boldsymbol{\rho}}(\mathbf{o})
    \right)
    \triangleq
    \inf
    \left\{
        \gamma:
        \text{\eqref{eq:pulsed_convergence} is satisfied}
    \right\},
    \label{eq:nonint_profile_threshold}
\end{equation}
where, as usual, \(\gamma=E_b/N_0\). Thus, in the non-interleaved case, the threshold is determined not only by the nominal duty cycle, but also by the way in which the pulse support overlaps the serialized variable-node types.

If the pulse offsets are random, then $\widehat{\boldsymbol{\rho}}(\mathbf{O})$, where $\mathbf{O}$ denotes the random offset vector, is a random profile, and $\gamma_{\rm th}(\widehat{\boldsymbol{\rho}}(\mathbf{O}))$ is a random threshold.  Section~\ref{sec:design} summarizes the upper tail of the random threshold over a finite set of sampled profiles, and Section~\ref{sec:designed}-\ref{subsec:iteration_budget} shows that, at the operating point of Section~\ref{sec:preli}-\ref{subsec:pulsed} and with $I_{\max}=2000$, all ten validation profiles of the designed rate-$1/2$ ensemble have thresholds above the homogeneous reference.

\section{PEXIT-based design of SC-LDPC code ensembles}
\label{sec:design}

In this section, we use the position-dependent \ac{PEXIT} analysis of Section~\ref{sec:pexit_noninterleaved} to design protograph-based \ac{SC-LDPC} code ensembles for the general case of non-interleaved pulsed jamming. Recall that the design is performed at the ensemble level. The method is described here for a generic set of design profiles, without numerical choices; its two instances, for a single deterministic pulse pattern and for random pulse realizations, and the resulting ensembles are presented in Section~\ref{sec:designed}.

\subsection{Admissible design space}
\label{subsec:design_space}

The design is performed for a fixed target rate and for a fixed chain length \(L\). The remaining structural parameters are selected from given finite sets. A candidate ensemble is specified by
\[
    \mathcal{C}
    =
    \left(
        r_0,
        n_0,
        m_s,
        \mathbf{B}_0,\ldots,\mathbf{B}_{m_s}
    \right),
\]
where \(r_0\) and \(n_0\) are the number of check-node types and the number of variable-node types in the uncoupled protograph, respectively, and \(m_s\) is the syndrome former memory order, i.e., the coupling memory of Section~\ref{sec:preli}-\ref{subsec:SC-LDPC}. The induced uncoupled base matrix is
\[
    \mathbf{B}
    =
    \sum_{\ell=0}^{m_s}
    \mathbf{B}_{\ell}.
\]

The admissible set \(\mathcal{F}\) is defined by imposing rate, degree, parallel-edge, and coupling-spread constraints. The terminated design rate, i.e., the rate $R$ of Section~\ref{sec:preli}-\ref{subsec:SC-LDPC} as a function of the candidate, is
\[
    R_L(\mathcal{C})
    =
    1-
    \frac{
        r_0(L+m_s)
    }{
        n_0L
    },
\]
and must satisfy
\begin{equation}
    \left|
        R_L(\mathcal{C})
        -
        R_{\rm des}
    \right|
    \leq
    \Delta_R,
    \label{eq:design_rate_constraint}
\end{equation}
where \(R_{\rm des}\) is the target uncoupled rate and \(\Delta_R=0.06\) is the allowed deviation of the terminated design rate from \(R_{\rm des}\).\footnote{In the search considered in this work, this constraint is nonbinding and acts only as a consistency check: since $n_0$ is selected such that $1-r_0/n_0=R_{\rm des}$, the rate loss is $R_{\rm des}-R_L(\mathcal{C})=m_s(1-R_{\rm des})/L\leq 3/146\approx0.021<\Delta_R$ for $L=73$, $m_s\leq3$, and $R_{\rm des}\geq1/2$.}

The variable-node degrees $\sum_{i=1}^{r_0}b_{i,j}$ and the check-node degrees $\sum_{j=1}^{n_0}b_{i,j}$ of $\mathbf{B}$ must lie in $[d_{v,\min},d_{v,\max}]$ and $[d_{c,\min},d_{c,\max}]$, respectively. Excessive parallelism is limited at the component-matrix level by requiring $[\mathbf{B}_{\ell}]_{i,j}\leq b_{\max}$. Finally, to prevent the edges incident to a variable-node type from being too localized in the coupled chain, we require them to occupy at least $s_{\min}$ distinct component matrices. The set of all candidates satisfying these constraints is denoted by $\mathcal{F}$. In the searches considered here, $s_{\min}=2$, which excludes candidates whose edges are all concentrated in a single component matrix. The selected ensembles also have nonzero entries at their largest given coupling shift, so their effective coupling memory coincides with $m_s$.

\subsection{Training profiles}
\label{subsec:design_training_profiles}

For a candidate \(\mathcal{C}\), the position-dependent \ac{PEXIT} recursion is run on the terminated coupled base matrix \(\mathbf{B}_{\rm SC}\). Therefore, the state profiles used in the design have length \(Ln_0\), equal to the number of variable-node columns of \(\mathbf{B}_{\rm SC}\). The design is performed over a training set
\[
    \mathcal{P}_{\rm tr}
    =
    \left\{
        \widehat{\boldsymbol{\rho}}^{(1)},
        \ldots,
        \widehat{\boldsymbol{\rho}}^{(N_{\rm p})}
    \right\},
\]
where each profile is obtained by computing, according to Section~\ref{sec:pexit_noninterleaved}, the overlap fractions induced by one pulse-offset vector $\mathbf{o}^{(r)}$. The choice of the offset vectors depends on the jamming scenario. If the pulse positions are not known in advance, the offsets are drawn at random and $N_{\rm p}>1$ profiles are used, so that the design accounts for the variability of the decoding threshold across different pulse realizations. If the pulse pattern is deterministic and known at the design stage, a single profile is used, i.e., $N_{\rm p}=1$ and $\widehat{\boldsymbol{\rho}}^{(1)}=\widehat{\boldsymbol{\rho}}(\mathbf{o})$ for the given offset vector $\mathbf{o}$. For a candidate $\mathcal{C}$ and a profile $\widehat{\boldsymbol{\rho}}^{(r)}$, let
\begin{equation}
    \gamma_r(\mathcal{C})
    \triangleq
    \gamma_{\rm th}
    \left(
        \mathcal{C};
        \widehat{\boldsymbol{\rho}}^{(r)}
    \right)
    \label{eq:design_profile_threshold}
\end{equation}
denote the threshold obtained from the position-dependent \ac{PEXIT} recursion.

\subsection{Design objective}
\label{subsec:design_objective}

The design objective is defined over the training profiles in $\mathcal{P}_{\rm tr}$. The most conservative metric would be the worst-profile threshold $\Gamma_{\rm wc}(\mathcal{C})=\max_{r}\gamma_r(\mathcal{C})$, which directly targets the most damaging training profile. However, a pure maximum can be overly sensitive to a single sampled profile realization and to small numerical fluctuations in the threshold computation. We therefore use a tail-average objective.\footnote{The motivation is that unfavorable pulse realizations may contribute disproportionately to the average BER. Focusing on the largest thresholds aims to improve robustness against these realizations without making the design depend on a single worst case.} Let $\gamma_{(1)}(\mathcal{C})\leq\cdots\leq\gamma_{(N_{\rm p})}(\mathcal{C})$ be the sorted values of $\gamma_1(\mathcal{C}),\ldots,\gamma_{N_{\rm p}}(\mathcal{C})$, and let $N_{\beta}=\max\{1,\lceil\beta N_{\rm p}\rceil\}$, with $\beta\in(0,1]$.
The tail-average objective is
\begin{equation}
    \Gamma_{\beta}(\mathcal{C})
    =
    \frac{1}{N_{\beta}}
    \sum_{r=N_{\rm p}-N_{\beta}+1}^{N_{\rm p}}
    \gamma_{(r)}(\mathcal{C}).
    \label{eq:design_tail_average_objective}
\end{equation}
Thus, \(\Gamma_{\beta}(\mathcal{C})\) averages the largest
\(N_{\beta}\) profile-dependent thresholds. In \eqref{eq:design_tail_average_objective}, the thresholds are expressed in dB, i.e., $\Gamma_\beta$ is the arithmetic mean of the $N_\beta$ largest thresholds in dB, which is equivalent to the geometric mean of the corresponding linear values. A candidate that fails the convergence test at the upper endpoint of the search interval for at least one training profile is assigned the penalty $\Gamma_\beta=+\infty$.  For small \(\beta\), the criterion remains concentrated on the most demanding profiles while reducing the sensitivity of the optimization to a single sampled realization. For a single deterministic profile, $N_{\rm p}=1$ gives $N_\beta=1$ for every $\beta$, and~\eqref{eq:design_tail_average_objective} reduces to the profile threshold, $\Gamma_\beta(\mathcal{C})=\gamma_1(\mathcal{C})=\gamma_{\rm th}(\mathcal{C};\widehat{\boldsymbol{\rho}}^{(1)})$; no further modification of the method is needed. The values of $N_{\rm p}$ and $\beta$ used in this work are reported in Section~\ref{sec:designed}-\ref{subsec:design_setup}. The design problem is
\begin{equation}
    \mathcal{C}^{\star}
    =
    \arg\min_{\mathcal{C}\in\mathcal{F}}
    \Gamma_{\beta}(\mathcal{C}).
    \label{eq:design_optimization_problem}
\end{equation}

\subsection{Constrained differential-evolution search}
\label{subsec:design_de}

The optimization problem in \eqref{eq:design_optimization_problem} is a constrained, discrete, and nonconvex problem. The structural parameters are selected from finite sets, the component matrices have integer entries, and the objective function is obtained through a \ac{PEXIT} threshold computation. Hence, the objective is not available in closed form. We solve the problem through a constrained differential-evolution search with deterministic repair.

The search is performed over predetermined finite sets for $r_0$, $n_0$, and $m_s$, while $L$ is fixed before each design run. A structural class is identified by $\mathcal{S}=(r_0,n_0,m_s)$, and it is retained only if its terminated design rate satisfies~\eqref{eq:design_rate_constraint}. Within each class, the check-node degrees are allowed to vary between $d_{c,\min}$ and $d_{c,\max}$, independently for each check-node type. Consequently, the total number of uncoupled protograph edges is also a design variable. Its admissible range is restricted by
$E_{\min}=\max\{r_0d_{c,\min},n_0\max(d_{v,\min},s_{\min})\}$ and $E_{\max}=\min\{r_0d_{c,\max},n_0d_{v,\max},r_0n_0(m_s+1)b_{\max}\}$. 
These bounds are necessary but not sufficient for feasibility, which is checked after the deterministic repair procedure.

For a fixed retained structural class, the differential-evolution search optimizes the placement of the \(E\) edges among the component matrices. The \(e\)-th edge is represented by the edge-placement tuple \((i_e,j_e,\ell_e),\) where \(i_e\) is the check-node type, \(j_e\) is the variable-node type,
and \(\ell_e\in\{0,\ldots,m_s\}\) is the coupling shift. The tuple \((i_e,j_e,\ell_e)\) increments the entry \([\mathbf{B}_{\ell_e}]_{i_e,j_e}\) by one. After all edge-placement tuples have been counted, the component matrices $\mathbf{B}_0,\ldots,\mathbf{B}_{m_s}$ and the uncoupled base matrix $\mathbf{B}=\sum_{\ell}\mathbf{B}_{\ell}$ are obtained.

The differential-evolution population consists of real vectors $\mathbf{x}^{(u)}\in[0,1]^{1+3E_{\max}}$, $u=1,\ldots,N_{\rm pop}$. The first coordinate selects the number of active edges $E\in\{E_{\min},\ldots,E_{\max}\}$ by uniform quantization. The next $3E$ coordinates are grouped into triples $(x^{(u)}_{3e-1},x^{(u)}_{3e},x^{(u)}_{3e+1})$, $e=1,\ldots,E$, which are mapped to the integer edge-placement tuples $(i_e,j_e,\ell_e)$ by uniform quantization into $r_0$, $n_0$, and $m_s+1$ levels, respectively. The other coordinates are inactive for the current candidate but remain part of the differential-evolution representation. Each entry $[\mathbf{B}_{\ell}]_{i,j}$ counts the active tuples with $(i_e,j_e,\ell_e)=(i,j,\ell)$.

After this deterministic mapping, the candidate might violate one or more of the constraints of Section~\ref{sec:design}-\ref{subsec:design_space}. A deterministic repair step, which operates directly on the integer edge counts of $\mathbf{B}_0,\ldots,\mathbf{B}_{m_s}$, is therefore applied before evaluating the candidate. Since the repair does not modify $r_0$, $n_0$, $m_s$, or $L$, a candidate whose structural class violates the rate constraint, or for which $s_{\min}$ cannot be satisfied with memory $m_s$, is rejected. Otherwise, entries exceeding $b_{\max}$ are clipped; variable-node and then check-node degree excesses are removed, without violating the lower degree bound of the adjacent node type or the coupling-spread condition; variable-node and then check-node degree deficits are corrected by adding edges in positions that respect the maximum degree and parallel-edge bounds, preferring unused coupling shifts; finally, for each variable-node type with fewer than $s_{\min}$ active shifts, an edge is added in an unused shift if degree capacity is available, or an existing edge is moved to an unused shift, while preserving all degree and parallel-edge constraints. The repaired candidate is checked again against all constraints and is passed to the \ac{PEXIT} evaluation only if it is feasible; otherwise, it is assigned an infeasible objective value.

For each retained structural class, we apply the standard DE/rand/1/bin strategy~\cite{StornPrice1997} to the real-valued representation of the candidate ensemble. Each trial vector is mapped to integer edge-placement tuples, repaired as described above, and accepted only if it is feasible and yields a lower value of $\Gamma_{\beta}$ than the corresponding parent candidate.

The numerical values of \(N_{\rm pop}\), the number of generations $N_{\rm gen}$, \(F_{\rm DE}\), \(C_{\rm DE}\), and the \ac{PEXIT} bisection parameters are reported in Section~\ref{sec:designed}-\ref{subsec:design_setup}.

\subsection{Candidate evaluation and final selection}
\label{subsec:design_candidate_evaluation}

Each feasible candidate generated by the optimizer is evaluated by first constructing the corresponding terminated coupled base matrix \(\mathbf{B}_{\rm SC}\). Then, for every training profile \(\widehat{\boldsymbol{\rho}}^{(r)}\in\mathcal{P}_{\rm tr}\), the position-dependent \ac{PEXIT} recursion of Section~\ref{sec:pexit_noninterleaved} is run, and the threshold \(\gamma_r(\mathcal{C})\) is computed by bisection over \(\gamma=E_b/N_0\). The resulting thresholds are combined according to \eqref{eq:design_tail_average_objective}.

For each target rate, the differential-evolution search is performed independently over six structural classes. The initial population combines randomly generated candidates with a feasible reference candidate and three local perturbations of it in its structural class. After the search, the three candidates with the lowest finite training objectives in each class, together with all additional candidates within $0.05$~dB of the class-best training objective, are retained for the final selection. All retained candidates and the reference candidate are evaluated on a selection set $\mathcal{P}_{\rm sel}$, and, for each rate, the candidate with the lowest objective~\eqref{eq:design_tail_average_objective} over $\mathcal{P}_{\rm sel}$ is selected. With random pulse offsets, $\mathcal{P}_{\rm sel}=\mathcal{P}_{\rm val}$ is a validation set, independent of the training set, of $N_{\rm val}$ profiles, generated according to the same non-interleaved pulse model but using independent pulse offsets, so that the selected candidate is not tuned to the particular training realizations. With a deterministic pulse pattern, the training set contains only the design profile, and no independent validation is needed, since the jamming pattern seen by the code is the one used in the design. The selection set therefore coincides with the training set, $\mathcal{P}_{\rm sel}=\mathcal{P}_{\rm tr}$, and the selected candidate is the one with the lowest threshold on the design profile.

\subsection{Finite-length lifting}
\label{subsec:finite_length_lifting}

After the ensemble-level design, the selected terminated coupled base matrix $\mathbf{B}_{\rm SC}^{\star}=[b_{i,j}^{\rm SC}]$ is instantiated using the lifting factor $M=M_{\rm d}^{\star}$ associated with the selected structural class; the present stage selects the permutation matrices defining one finite-length Tanner graph. If $b_{i,j}^{\rm SC}=0$, the corresponding $M\times M$ block $\mathbf{H}_{i,j}$ of the lifted parity-check matrix is the all-zero matrix. If $b_{i,j}^{\rm SC}>0$, then $\mathbf{H}_{i,j}=\sum_{a=1}^{b_{i,j}^{\rm SC}}\mathbf{P}_{i,j}^{[a]}$, where the $\mathbf{P}_{i,j}^{[a]}$ are $M\times M$ permutation matrices with pairwise disjoint supports. This condition maps parallel protograph edges to distinct lifted edges and ensures that every lifted block remains binary.

The disjoint permutation sets are generated through a randomized Latin-rectangle construction. Let \(\mathbf{\Lambda}_{\delta}\) denote the cyclic-shift permutation matrix with shift \(\delta\). For a block with \(b_{i,j}^{\rm SC}>0\), distinct shifts \(\delta_1,\ldots,\delta_{b_{i,j}^{\rm SC}}\) are selected, and a candidate set of permutation matrices is generated as
\[
\mathbf{P}_{i,j}^{[a]}=\mathbf{\Pi}_{i,j}^{\rm row}\mathbf{\Lambda}_{\delta_a}\mathbf{\Pi}_{i,j}^{\rm col}, \qquad a=1,\ldots,b_{i,j}^{\rm SC},
\]
where \(\mathbf{\Pi}_{i,j}^{\rm row}\) and \(\mathbf{\Pi}_{i,j}^{\rm col}\) are randomly selected row and column permutation matrices, common to all parallel edges in the same lifted block. Distinct cyclic shifts have disjoint supports, and common row and column relabelings preserve this property. Hence, the construction guarantees non-overlapping permutation matrices without rejection sampling. For \(b_{i,j}^{\rm SC}\geq 3\), any three distinct cyclic shifts in the same lifted block necessarily form length-\(6\) cycles, which are preserved by the common row and column relabelings. This issue does not arise for the codes reported in Section~\ref{sec:results}, whose coupled base matrices have entries at most two. Although each individual permutation block is obtained from a cyclic shift through row and column relabeling, the overall lift is not quasi-cyclic, since these relabelings are chosen independently across the nonzero blocks of \(\mathbf{B}_{\rm SC}^{\star}\). Consequently, there is no common cyclic indexing of the lifted node copies across the Tanner graph, and the copies of each type are scattered as required by Lemma~\ref{lem:nonint_sampling}.

The lifting is time-varying: independent candidate blocks are generated for the different nonzero entries of \(\mathbf{B}_{\rm SC}^{\star}\), including entries corresponding to different spatial positions. To improve the local Tanner-graph structure, several admissible candidate blocks are generated for each nonzero entry \(b_{i,j}^{\rm SC}\), and the candidate creating the smallest local short-cycle score is retained. The score is
\[
    \Omega
    =
    \lambda_4 N_{4}^{\rm new}
    +
    \lambda_6 N_{6}^{\rm new},
\]
where \(N_{4}^{\rm new}\) and \(N_{6}^{\rm new}\) denote, respectively, the number of new cycles of length four and the number of new cycles of length six that would be created by inserting the candidate block into the partially constructed Tanner graph. They are computed incrementally, by counting the existing paths of length three and five between the endpoints of each candidate lifted edge, so that only local neighborhoods are explored and no global cycle enumeration is required.

In the finite-length simulations, the algorithm tests \(N_{\rm cand}=32\) admissible candidate blocks for each nonzero lifted block. The weights are set to \(
    \lambda_4=10^6, \,
    \lambda_6=1,\)
so that avoiding newly created 4-cycles has dominant priority over reducing newly created 6-cycles. The lifted blocks are processed in randomized order.

\section{Design examples and selected ensembles}
\label{sec:designed}

In this section, the method of Section~\ref{sec:design} is applied with the numerical setup that will be described in Section~\ref{sec:designed}-\ref{subsec:design_setup}. We consider two design scenarios, depending on whether or not the pulse positions are known in advance. The first scenario assumes a fixed, known pulse pattern and serves primarily as a controlled benchmark for validating the proposed design methodology. By using the same jamming pattern in the design and finite-length simulations, we can assess whether the PEXIT thresholds correctly predict the relative performance of the codes, as done in Section~\ref{sec:results}-\ref{subsec:results_single}. This scenario also represents the case in which the interference pattern is known or predictable, allowing the code to be optimized specifically for it. In real settings, however, pulse positions may be unknown or vary between transmissions, making a design tailored to a single pattern less suitable. We therefore consider a second scenario in which the optimization accounts for multiple random pulse realizations, aiming to obtain codes that are robust to variations in the jamming pattern. The resulting ensembles constitute the main outcome of the proposed design and are evaluated under random pulsed jamming in Section~\ref{sec:results}-\ref{subsec:main_results}.

\subsection{Design setup}
\label{subsec:design_setup}

For reproducibility, the numerical parameters of the \ac{PEXIT}-based design are reported in Table~\ref{tab:design_setup}. The same search procedure is applied independently to the three target rates $R_{\rm des}\in\{1/2,2/3,4/5\}$. For each structural class, $n_0=r_0/(1-R_{\rm des})$, so that the uncoupled protograph rate satisfies $1-r_0/n_0=R_{\rm des}$. The common optimization parameters are kept fixed across rates, while the admissible check-node degree ranges and the target constraint lengths are selected according to the given code rate. The two design instances use the same parameters and differ only in the training and selection profiles, as indicated in the corresponding rows of Table~\ref{tab:design_setup}. All the reported results are obtained with $\beta=0.34$, i.e., $N_\beta=4$ for $N_{\rm p}=10$; we have also verified that any $\beta\in[0.34,1]$, from the average of the four largest thresholds to the average of all of them, leads to the same selected ensembles.

\begin{table}[t]
\centering
\caption{Parameters of the \ac{PEXIT}-based design.}
\label{tab:design_setup}
\footnotesize
\setlength{\tabcolsep}{4pt}
\begin{tabular}{@{}ll@{}}
\hline
Target rates & $R_{\rm des}\in\{1/2,2/3,4/5\}$, $L=73$\\
Structural classes & $r_0\in\{1,2\}$, $m_s\in\{1,2,3\}$\\
Variable-node degrees & $d_{v,\min}=2$, $d_{v,\max}=5$\\
Check-node degree range & $[5,9]$, $[7,11]$, $[12,18]$ for the three rates\\
Edge constraints & $b_{\max}=2$, $s_{\min}=2$\\
Rate tolerance & $\Delta_R=0.06$\\
Constraint-length target & $\nu_{\rm target}=2048,\ 1536,\ 1280$\\
Pulsed jammer & $L_{\rm p}=1280$, $L_{\rm b}=128$, $E_b/J_0=5$~dB\\
Profiles, random offsets & $N_{\rm p}=N_{\rm val}=10$\\
Profiles, single pattern & $N_{\rm p}=1$, selection on the same profile\\
Objective & $\Gamma_\beta$, $\beta=0.34$, $N_\beta=4$ (random), $1$ (fixed)\\
Differential evolution & $N_{\rm pop}=12$, $N_{\rm gen}=20$\\
Mutation and crossover & $F_{\rm DE}=0.7$, $C_{\rm DE}=0.8$\\
Initialization & Feasible seed, three local perturbations\\
Selection shortlist & Top three per class and $0.05$~dB margin\\
\ac{PEXIT} & $I_{\max}=2000$, APP target $1-10^{-3}$\\
Stagnation tolerance & $10^{-4}$\\
Bisection interval & $E_b/N_0\in[-1,10]$~dB\\
Bisection tolerance & $0.005$~dB\\
\hline
\end{tabular}
\end{table}

For each structural class $\mathcal{S}=(r_0,n_0,m_s)$, the lifting factor used during the design is selected according to
\begin{equation}
    M_{\rm d}(\mathcal{S})
    =
    \min
    \left\{
        p\in\mathbb{P}:
        p
        \geq
        \left\lceil
            \frac{
                \nu_{\rm target}
            }{
                (m_s+1)n_0
            }
        \right\rceil
    \right\},
    \label{eq:design_lifting_factor}
\end{equation}
where $\mathbb{P}$ denotes the set of prime numbers and $\nu_{\rm target}=n_{\rm AR4JA}(R_{\rm des})$ is the transmitted block length of the corresponding standardized \ac{AR4JA} code, equal to $2048$, $1536$, and $1280$ for the three target rates, respectively. The resulting constraint length is $\nu_s(\mathcal{S})=(m_s+1)n_0M_{\rm d}(\mathcal{S})$.

In the random-offset instance, the pulse offsets of the training and validation profiles are generated independently and uniformly over $\{0,\ldots,L_{\rm p}-L_{\rm b}\}$, using independent realizations for the two sets. For each structural class and each pulse-offset vector, random or deterministic, $\mathbf{o}^{(r)}$, the profile is evaluated at $M_{\rm d}(\mathcal{S})$, i.e., $\widehat{\rho}_j^{(r)}=\rho_{j,M_{\rm d}(\mathcal{S})}(\mathbf{o}^{(r)})$, $j=1,\ldots,Ln_0$. Since $L_{\rm p}$ and $L_{\rm b}$ are fixed in coded-symbol intervals, the physical period $T=L_{\rm p}T_e$ and pulse duration $D=L_{\rm b}T_e$ change with the code rate when the information bit rate is kept fixed. The comparison across rates is therefore carried out for a fixed symbol-domain jamming pattern, not for a fixed physical pulse duration.\footnote{For reproducibility, the profile-generation base seed is $4200$, with offsets $17$ and $100017$ for the training and validation random streams, respectively. The differential-evolution base seed is $42$, incremented by the structural-class index minus one. Population updates are performed sequentially, and mutant coordinates outside $[0,1]$ are mapped back into the interval by reflection.}

With the search interval and tolerance of Table~\ref{tab:design_setup}, bisection stops after twelve steps, when the residual interval has width $11/2^{12}\approx0.00269$~dB; its upper end is returned. Besides the limit $I_{\max}=2000$, the recursion is stopped when the largest change of $I_{{\rm E}c}(i,j)$ over all edge types between two consecutive iterations is below $10^{-4}$, and the convergence test~\eqref{eq:pulsed_convergence} is then applied.

\subsection{Design for a single deterministic pulse pattern}
\label{subsec:single_profile_design}

In the first instance, the jamming pattern is deterministic and known at the design stage. The pulse offsets are fixed to the vector $\mathbf{o}^{\star}=(o^{\star}_0,\ldots,o^{\star}_{Q_M-1})$ listed in Table~\ref{tab:single_profile_offsets}, and the same pattern is used in the finite-length simulations of Section~\ref{sec:results}-\ref{subsec:results_single}. As in Section~\ref{sec:preli}-\ref{subsec:pulsed}, the frame starts at the beginning of a jamming period, so that the $q$-th pulse occupies the coded symbols $qL_{\rm p}+o^{\star}_q,\ldots,qL_{\rm p}+o^{\star}_q+L_{\rm b}-1$. The offsets are a single realization of independent, uniformly distributed offsets over $\{0,\ldots,L_{\rm p}-L_{\rm b}\}$, which is then kept fixed.  The frames of the selected ensembles span $Q_M=40$, $23$, and $20$ jamming periods at rates $1/2$, $2/3$, and $4/5$, respectively, so that the three patterns are nested, and the jammed fraction of each frame, which ranges from $9.85\%$ to $10.26\%$ depending on the rate, is close to $\rho$.  Since only one fixed jamming pattern is used for training at each rate, $N_{\rm p}=1$, and the training objective reduces to the threshold on the corresponding profile $\widehat{\boldsymbol{\rho}}(\mathbf{o}^{\star})$. Since the pattern is known, the final selection of Section~\ref{sec:design}-\ref{subsec:design_candidate_evaluation} is performed on the same profile, $\mathcal{P}_{\rm sel}=\mathcal{P}_{\rm tr}$, so that, at each rate, the selected ensemble is the candidate with the lowest threshold on $\mathbf{o}^{\star}$ among those evaluated in the search. Its threshold is reported in Table~\ref{tab:single_profile_thresholds}. All the other parameters are those of Table~\ref{tab:design_setup}.

The selected ensembles have the following structural parameters and component base matrices:
\begin{itemize}

\item The rate-$1/2$ ensemble has $r_0=1$, $n_0=2$, $m_s=2$, and $M=347$, yielding a frame length of $50\,662$ and a terminated design rate $R_L=71/146\approx0.4863$. Its component base matrices are $\mathbf{B}_0=[2\;1]$, $\mathbf{B}_1=[0\;1]$, and $\mathbf{B}_2=[2\;2]$.
\item The rate-$2/3$ ensemble has $r_0=1$, $n_0=3$, $m_s=3$, and $M=131$, yielding a frame length of $28\,689$ and a terminated design rate $R_L=143/219\approx0.6530$. Its component base matrices are $\mathbf{B}_0=[1\;1\;1]$, $\mathbf{B}_1=[0\;1\;1]$, $\mathbf{B}_2=[0\;1\;1]$, and $\mathbf{B}_3=[1\;2\;1]$.
\item The rate-$4/5$ ensemble has $r_0=1$, $n_0=5$, $m_s=3$, and $M=67$, yielding a frame length of $24\,455$ and a terminated design rate $R_L=289/365\approx0.7918$. Its component base matrices are $\mathbf{B}_0=[2\;1\;1\;0\;2]$, $\mathbf{B}_1=[1\;1\;1\;0\;1]$, $\mathbf{B}_2=[0\;0\;1\;1\;0]$, and $\mathbf{B}_3=[2\;0\;2\;1\;1]$.
\end{itemize}
Their syndrome former constraint lengths are $\nu_s=2082$, $1572$, and $1340$, respectively, and, with $W=5(m_s+1)$, the sizes $Wn_0M$ of their \ac{SWD} windows are  $10\,410$, $7860$, and $6700$ coded symbols.

Table~\ref{tab:single_profile_thresholds} reports the thresholds on the design pattern of the selected ensembles and of three benchmarks, the non-optimized \ac{SC-LDPC} code ensembles of~\cite{GiulianiSoftCOM2026} and the random-lift ensembles of the \ac{AR4JA} protographs with $k=1024$ and $k=4096$, whose block lengths are approximately equal to the constraint length of the \ac{SC-LDPC} codes and comparable with the maximum arrival-to-decision latency of \ac{SWD}, respectively (see Section~\ref{sec:results}-\ref{subsec:results_latency}). For each ensemble, the profile is obtained from the same offset sequence through its own serialization and lifting factor. Since the \ac{AR4JA} codewords are shorter than the frames of the coupled codes, different \ac{AR4JA} codewords see different portions of $\mathbf{o}^{\star}$, and their thresholds vary from codeword to codeword. Moreover, since the frame length $n$ of the coupled code is not a multiple of the \ac{AR4JA} block length $n_{\rm A}$, the position of the \ac{AR4JA} codewords relative to the jamming sequence drifts from one repetition of the sequence to the next (see Section~\ref{sec:results}-\ref{subsec:results_setup}), and the initial alignment recurs only after $n/\gcd(n,n_{\rm A})$ codewords, i.e., $25\,331$, $9563$, and $4891$ codewords at rates $1/2$, $2/3$, and $4/5$, respectively. We compute the threshold of every codeword of this cycle, i.e., for every profile seen by the \ac{AR4JA} codewords in the simulations, and Table~\ref{tab:single_profile_thresholds} reports the median and, between round brackets, the $90$th percentile of these thresholds. For $k=4096$, each codeword spans several jamming periods, its jammed fraction remains close to $\rho$, and the thresholds are concentrated around the median. For $k=1024$, instead, a codeword may intersect from zero to three pulses, possibly truncated (at most two at rate $4/5$), and the thresholds are spread; at rate $4/5$, a small fraction of the codewords, all containing two pulses, do not converge at all. The designed ensembles have the lowest thresholds. At rates $1/2$ and $2/3$, their margin over both the ensembles of~\cite{GiulianiSoftCOM2026} and the median \ac{AR4JA} codeword with $k=4096$ is small, of the order of $0.1$~dB; at rate $4/5$, it is large over the ensemble of~\cite{GiulianiSoftCOM2026} but only $0.05$~dB over the same \ac{AR4JA} ensemble. The only exception is the median \ac{AR4JA} codeword with $k=1024$ at rate $2/3$, whose threshold is $0.01$~dB lower than that of the designed ensemble. However, its $90$th percentile is $0.75$~dB higher.

To check the Gaussian approximation of Section~\ref{sec:pexitjam}-\ref{subsec:pexit_pulsed}, including the Gaussian representation of the state mixture, we also computed the thresholds of the ensembles designed for $\mathbf{o}^{\star}$ through Monte Carlo density evolution, i.e., by propagating populations of $5000$ message samples per protograph edge on the same terminated chain, with the same type-dependent profile, the same two-state channel, and $I_{\max}=2000$, without any assumption on the message densities.\footnote{The populations are resampled independently for each edge at every update, and the \acp{LLR} are clipped at $\pm60$. A run is successful if every variable-node type reaches an a posteriori \ac{MI}, estimated from the samples, of at least $1-10^{-3}$ within $I_{\max}$ iterations, and it is stopped as unsuccessful if no further variable-node type converges for $300$ iterations. } With a bisection resolution of $0.0125$~dB, the resulting thresholds at rates $1/2$, $2/3$, and $4/5$ are $1.02$, $2.07$, and $3.39$~dB, respectively, i.e., between $0.01$ and $0.03$~dB above the \ac{PEXIT} thresholds in Table~\ref{tab:single_profile_thresholds}.

\begin{table}[t]
\centering
\caption{Pulse offsets $o^{\star}_q$, in coded-symbol intervals, of the deterministic design pattern $\mathbf{o}^{\star}$, for the jamming periods $q=0,\ldots,39$, read row by row. At rates $2/3$ and $4/5$, only the first $23$ and $20$ entries are used, respectively.}
\label{tab:single_profile_offsets}
\footnotesize
\setlength{\tabcolsep}{2.5pt}
\begin{tabular}{l|cccccccccc}
\hline
0--9 & 526 & 707 & 1093 & 743 & 976 & 1152 & 153 & 586 & 341 & 952\\
10--19 & 661 & 352 & 621 & 1080 & 1135 & 885 & 502 & 528 & 1003 & 624\\
20--29 & 444 & 982 & 351 & 654 & 840 & 582 & 1151 & 242 & 69 & 198\\
30--39 & 647 & 480 & 45 & 154 & 57 & 474 & 180 & 34 & 144 & 1021\\
\hline
\end{tabular}
\end{table}

\begin{table}[t]
\centering
\caption{\ac{PEXIT} thresholds [dB] on the deterministic design pattern $\mathbf{o}^{\star}$, with ideal \ac{JSI}, $E_b/J_0=5$~dB, $\rho=0.1$, $I_{\max}=2000$, and flooding decoding of the random-lift ensembles.}
\label{tab:single_profile_thresholds}
\footnotesize
\setlength{\tabcolsep}{4pt}
\begin{tabular}{lccc}
\hline
Ensemble &  $R=1/2$  &  $R=2/3$  &  $R=4/5$ \\
\hline
 Designed for $\mathbf{o ^{\star}}$ &  1.01  &  2.04  &  3.38 \\
 \ac{SC-LDPC}~\cite{GiulianiSoftCOM2026}  &  1.07  &  2.16  &  4.65 \\
 \ac{AR4JA}, $k=1024$  &  1.25 (1.33)  &  2.03 (2.79)  & 3.49 (5.06) \\
 \ac{AR4JA}, $k=4096$  &  1.14 (1.22)  & 2.15 (2.29)  & 3.43 (3.70) \\
\hline
\end{tabular}
\end{table}

\subsection{Ensembles designed for random pulse offsets}
\label{subsec:selected_ensembles}

In the second instance, the training and validation profiles are obtained from random pulse offsets, as described in Section~\ref{sec:designed}-\ref{subsec:design_setup}. The searches were completed for all six structural classes at each target rate. Table~\ref{tab:design_validation} reports the number of distinct feasible candidates evaluated during the searches, the number of candidates evaluated on the independent validation profiles, and the validation objectives of the ensembles selected for finite-length simulations. The selected ensembles attain the lowest validation objectives among the evaluated candidates; at rate $2/3$, the selected ensemble shares this lowest value, within the bisection resolution, with other candidates.

\begin{table}[t]
\centering
\caption{Results of the \ac{PEXIT}-based searches with $I_{\max}=2000$.}
\label{tab:design_validation}
\footnotesize
\setlength{\tabcolsep}{4pt}
\begin{tabular}{cccc}
\hline
Rate & Search candidates & Validated candidates & $\Gamma_\beta$ [dB]\\
\hline
$1/2$ & 1149 & 128 & 1.0538\\
$2/3$ & 1295 & 110 & 2.1414\\
$4/5$ & 1459 & 44 & 3.4802\\
\hline
\end{tabular}
\end{table}

The three ensembles selected for the finite-length assessment under random pulsed jamming  have the following structural parameters and component base matrices:
\begin{itemize}
\item The rate-$1/2$ ensemble has $r_0=2$, $n_0=4$, $m_s=1$, and $M=257$, yielding a frame length of $75\,044$ and a terminated design rate $R_L=36/73\approx0.4932$. Its component base matrices are
\[
\mathbf{B}_0=
\begin{bmatrix}
0 & 1 & 1 & 2\\
1 & 0 & 2 & 1
\end{bmatrix},
\qquad
\mathbf{B}_1=
\begin{bmatrix}
1 & 0 & 1 & 1\\
0 & 1 & 1 & 1
\end{bmatrix}.
\]

\item The rate-$2/3$ ensemble has $r_0=1$, $n_0=3$, $m_s=3$, and $M=131$, yielding a frame length of $28\,689$ and a terminated design rate $R_L=143/219\approx0.6530$. Its component base matrices are $\mathbf{B}_0=[1\;1\;1]$, $\mathbf{B}_1=[0\;1\;2]$, $\mathbf{B}_2=[0\;1\;1]$, and $\mathbf{B}_3=[1\;1\;1]$.

\item The rate-$4/5$ ensemble has $r_0=1$, $n_0=5$, $m_s=3$, and $M=67$, yielding a frame length of $24\,455$ and a terminated design rate $R_L=289/365\approx0.7918$. Its component base matrices are $\mathbf{B}_0=[2\;1\;1\;1\;1]$, $\mathbf{B}_1=[0\;1\;1\;1\;0]$, $\mathbf{B}_2=[1\;1\;0\;2\;0]$, and $\mathbf{B}_3=[1\;2\;0\;1\;1]$.
\end{itemize}
Their syndrome former constraint lengths are $\nu_s=2056$, $1572$, and $1340$, respectively, close to the transmitted block lengths $2048$, $1536$, and $1280$ of the corresponding \ac{AR4JA} codes, as dictated by~\eqref{eq:design_lifting_factor}.

The terminated design rates are slightly lower than the corresponding nominal rates. Since the channel normalization uses the terminated design rate $R_L$, the difference with respect to a normalization based on the nominal rate, $10\log_{10}(R_{\rm des}/R_L)$, is below $0.1$~dB for all the ensembles designed for random pulse realizations and reaches approximately $0.12$~dB for the rate-$1/2$ ensemble designed for the fixed pulse pattern.

\subsection{Effect of the maximum number of iterations}
\label{subsec:iteration_budget}

\begin{table}[t]
\centering
\caption{Effect of the maximum number of iterations on the position-dependent \ac{PEXIT} thresholds of a regular $(3,6)$ ensemble and the rate-$1/2$ ensemble, for $L=73$, $E_b/J_0=5$~dB, and $\rho=0.1$.}
\label{tab:iteration_budget}
\footnotesize
\setlength{\tabcolsep}{4pt}
\begin{tabular}{lcc}
\hline
Ensemble, profile & \multicolumn{2}{c}{$I_{\max}$}\\
 & 100 & 2000\\
\hline
$(3,6)$, $m_s=2$, no jamming & 1.08 & 0.60\\
$(3,6)$, $m_s=2$, homogeneous & 1.62 & 1.10\\
Designed $R=1/2$, no jamming & 1.04 & 0.44\\
Designed $R=1/2$, homogeneous & 1.58 & 0.93\\
Designed $R=1/2$, 10 rand.\ offsets & 1.55--1.58 & 1.00--1.07\\
\hline
\end{tabular}
\end{table}

Table~\ref{tab:iteration_budget} reports the position-dependent \ac{PEXIT} thresholds for $L=73$, computed with $I_{\max}=100$ and $I_{\max}=2000$, a bisection tolerance of $0.005$~dB, and the APP convergence target and stagnation tolerance of Table~\ref{tab:design_setup}. The rate-$1/2$ ensemble is the one in Section~\ref{sec:designed}-\ref{subsec:selected_ensembles}, and the random-offset entry reports the minimum and maximum thresholds over the ten validation profiles. The regular $(3,6)$ chain has $\mathbf{B}_0=\mathbf{B}_1=\mathbf{B}_2=[1\;1]$. For both ensembles, with and without jamming, increasing $I_{\max}$ from $100$ to $2000$ lowers the threshold by $0.48$ to $0.65$~dB, which shows that the smaller maximum number of iterations captures only part of the coupling gain.

The maximum number of iterations also changes the sensitivity of the threshold to the pulse positions. For the designed rate-$1/2$ ensemble, with $I_{\max}=2000$, all ten non-interleaved validation profiles have thresholds above the homogeneous reference, whereas with $I_{\max}=100$ they are within a few hundredths of a dB of it, some even slightly below, i.e., the smaller value hides the effect of the burst positions. This justifies the value $I_{\max}=2000$ adopted in Table~\ref{tab:design_setup}.

Table~\ref{tab:iteration_budget} also allows comparing the threshold penalties due to jamming. At $I_{\max}=2000$, homogeneous jamming increases the threshold with respect to the jamming-free case by about $0.5$~dB for both ensembles, whereas the ten non-interleaved profiles increase the threshold of the designed ensemble by a further $0.07$--$0.14$~dB, depending on the pulse offsets. Hence, for these ensembles, the homogeneous model shifts the thresholds by nearly the same amount and adds little information to the jamming-free threshold, while the position-dependent profiles capture the effect of the burst localization along the chain. This motivates the use of the non-interleaved profiles in the design objective.

\section{Symbol reception and decoding latency}\label{sec:decoder}
The designed and the benchmark \ac{SC-LDPC} codes are decoded through \ac{SWD}, whereas the \ac{AR4JA} codes are decoded through the conventional flooding \ac{SPA}. In this section, we characterize the symbol-reception requirements and the decoding latency.

To reduce the latency of flooding-schedule decoding, \ac{SWD} applies the \ac{SPA} only to a portion of the parity-check matrix at each decoding step~\cite{Iyengar2012}. Let $W$ denote the window size, measured in spatial positions. The first decoding window contains $Wn_0M$ coded symbols. After decoding its target symbols, the window shifts by one spatial position, and $n_0M$ additional coded symbols are received before the next window can be processed. When the window reaches the right boundary of the parity-check matrix, it is extended to include the remaining parity-check equations~\cite{Ali2017}, and all the remaining $W$ spatial positions are decided in this final window.   Overall, the decoder processes $N_{\rm w}=L-W+1$ window positions.

  To compare \ac{SWD} with block decoding, we define the average number of channel symbols received per decoding step as
\begin{equation}
    \overline{\tau}
    =
    \frac{Wn_0M+(N_{\rm w}-1)n_0M}{N_{\rm w}}
    =
    \frac{L}{L-W+1}n_0M.
    \label{eq:average_reception_latency}
\end{equation}

 For a block code, the same definition gives the transmitted block length $n$, since one decoding step is performed per codeword.  In contrast, the arrival-to-decision latency of an individual coded symbol is the number of coded symbols received between its arrival and the decision on it. This latency is at most $Wn_0M$ for \ac{SWD} and at most $n$ for a block decoder~\cite{ulHassanISTC2012}.

\section{Numerical assessment of error-rate performance}
\label{sec:results}

In this section, we assess the finite-length performance of the considered coding schemes under pulsed jamming.  We evaluate the standardized \ac{AR4JA} \ac{LDPC} codes against spatially coupled alternatives at rates $R=1/2$, $R=2/3$, and $R=4/5$. We also examine how the observed behavior changes with the availability of \ac{JSI}, the jammer duty cycle, and symbol interleaving. We report the \ac{BER} rather than the \ac{FER}, which would be computed on the whole frame length $Ln_0M$. Since the terminated \ac{SC-LDPC} frames contain many spatial positions and are substantially longer than the \ac{AR4JA} blocks, a whole-chain \ac{FER} depends strongly on the chain length (it increases with $L$ and tends to one as $L\to\infty$, even when the \ac{BER} remains small, if error events at distant positions are weakly dependent) and is not directly comparable with the block \ac{FER} of an \ac{AR4JA} code. The \ac{BER} is therefore adopted as a length-normalized measure of error-correction performance; it is not intended as a measure of packet-delivery probability.

\subsection{Code families and simulation setup}
\label{subsec:results_setup}

For each of the considered rates, the reference block codes are the corresponding \ac{AR4JA} \ac{LDPC} codes specified in the \ac{CCSDS} telemetry standard~\cite{TMst,CCSDS130G3}, which provide the standardized telemetry baseline for the comparison. We consider the \ac{AR4JA} codes with $k=1024$ information bits, whose transmitted block lengths, $2048$, $1536$, and $1280$, are approximately equal to the syndrome former constraint lengths of the designed \ac{SC-LDPC} codes. We also consider the \ac{AR4JA} codes with $k=4096$, whose transmitted block lengths, $8192$, $6144$, and $5120$, are comparable with the size of the sliding window decoder (see Section~\ref{sec:results}-\ref{subsec:results_latency}).

For the spatially coupled case, we consider two classes of terminated \ac{SC-LDPC} codes. The first consists of the non-optimized ensembles of~\cite{GiulianiSoftCOM2026}, included to assess the benefits of spatial coupling independently of the proposed optimization, and decoded with the same \ac{SWD} settings. The second consists of the ensembles selected through the \ac{PEXIT}-based design procedure of Section~\ref{sec:design}, in the two instances of Section~\ref{sec:designed}: the ensembles designed for the single deterministic pattern $\mathbf{o}^{\star}$ (Section~\ref{sec:designed}-\ref{subsec:single_profile_design}), which are simulated only under that pattern, and the ensembles designed for random pulse offsets (Section~\ref{sec:designed}-\ref{subsec:selected_ensembles}), which are simulated under random pulsed jamming. The finite-length lifting of the selected \ac{SC-LDPC} code ensembles is performed as described in Section~\ref{sec:design}-\ref{subsec:finite_length_lifting}, with the lifting factor given by~\eqref{eq:design_lifting_factor}.

All transmissions use \ac{BPSK} modulation over an \ac{AWGN} channel affected by pulsed jamming. The main comparison is carried out at the operating point discussed in Section~\ref{sec:preli}-\ref{subsec:pulsed}, i.e., $E_b/J_0=5$~dB and $\rho=0.1$, which is also the one used in the design (Table~\ref{tab:design_setup}), and without symbol interleaving. The absence of interleaving is the most relevant setting for the proposed design, since the optimized ensembles are selected using non-interleaved pulsed-jamming profiles; the effects of $\rho$ and of interleaving are assessed in Sections~\ref{sec:results}-\ref{subsec:rho_results} and~\ref{sec:results}-\ref{subsec:interleaving_results}, respectively. Unless otherwise stated, the optimized code is the one designed for random pulse offsets. We consider both ideal \ac{JSI}, where the demodulator uses the state-dependent channel reliability, and the case without \ac{JSI}, where the receiver does not know which symbols are affected by the jammer. In particular, with ideal \ac{JSI}, the demodulator uses variance $N_0/2$ for unjammed symbols and \eqref{eq:pulsed_total_variance} for jammed symbols. Without \ac{JSI}, the receiver uses the same Gaussian reliability for all received symbols, based on the average disturbance power, and thus requires knowledge of neither the instantaneous jamming state nor the jammer duty cycle. This corresponds to a mismatched demodulation metric, since the actual disturbance variance remains state dependent. The results without \ac{JSI} therefore refer to the adopted average-power Gaussian metric and, obviously, are not predicted by the ideal-\ac{JSI} analysis.

In the non-interleaved finite-length simulations, the number of coded symbols is $L_{\rm p}=1280$ in one jamming period and $L_{\rm b}=\rho L_{\rm p}$ in one active pulse. The jamming signal is a sequence of exactly $n=Ln_0M$ coded symbols, the frame length of the optimized code: it consists of the $Q_M=\lceil n/L_{\rm p}\rceil$ jamming periods in that frame, the last of which is truncated at the end of the frame, and it is repeated cyclically. Codewords and frames are transmitted back to back. Hence, every frame of the optimized code starts at the beginning of the jamming sequence, as assumed in the design profiles (see Section~\ref{sec:preli}-\ref{subsec:pulsed}), whereas the benchmark codes, whose lengths differ, see different portions of the sequence from codeword to codeword, and a pulse may hit two consecutive codewords. In particular, the jammed fraction of the \ac{AR4JA} codewords, which are shorter than the \ac{SC-LDPC} frames, varies from codeword to codeword, with an average value close to $\rho$. Two jamming patterns are considered. Under random pulsed jamming, the offset of each pulse within its period is drawn independently and uniformly over $\{0,\ldots,L_{\rm p}-L_{\rm b}\}$, as in the design profiles of Section~\ref{sec:designed}-\ref{subsec:selected_ensembles}, and new offsets are drawn at every repetition of the sequence. In the single-pattern simulations of Section~\ref{sec:results}-\ref{subsec:results_single}, the offsets are those of the design pattern $\mathbf{o}^{\star}$ at every repetition, so that every frame of the optimized code exactly experiences the design profile $\widehat{\boldsymbol{\rho}}(\mathbf{o}^{\star})$.

The maximum number of \ac{SPA} iterations is 100 per block for the \ac{AR4JA} codes and 100 per decoding window for the \ac{SC-LDPC} codes. The \ac{AR4JA} decoder terminates earlier when the syndrome of the decoded block is zero. For \ac{SWD}, early termination within each window is based on the satisfaction of the parity-check equations associated with the target symbols. The window size is $W=5(m_s+1)$. Each $E_b/N_0$ point was simulated until 100 erroneous frames were observed.

\subsection{Latency comparison}
\label{subsec:results_latency}

Table~\ref{tab:reception} reports, for the three rates, the average number of channel symbols received per decoding step, $\overline{\tau}$, defined in~\eqref{eq:average_reception_latency}, the maximum arrival-to-decision latency $Wn_0M$, and the transmitted block length of the corresponding \ac{AR4JA} codes. At all rates, \ac{SWD} requires fewer received symbols per decoding step than the \ac{AR4JA} block decoder, e.g., $\overline{\tau}=1172.6$ instead of $2048$ at rate $1/2$, at the cost of a longer arrival-to-decision latency. The maximum arrival-to-decision latency $Wn_0M$ is about five times the block length of the \ac{AR4JA} codes with $k=1024$ and about $1.3$ times that of the \ac{AR4JA} codes with $k=4096$, which are therefore the closest block-code benchmarks in terms of latency. Since $\overline{\tau}$ increases with $W$ only through the factor $L/(L-W+1)$, it remains below the \ac{AR4JA} block length for windows much larger than those adopted here, at least up to $W=37$.

\begin{table}[t]
\centering
\caption{Average number of channel symbols received per decoding step, $\overline{\tau}$, and maximum arrival-to-decision latency, $Wn_0M$, of the optimized \ac{SC-LDPC} codes ($L=73$, $W=5(m_s+1)$), and transmitted block length of the corresponding \ac{AR4JA} codes.}
\label{tab:reception}
\begin{tabular}{cccccc}
\hline
Rate & $W$ & $n_0M$ & $\overline{\tau}$ & $Wn_0M$ & $n_{\rm AR4JA}$\\
\hline
$1/2$ & 10 & 1028 & 1172.6 & 10\,280 & 2048\\
$2/3$ & 20 & 393 & 531.3 & 7\,860 & 1536\\
$4/5$ & 20 & 335 & 452.9 & 6\,700 & 1280\\
\hline
\end{tabular}
\end{table}

\subsection{Codes designed for a single deterministic jamming pattern}
\label{subsec:results_single}

\begin{figure}[t]
    \centering
    \resizebox{0.80\columnwidth}{!}{\definecolor{mycolor1}{rgb}{0.31373,0.78431,0.47059}
\definecolor{mycolor2}{rgb}{1.00000,0.00000,1.00000}

\begin{tikzpicture}

\begin{axis}[
width=4.521in,
height=3.566in,
at={(0.758in,0.481in)},
scale only axis,
xmin=0,
xmax=4,
xlabel style={font=\color{white!15!black}},
xlabel={$E_b/N_0$ [dB]},
ymode=log,
ymin=1e-8,
ymax=1,
yminorticks=true,
ylabel style={font=\color{white!15!black}},
ylabel={BER},
axis background/.style={fill=white},
xmajorgrids,
ymajorgrids,
yminorgrids,
legend style={at={(0.01,0.01)}, anchor=south west, legend cell align=left, align=left, draw=white!15!black}
]

\addplot [color=red, line width=0.6pt, mark=diamond, mark options={red}, forget plot]
  table[row sep=crcr]{
0.00 1.4936E-01\\
0.50 1.2224E-01\\
1.00 8.1767E-02\\
1.50 3.1606E-02\\
2.00 3.0902E-03\\
2.50 1.1425E-04\\
3.00 1.1685E-06\\
};

\addplot [color=red, dashed, line width=0.6pt, mark=diamond, mark options={red}, forget plot]
  table[row sep=crcr]{
0.00 1.6136E-01\\
0.50 1.4067E-01\\
1.00 1.1648E-01\\
1.50 9.0486E-02\\
2.00 5.8038E-02\\
2.50 2.9664E-02\\
3.00 1.2511E-02\\
3.50 4.0840E-03\\
4.00 1.6392E-03\\
4.50 8.0434E-04\\
5.00 4.1412E-04\\
5.50 2.5227E-04\\
6.00 1.7931E-04\\
};

\addplot [color=orange, line width=0.6pt, mark=diamond, mark options={orange}, forget plot]
  table[row sep=crcr]{
0.00 1.4847E-01\\
0.50 1.2415E-01\\
1.00 9.1646E-02\\
1.50 1.7620E-02\\
2.00 3.6793E-06\\
};

\addplot [color=orange, dashed, line width=0.6pt, mark=diamond, mark options={orange}, forget plot]
  table[row sep=crcr]{
0.00 1.5955E-01\\
0.50 1.4120E-01\\
1.00 1.2009E-01\\
1.50 9.5977E-02\\
2.00 6.5128E-02\\
2.50 1.1667E-02\\
3.00 3.3931E-04\\
3.50 1.9474E-06\\
};

\addplot [color=green, line width=0.6pt, mark=diamond, mark options={green}, forget plot]
  table[row sep=crcr]{
0.00 1.6215E-01\\
0.50 1.4396E-01\\
1.00 1.2245E-01\\
1.50 6.5928E-02\\
2.00 1.2179E-05\\
2.50 3.3496E-06\\
3.00 1.2857E-07\\
3.50 5.1716E-08\\
};

\addplot [color=green, dashed, line width=0.6pt, mark=diamond, mark options={green}, forget plot]
  table[row sep=crcr]{
0.00 1.6593E-01\\
0.50 1.4993E-01\\
1.00 1.3365E-01\\
1.50 1.1576E-01\\
2.00 9.5577E-02\\
2.50 5.5874E-02\\
3.00 1.1639E-03\\
3.50 1.3278E-05\\
4.00 7.7142E-07\\
};

\addplot [color=blue, line width=0.6pt, mark=diamond, mark options={blue}, forget plot]
  table[row sep=crcr]{
0.00 1.6676E-01\\
0.50 1.4953E-01\\
1.00 1.2467E-01\\
1.50 2.7868E-02\\
2.00 2.8367E-05\\
2.1 1.07E-05\\
2.2 1.8E-06\\
};

\addplot [color=blue, dashed, line width=0.6pt, mark=diamond, mark options={blue}, forget plot]
  table[row sep=crcr]{
0.00 1.6959E-01\\
0.50 1.5449E-01\\
1.00 1.3787E-01\\
1.50 1.1773E-01\\
2.00 9.1185E-02\\
2.50 3.0295E-02\\
3.00 1.3301E-03\\
3.50 5.6279E-05\\
3.6 1.2708E-05\\
};

\draw[
    thick,
    color=blue,
    densely dotted,
    opacity=0.8
]
    (axis cs: 1.006104, 1E-8) -- (axis cs: 1.006104, 1E0);
\draw [ 
    thick, 
    color=green,
    densely dotted,
    opacity=0.8
] 
    (axis cs: 1.071289, 1E-7) -- (axis cs: 1.071289, 1E0);

 \draw [ 
    thick, 
    color=orange,
    densely dotted,
    opacity=0.8
] 
(axis cs: 1.135010, 1E-7) -- (axis cs: 1.135010, 1E0);

 \draw [ 
    thick, 
    color=red,
    densely dotted,
    opacity=0.8
] 
(axis cs: 1.247803, 1E-7) -- (axis cs: 1.247803, 1E0);

\addlegendimage{red, thick} \addlegendentry{AR4JA code, $k = 1024$ \cite{TMst}}
\addlegendimage{orange, thick} \addlegendentry{AR4JA code, $k = 4096$ \cite{TMst}}
\addlegendimage{green, thick}  \addlegendentry{SC-LDPC code \cite{GiulianiSoftCOM2026}}
\addlegendimage{blue, thick}  \addlegendentry{Optimized SC-LDPC code}
\addlegendimage{black, solid}
\addlegendentry{With JSI}
\addlegendimage{black, dashed}
\addlegendentry{No JSI}
\addlegendimage{black, dotted}
\addlegendentry{PEXIT thresholds}

\end{axis}

\begin{axis}[
width=5.833in,
height=4.375in,
at={(0in,0in)},
scale only axis,
xmin=0,
xmax=1,
ymin=0,
ymax=1,
axis line style={draw=none},
ticks=none,
axis x line*=bottom,
axis y line*=left
]
\end{axis}
\end{tikzpicture}}
    \caption{\ac{BER} comparison at $R=1/2$, $E_b/J_0=5$~dB, $\rho=0.1$, and no interleaving, under the pulse pattern $\mathbf{o}^{\star}$.}
    \label{fig:EbJ05dB_12_SingleProfile}
\end{figure}
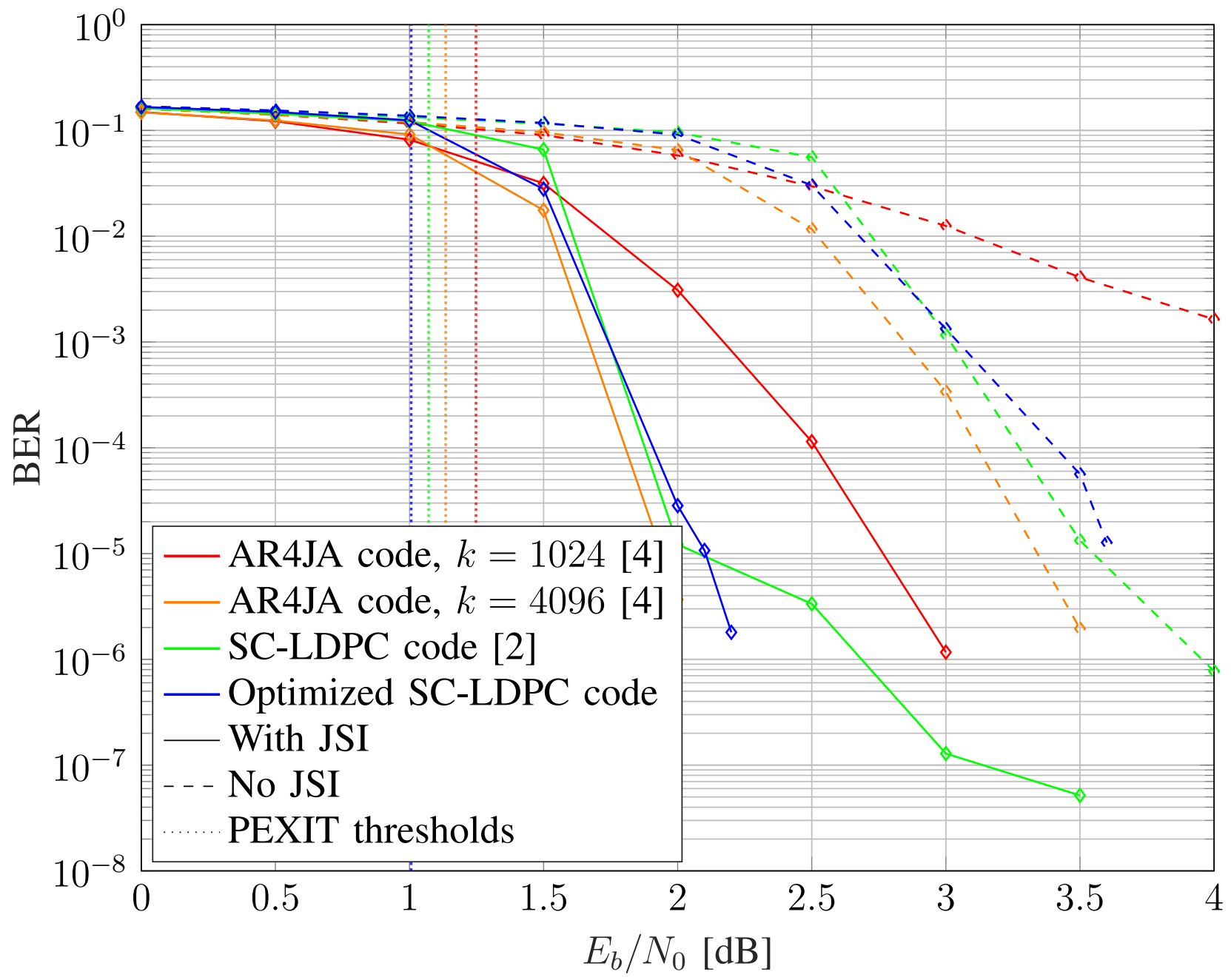

\begin{figure}[t]
    \centering
    \resizebox{0.80\columnwidth}{!}{\definecolor{mycolor1}{rgb}{0.31373,0.78431,0.47059}
\definecolor{mycolor2}{rgb}{1.00000,0.00000,1.00000}

\begin{tikzpicture}

\begin{axis}[
width=4.521in,
height=3.566in,
at={(0.758in,0.481in)},
scale only axis,
xmin=1,
xmax=5,
xlabel style={font=\color{white!15!black}},
xlabel={$E_b/N_0$ [dB]},
ymode=log,
ymin=1e-8,
ymax=1,
yminorticks=true,
ylabel style={font=\color{white!15!black}},
ylabel={BER},
axis background/.style={fill=white},
xmajorgrids,
ymajorgrids,
yminorgrids,
legend style={at={(0.01,0.01)}, anchor=south west, legend cell align=left, align=left, draw=white!15!black, text opacity=1}
]

\addplot [color=red, line width=0.6pt, mark=square, mark options={red}, forget plot]
  table[row sep=crcr]{
1.00 9.9983E-02\\
1.50 8.4291E-02\\
2.00 5.7747E-02\\
2.50 2.7181E-02\\
3.00 6.4500E-03\\
3.50 1.9901E-03\\
4.00 3.3580E-04\\
4.50 8.2797E-05\\
5.00 1.1992E-05\\
5.50 1.2852E-06\\
6.00 1.0972E-07\\
};

\addplot [color=red, dashed, line width=0.6pt, mark=square, mark options={red}, forget plot]
  table[row sep=crcr]{
1.00 1.1148E-01\\
1.50 9.6730E-02\\
2.00 8.0742E-02\\
2.50 6.7463E-02\\
3.00 4.9803E-02\\
3.50 4.2780E-02\\
4.00 3.0725E-02\\
4.50 2.3001E-02\\
5.00 1.7484E-02\\
5.50 1.9050E-02\\
6.00 1.2277E-02\\
6.50 1.4388E-02\\
7.00 8.4140E-03\\
};

\addplot [color=orange, line width=0.6pt, mark=square, mark options={orange}, forget plot]
  table[row sep=crcr]{
1.00 9.9519E-02\\
1.50 8.1406E-02\\
2.00 6.3527E-02\\
2.50 2.2266E-02\\
3.00 1.5075E-04\\
3.25 1.3949E-06\\
};

\addplot [color=orange, dashed, line width=0.6pt, mark=square, mark options={orange}, forget plot]
  table[row sep=crcr]{
1.00 1.0916E-01\\
1.50 9.5020E-02\\
2.00 8.2642E-02\\
2.50 7.0960E-02\\
3.00 5.6539E-02\\
3.50 4.3987E-02\\
4.00 3.4104E-02\\
4.50 2.4629E-02\\
5.00 1.1242E-02\\
};

\addplot [color=green, line width=0.6pt, mark=square, mark options={green}, forget plot]
  table[row sep=crcr]{
1.00 1.0591E-01\\
1.50 8.8563E-02\\
2.00 6.7535E-02\\
2.50 3.9370E-03\\
3.00 1.6624E-05\\
3.50 3.9307E-06\\
4.00 1.6400E-06\\
4.50 5.6510E-07\\
5.00 2.7720E-07\\
};

\addplot [color=green, dashed, line width=0.6pt, mark=square, mark options={green}, forget plot]
  table[row sep=crcr]{
1.00 1.1073E-01\\
1.50 9.6946E-02\\
2.00 8.3316E-02\\
2.50 7.0662E-02\\
3.00 6.0601E-02\\
3.50 5.3707E-02\\
4.00 4.0765E-02\\
4.50 2.3533E-02\\
5.00 1.0579E-02\\
};

\addplot [color=blue, line width=0.6pt, mark=square, mark options={blue}, forget plot]
  table[row sep=crcr]{
1.00 1.0804E-01\\
1.50 9.0401E-02\\
2.00 6.5652E-02\\
2.50 6.7584E-03\\
3.00 1.7110E-06\\
3.50 1.9704E-07\\
4.00 1.5007E-07\\
};

\addplot [color=blue, dashed, line width=0.6pt, mark=square, mark options={blue}, forget plot]
  table[row sep=crcr]{
1.00 1.1321E-01\\
1.50 9.9810E-02\\
2.00 8.6655E-02\\
2.50 7.3860E-02\\
3.00 6.0389E-02\\
3.50 5.2223E-02\\
4.00 3.4043E-02\\
4.50 1.5580E-02\\
5.00 6.5499E-03\\
};

\draw[
    thick,
    color=blue,
    densely dotted,
    opacity=0.8
] 
    (axis cs: 2.042725, 1E-8) -- (axis cs: 2.042725, 1E0);
\draw [ 
    thick, 
    color=green,
    densely dotted,
    opacity=0.8
] 

    (axis cs: 2.156738, 1E-7) -- (axis cs: 2.156738, 1E0);

 \draw [ 
    thick, 
    color=orange,
    densely dotted,
    opacity=0.8
] 
    (axis cs: 2.152832, 1E-7) -- (axis cs: 2.152832, 1E0);
 \draw [ 
    thick, 
    color=red,
    densely dotted,
    opacity=0.8
] 
    (axis cs: 2.031982, 1E-7) -- (axis cs: 2.031982, 1E0);

\addlegendimage{red, thick} \addlegendentry{AR4JA code, $k = 1024$ \cite{TMst}}
\addlegendimage{orange, thick} \addlegendentry{AR4JA code, $k = 4096$ \cite{TMst}}
\addlegendimage{green, thick}  \addlegendentry{SC-LDPC code \cite{GiulianiSoftCOM2026}}
\addlegendimage{blue, thick}  \addlegendentry{Optimized SC-LDPC code}
\addlegendimage{black, solid}
\addlegendentry{With JSI}
\addlegendimage{black, dashed}
\addlegendentry{No JSI}
\addlegendimage{black, dotted}
\addlegendentry{PEXIT thresholds}

\end{axis}

\begin{axis}[
width=5.833in,
height=4.375in,
at={(0in,0in)},
scale only axis,
xmin=0,
xmax=1,
ymin=0,
ymax=1,
axis line style={draw=none},
ticks=none,
axis x line*=bottom,
axis y line*=left
]
\end{axis}
\end{tikzpicture}}
    \caption{\ac{BER} comparison at $R=2/3$, $E_b/J_0=5$~dB, $\rho=0.1$, and no interleaving, under the pulse pattern $\mathbf{o}^{\star}$.}
    \label{fig:EbJ05dB_23_SingleProfile}
\end{figure}
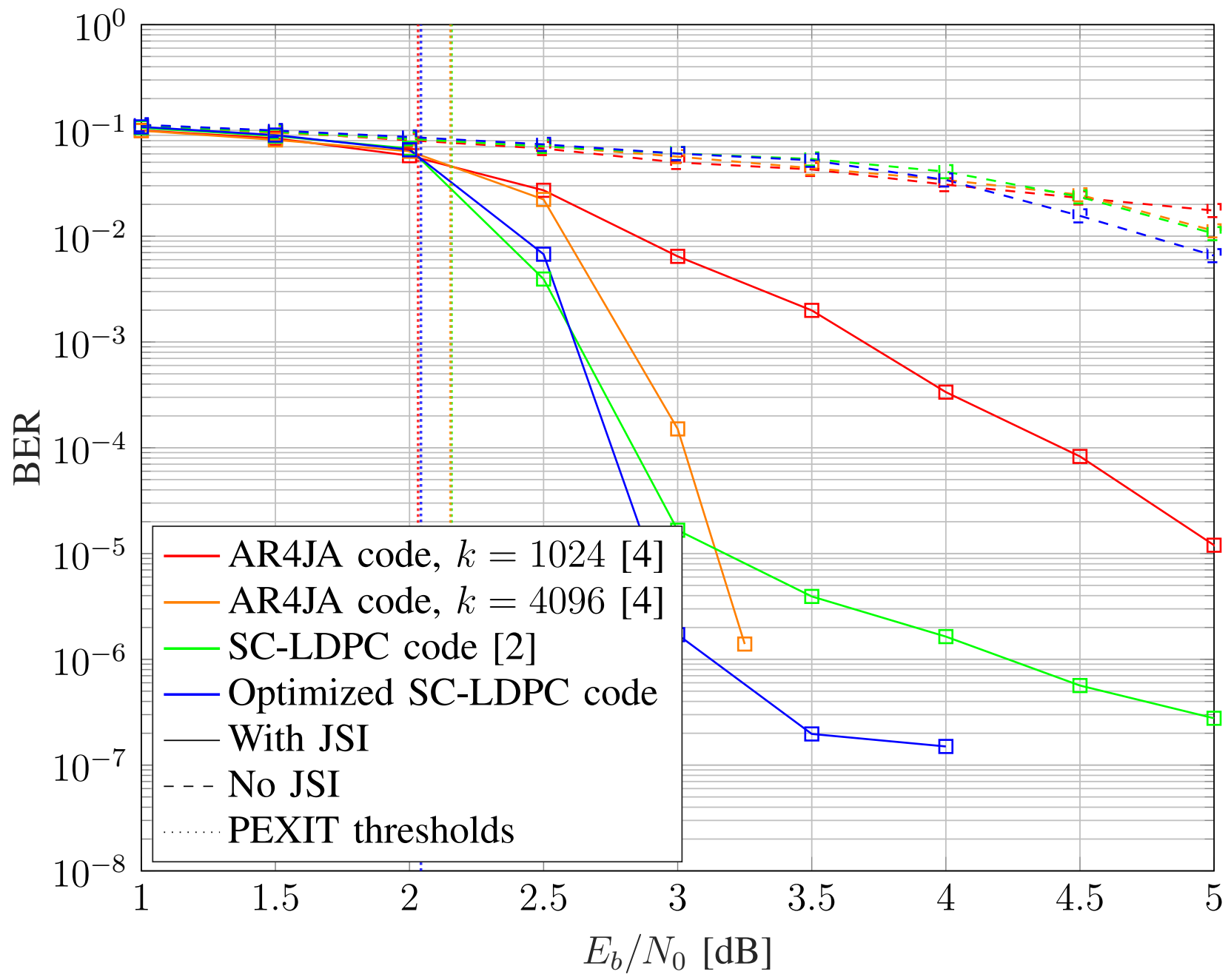

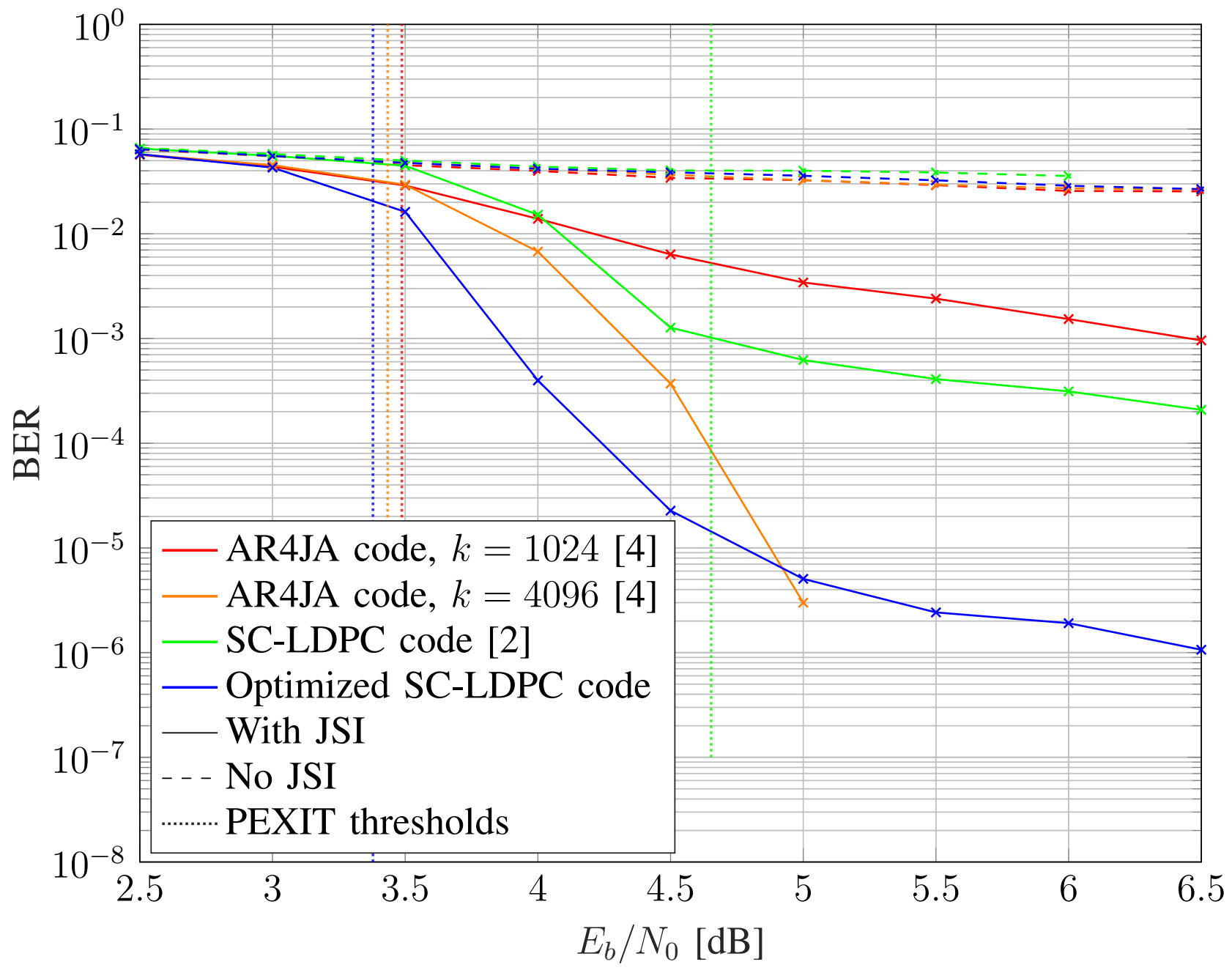
\begin{figure}[t]
    \centering
    \resizebox{0.80\columnwidth}{!}{\definecolor{mycolor1}{rgb}{0.31373,0.78431,0.47059}
\definecolor{mycolor2}{rgb}{1.00000,0.00000,1.00000}

\begin{tikzpicture}

\begin{axis}[
width=4.521in,
height=3.566in,
at={(0.758in,0.481in)},
scale only axis,
xmin=2.5,
xmax=6.5,
xlabel style={font=\color{white!15!black}},
xlabel={$E_b/N_0$ [dB]},
ymode=log,
ymin=1e-8,
ymax=1,
yminorticks=true,
ylabel style={font=\color{white!15!black}},
ylabel={BER},
axis background/.style={fill=white},
xmajorgrids,
ymajorgrids,
yminorgrids,
legend style={at={(0.01,0.01)}, anchor=south west, legend cell align=left, align=left, draw=white!15!black, text opacity=1}
]

\addplot [color=red, line width=0.6pt, mark=x, mark options={red}, forget plot]
  table[row sep=crcr]{
2.50 5.6867E-02\\
3.00 4.3755E-02\\
3.50 2.8947E-02\\
4.00 1.3905E-02\\
4.50 6.3632E-03\\
5.00 3.4299E-03\\
5.50 2.4032E-03\\
6.00 1.5336E-03\\
6.50 9.5880E-04\\
7.00 8.2291E-04\\
7.50 6.2903E-04\\
8.00 4.8156E-04\\
8.50 3.8944E-04\\
};

\addplot [color=red, dashed, line width=0.6pt, mark=x, mark options={red}, forget plot]
  table[row sep=crcr]{%
2.50 6.4219E-02\\
3.00 5.5481E-02\\
3.50 4.5205E-02\\
4.00 3.9872E-02\\
4.50 3.4216E-02\\
5.00 3.2525E-02\\
5.50 2.9160E-02\\
6.00 2.5651E-02\\
6.50 2.5406E-02\\
7.00 2.5436E-02\\
7.50 2.3007E-02\\
8.00 2.1889E-02\\
8.50 2.2027E-02\\
};

\addplot [color=orange, line width=0.6pt, mark=x, mark options={orange}, forget plot]
  table[row sep=crcr]{%
2.50 5.7282E-02\\
3.00 4.5343E-02\\
3.50 2.9418E-02\\
4.00 6.7629E-03\\
4.50 3.7085E-04\\
5.00 2.9981E-06\\
};

\addplot [color=orange, dashed, line width=0.6pt, mark=x, mark options={orange}, forget plot]
  table[row sep=crcr]{%
2.50 6.3478E-02\\
3.00 5.4991E-02\\
3.50 4.7710E-02\\
4.00 4.1667E-02\\
4.50 3.6664E-02\\
5.00 3.2534E-02\\
5.50 2.9464E-02\\
6.00 2.7170E-02\\
6.50 2.6530E-02\\
};

\addplot [color=green, line width=0.6pt, mark=x, mark options={green}, forget plot]
  table[row sep=crcr]{%
2.50 6.5138E-02\\
3.00 5.5562E-02\\
3.50 4.4670E-02\\
4.00 1.5206E-02\\
4.50 1.2701E-03\\
5.00 6.2252E-04\\
5.50 4.0958E-04\\
6.00 3.1258E-04\\
6.50 2.0812E-04\\
};

\addplot [color=green, dashed, line width=0.6pt, mark=x, mark options={green}, forget plot]
  table[row sep=crcr]{%
2.50 6.6235E-02\\
3.00 5.7852E-02\\
3.50 5.0123E-02\\
4.00 4.3865E-02\\
4.50 4.0283E-02\\
5.00 4.0076E-02\\
5.50 3.8515E-02\\
6.00 3.5667E-02\\
};

\addplot [color=blue, line width=0.6pt, mark=x, mark options={blue}, forget plot]
  table[row sep=crcr]{%
2.50 5.7598E-02\\
3.00 4.2886E-02\\
3.50 1.6209E-02\\
4.00 3.9728E-04\\
4.50 2.2691E-05\\
5.00 5.0585E-06\\
5.50 2.4204E-06\\
6.00 1.9098E-06\\
6.50 1.0649E-06\\
};

\addplot [color=blue, dashed, line width=0.6pt, mark=x, mark options={blue}, forget plot]
  table[row sep=crcr]{%
2.50 6.4260E-02\\
3.00 5.5376E-02\\
3.50 4.7598E-02\\
4.00 4.1829E-02\\
4.50 3.8647E-02\\
5.00 3.5829E-02\\
5.50 3.2421E-02\\
6.00 2.8763E-02\\
6.50 2.6692E-02\\
};
\draw [ 
    thick, 
    color=green,
    densely dotted,
    opacity=0.8
] 
    (axis cs: 4.65283, 1E-7) -- (axis cs: 4.65283, 1E0);
\draw [ 
    thick, 
    color=blue,
    densely dotted,
    opacity=0.8
] 
    (axis cs: 3.378418, 1E-8) -- (axis cs: 3.378418, 1E0);
 \draw [ 
    thick,
    densely dotted, 
    color=orange,
    opacity=0.8
] 

    (axis cs: 3.433838, 1E-7) -- (axis cs: 3.433838, 1E0);

 \draw [ 
    thick,
    densely dotted, 
    color=red,
    opacity=0.8
] 
    (axis cs: 3.487549, 1E-7) -- (axis cs: 3.487549, 1E0);

\addlegendimage{red, thick} \addlegendentry{AR4JA code, $k = 1024$ \cite{TMst}}
\addlegendimage{orange, thick} \addlegendentry{AR4JA code, $k = 4096$ \cite{TMst}}
\addlegendimage{green, thick}  \addlegendentry{SC-LDPC code \cite{GiulianiSoftCOM2026}}
\addlegendimage{blue, thick}  \addlegendentry{Optimized SC-LDPC code}
\addlegendimage{black, solid}
\addlegendentry{With JSI}
\addlegendimage{black, dashed}
\addlegendentry{No JSI}
\addlegendimage{densely dotted, thick} \addlegendentry{PEXIT thresholds}

\end{axis}

\begin{axis}[
width=5.833in,
height=4.375in,
at={(0in,0in)},
scale only axis,
xmin=0,
xmax=1,
ymin=0,
ymax=1,
axis line style={draw=none},
ticks=none,
axis x line*=bottom,
axis y line*=left
]
\end{axis}
\end{tikzpicture}}
    \caption{\ac{BER} comparison at $R=4/5$, $E_b/J_0=5$~dB, $\rho=0.1$, and no interleaving, under the pulse pattern $\mathbf{o}^{\star}$.}
    \label{fig:EbJ05dB_45_SingleProfile}
\end{figure}

We first consider the ensembles designed for the deterministic pattern $\mathbf{o}^{\star}$ of Section~\ref{sec:designed}-\ref{subsec:single_profile_design}, simulated under the same pattern. Figs.~\ref{fig:EbJ05dB_12_SingleProfile}, \ref{fig:EbJ05dB_23_SingleProfile}, and~\ref{fig:EbJ05dB_45_SingleProfile} report the results for rates $R=1/2$, $R=2/3$, and $R=4/5$, respectively, with $E_b/J_0=5$~dB, $\rho=0.1$, and no interleaving, with ideal \ac{JSI} (solid lines) and without \ac{JSI} (dashed lines). In these figures, the optimized code is the one designed for $\mathbf{o}^{\star}$ in Section~\ref{sec:designed}-\ref{subsec:single_profile_design}, and the dotted vertical lines are the thresholds of Table~\ref{tab:single_profile_thresholds}, i.e., the medians for the \ac{AR4JA} codes. Since the design and the simulated jamming patterns coincide, these results test how well the thresholds predict the ranking of the finite-length codes; these codes are optimized for one fixed pulse pattern, without constraints aimed at the error floor, and are not meant to achieve the best error rate.

With ideal \ac{JSI}, the designed codes have the best finite-length performance in the waterfall region at rates $2/3$ and $4/5$, consistently with their thresholds, but their \ac{BER} curves flatten at lower error rates. At rate $2/3$, the designed code has a steeper waterfall than the code of~\cite{GiulianiSoftCOM2026}: its gain is below $0.1$~dB at a \ac{BER} of $10^{-4}$ but grows to about $1$~dB at $10^{-6}$, and it outperforms the \ac{AR4JA} code with $k=4096$ by about $0.2$~dB at $10^{-3}$. Below about $2\cdot10^{-7}$, however, its \ac{BER} decreases only slightly, from $2.0\cdot10^{-7}$ at $3.5$~dB to $1.5\cdot10^{-7}$ at $4$~dB. At rate $4/5$, the code of~\cite{GiulianiSoftCOM2026} is far behind, consistently with its much higher threshold, and the designed code reaches $10^{-4}$ about $0.4$~dB before the \ac{AR4JA} code with $k=4096$. Below $10^{-5}$, however, the \ac{BER} of the designed rate-$4/5$ code decreases slowly, and the \ac{AR4JA} code with $k=4096$ becomes better. This behavior is consistent with a distance-limited floor, as for the ensemble designed for random pulse offsets in Section~\ref{sec:results}-\ref{subsec:main_results}: two degree-two variable-node types of this ensemble also form cycles along the coupled chain.

At rate $1/2$, instead, the finite-length ranking does not strictly follow the threshold ranking. The designed code and the code of~\cite{GiulianiSoftCOM2026}, whose thresholds differ by $0.06$~dB, have close waterfalls, with the code of~\cite{GiulianiSoftCOM2026} slightly ahead in the upper part of the waterfall. The \ac{AR4JA} code with $k=4096$, whose median threshold is $0.13$~dB higher than that of the designed code, has a steeper waterfall and reaches $10^{-4}$ about $0.1$~dB earlier. Hence, at this rate, threshold differences below $0.2$~dB are smaller than the differences among the finite-length losses with respect to the thresholds, which depend on the block length, the lifting factor, and the decoding schedule of each code.

The \ac{AR4JA} codes with $k=1024$ are not competitive at any rate under this pattern, although their median thresholds are close to those of the \ac{AR4JA} codes with $k=4096$. As noted in Section~\ref{sec:results}-\ref{subsec:results_setup}, the jammed fraction of these short codewords varies from codeword to codeword, so that a single threshold does not describe all of them, and the most heavily jammed codewords are expected to dominate the error rate. Accordingly, their $90$th percentiles in Table~\ref{tab:single_profile_thresholds} exceed those of the \ac{AR4JA} codes with $k=4096$ by an amount that grows with the rate, and, at rate $4/5$, the codewords that do not converge at all are consistent with the error floor of the corresponding curve in Fig.~\ref{fig:EbJ05dB_45_SingleProfile}.

Without \ac{JSI}, the receiver uses the mismatched metric of Section~\ref{sec:results}-\ref{subsec:results_setup}, which is not covered by the analysis, so these curves are not useful for testing the thresholds.

\subsection{BER comparison under random pulsed jamming}
\label{subsec:main_results}

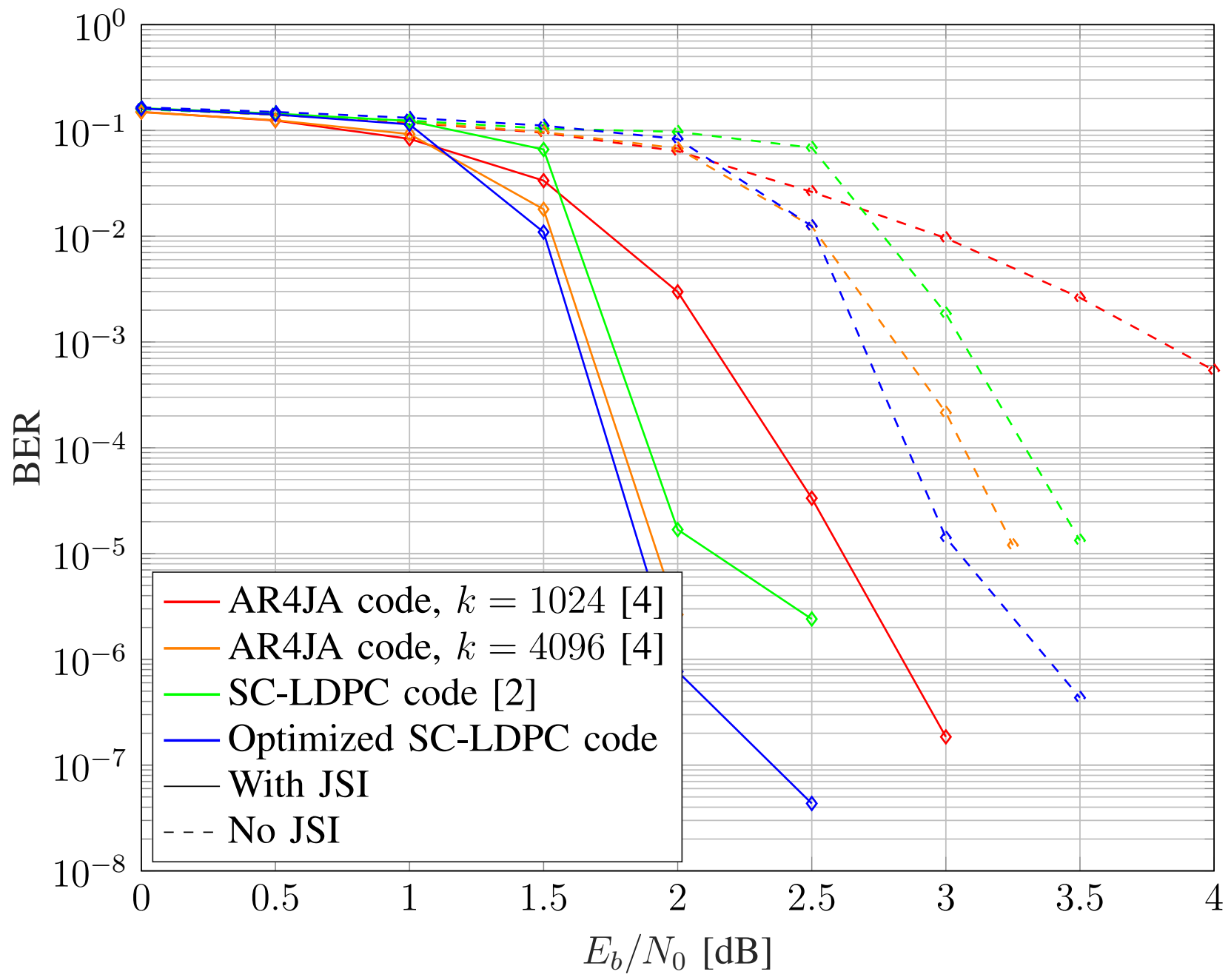
\begin{figure}[t]
    \centering
    \resizebox{0.80\columnwidth}{!}{\definecolor{mycolor1}{rgb}{0.31373,0.78431,0.47059}
\definecolor{mycolor2}{rgb}{1.00000,0.00000,1.00000}

\begin{tikzpicture}

\begin{axis}[
width=4.521in,
height=3.566in,
at={(0.758in,0.481in)},
scale only axis,
xmin=0,
xmax=4,
xlabel style={font=\color{white!15!black}},
xlabel={$E_b/N_0$ [dB]},
ymode=log,
ymin=1e-8,
ymax=1,
yminorticks=true,
ylabel style={font=\color{white!15!black}},
ylabel={BER},
axis background/.style={fill=white},
xmajorgrids,
ymajorgrids,
yminorgrids,
legend style={at={(0.01,0.01)}, anchor=south west, legend cell align=left, align=left, draw=white!15!black}
]

\addplot [color=red, line width=0.6pt, mark=diamond, mark options={red}, forget plot]
  table[row sep=crcr]{
0.00 1.4970E-01\\
0.50 1.2405E-01\\
1.00 8.3267E-02\\
1.50 3.3540E-02\\
2.00 2.9807E-03\\
2.50 3.3353E-05\\
3.00 1.8489E-07\\
};

\addplot [color=red, dashed, line width=0.6pt, mark=diamond, mark options={red}, forget plot]
  table[row sep=crcr]{
0.00 1.6001E-01\\
0.50 1.4051E-01\\
1.00 1.1841E-01\\
1.50 9.4984E-02\\
2.00 6.3560E-02\\
2.50 2.6238E-02\\
3.00 9.5949E-03\\
3.50 2.6207E-03\\
4.00 5.3855E-04\\
4.50 1.4878E-04\\
5.00 4.8933E-05\\
5.50 2.1671E-05\\
6.00 9.2693E-06\\
};

\addplot [color=orange, line width=0.6pt, mark=diamond, mark options={orange}, forget plot]
  table[row sep=crcr]{
0.00 1.4884E-01\\
0.50 1.2491E-01\\
1.00 9.2373E-02\\
1.50 1.7973E-02\\
2.00 2.6454E-06\\
};

\addplot [color=orange, dashed, line width=0.6pt, mark=diamond, mark options={orange}, forget plot]
  table[row sep=crcr]{
0.00 1.6058E-01\\
0.50 1.4128E-01\\
1.00 1.2031E-01\\
1.50 9.6643E-02\\
2.00 6.7707E-02\\
2.50 1.2400E-02\\
3.00 2.1462E-04\\
3.25 1.2016E-05\\
};

\addplot [color=green, line width=0.6pt, mark=diamond, mark options={green}, forget plot]
  table[row sep=crcr]{
0.00 1.6247E-01\\
0.50 1.4437E-01\\
1.00 1.2292E-01\\
1.50 6.6073E-02\\
2.00 1.6845E-05\\
2.50 2.4037E-06\\
};

\addplot [color=green, dashed, line width=0.6pt, mark=diamond, mark options={green}, forget plot]
  table[row sep=crcr]{
0.0 1.5842E-01\\
0.5 1.4167E-01\\
1.0 1.2383E-01\\
1.5 1.0421E-01\\
2.0 9.6975E-02\\
2.5 6.8704E-02\\
3.0 1.8591E-03\\
3.5 1.3296E-05\\
};

\addplot [color=blue, line width=0.6pt, mark=diamond, mark options={blue}, forget plot]
  table[row sep=crcr]{
0.00 1.6105E-01\\
0.50 1.4148E-01\\
1.00 1.1411E-01\\
1.50 1.0934E-02\\
2.00 7.6594E-07\\
2.50 4.3367E-08\\
};

\addplot [color=blue, dashed, line width=0.6pt, mark=diamond, mark options={blue}, forget plot]
  table[row sep=crcr]{
0.00 1.6555E-01\\
0.50 1.4954E-01\\
1.00 1.3162E-01\\
1.50 1.1111E-01\\
2.00 8.3034E-02\\
2.50 1.2454E-02\\
3.00 1.4126E-05\\
3.50 4.3348E-07\\
};
\addlegendimage{red, thick} \addlegendentry{AR4JA code, $k = 1024$ \cite{TMst}}
\addlegendimage{orange, thick} \addlegendentry{AR4JA code, $k = 4096$ \cite{TMst}}
\addlegendimage{green, thick}  \addlegendentry{SC-LDPC code \cite{GiulianiSoftCOM2026}}
\addlegendimage{blue, thick}  \addlegendentry{Optimized SC-LDPC code}
\addlegendimage{black, solid}
\addlegendentry{With JSI}
\addlegendimage{black, dashed}
\addlegendentry{No JSI}

\end{axis}

\begin{axis}[
width=5.833in,
height=4.375in,
at={(0in,0in)},
scale only axis,
xmin=0,
xmax=1,
ymin=0,
ymax=1,
axis line style={draw=none},
ticks=none,
axis x line*=bottom,
axis y line*=left
]
\end{axis}
\end{tikzpicture}}
    \caption{\ac{BER} comparison at $R=1/2$, $E_b/J_0=5$~dB, $\rho=0.1$, and no interleaving, under random pulsed jamming.}
    \label{fig:EbJ05dB_12_RandJam}
\end{figure}

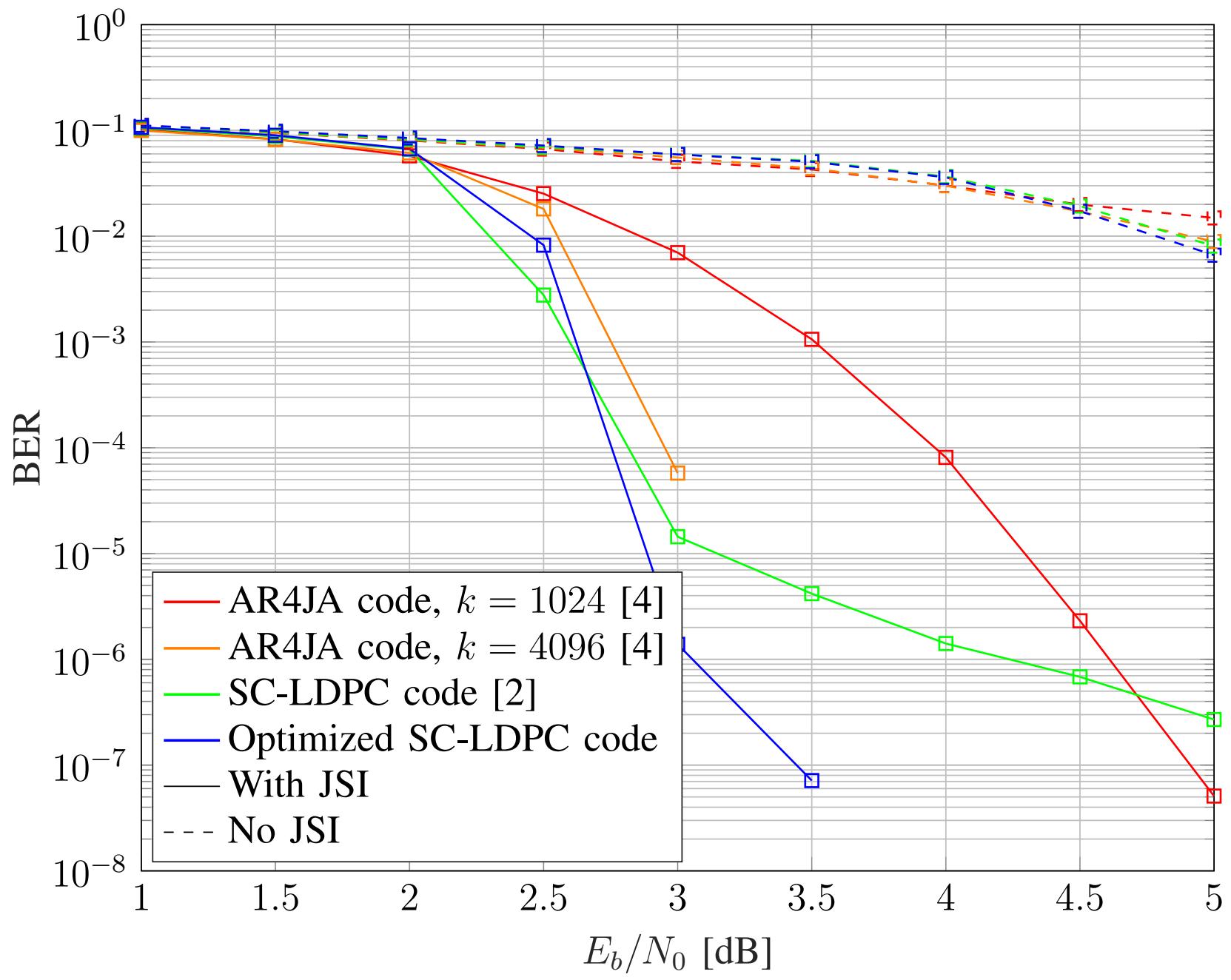
\begin{figure}[t]
    \centering
    \resizebox{0.80\columnwidth}{!}{\definecolor{mycolor1}{rgb}{0.31373,0.78431,0.47059}
\definecolor{mycolor2}{rgb}{1.00000,0.00000,1.00000}

\begin{tikzpicture}

\begin{axis}[
width=4.521in,
height=3.566in,
at={(0.758in,0.481in)},
scale only axis,
xmin=1,
xmax=5,
xlabel style={font=\color{white!15!black}},
xlabel={$E_b/N_0$ [dB]},
ymode=log,
ymin=1e-8,
ymax=1,
yminorticks=true,
ylabel style={font=\color{white!15!black}},
ylabel={BER},
axis background/.style={fill=white},
xmajorgrids,
ymajorgrids,
yminorgrids,
legend style={at={(0.01,0.01)}, anchor=south west, legend cell align=left, align=left, draw=white!15!black, text opacity=1}
]

\addplot [color=red, line width=0.6pt, mark=square, mark options={red}, forget plot]
  table[row sep=crcr]{
1.00 1.0179E-01\\
1.50 8.2622E-02\\
2.00 5.7519E-02\\
2.50 2.5271E-02\\
3.00 6.9973E-03\\
3.50 1.0606E-03\\
4.00 8.1071E-05\\
4.50 2.3135E-06\\
5.00 5.0965E-08\\
};

\addplot [color=red, dashed, line width=0.6pt, mark=square, mark options={red}, forget plot]
  table[row sep=crcr]{
1.00 1.1074E-01\\
1.50 9.4542E-02\\
2.00 8.0430E-02\\
2.50 6.6677E-02\\
3.00 5.1003E-02\\
3.50 4.2736E-02\\
4.00 3.0104E-02\\
4.50 1.9924E-02\\
5.00 1.4892E-02\\
5.50 1.1988E-02\\
6.00 1.1808E-02\\
6.50 8.6292E-03\\
7.00 7.2174E-03\\
};

\addplot [color=orange, line width=0.6pt, mark=square, mark options={orange}, forget plot]
  table[row sep=crcr]{
1.00 9.9848E-02\\
1.50 8.2200E-02\\
2.00 6.1283E-02\\
2.50 1.8113E-02\\
3.00 5.7642E-05\\
};

\addplot [color=orange, dashed, line width=0.6pt, mark=square, mark options={orange}, forget plot]
  table[row sep=crcr]{
1.00 1.0867E-01\\
1.50 9.4880E-02\\
2.00 8.1359E-02\\
2.50 6.8556E-02\\
3.00 5.5483E-02\\
3.50 4.3737E-02\\
4.00 3.0018E-02\\
4.50 1.7106E-02\\
5.00 8.9110E-03\\
};

\addplot [color=green, line width=0.6pt, mark=square, mark options={green}, forget plot]
  table[row sep=crcr]{
1.00 1.0520E-01\\
1.50 8.7796E-02\\
2.00 6.6312E-02\\
2.50 2.7753E-03\\
3.00 1.4419E-05\\
3.50 4.1776E-06\\
4.00 1.4103E-06\\
4.50 6.8253E-07\\
5.00 2.6972E-07\\
};

\addplot [color=green, dashed, line width=0.6pt, mark=square, mark options={green}, forget plot]
  table[row sep=crcr]{
1.00 1.0995E-01\\
1.50 9.6111E-02\\
2.00 8.2391E-02\\
2.50 6.9440E-02\\
3.00 5.9380E-02\\
3.50 5.1571E-02\\
4.00 3.6617E-02\\
4.50 1.9294E-02\\
5.00 8.0658E-03\\
};

\addplot [color=blue, line width=0.6pt, mark=square, mark options={blue}, forget plot]
  table[row sep=crcr]{
1.00 1.0709E-01\\
1.50 8.9995E-02\\
2.00 6.7058E-02\\
2.50 8.2378E-03\\
3.00 1.3862E-06\\
3.50 7.1431E-08\\
};

\addplot [color=blue, dashed, line width=0.6pt, mark=square, mark options={blue}, forget plot]
  table[row sep=crcr]{
1.00 1.1166E-01\\
1.50 9.8283E-02\\
2.00 8.4810E-02\\
2.50 7.1944E-02\\
3.00 5.9258E-02\\
3.50 5.0588E-02\\
4.00 3.6000E-02\\
4.50 1.7244E-02\\
5.00 6.6167E-03\\
};

\addlegendimage{red, thick} \addlegendentry{AR4JA code, $k = 1024$ \cite{TMst}}
\addlegendimage{orange, thick} \addlegendentry{AR4JA code, $k = 4096$ \cite{TMst}}
\addlegendimage{green, thick}  \addlegendentry{SC-LDPC code \cite{GiulianiSoftCOM2026}}
\addlegendimage{blue, thick}  \addlegendentry{Optimized SC-LDPC code}
\addlegendimage{black, solid}
\addlegendentry{With JSI}
\addlegendimage{black, dashed}
\addlegendentry{No JSI}

\end{axis}

\begin{axis}[
width=5.833in,
height=4.375in,
at={(0in,0in)},
scale only axis,
xmin=0,
xmax=1,
ymin=0,
ymax=1,
axis line style={draw=none},
ticks=none,
axis x line*=bottom,
axis y line*=left
]
\end{axis}
\end{tikzpicture}}
    \caption{\ac{BER} comparison at $R=2/3$, $E_b/J_0=5$~dB, $\rho=0.1$, and no interleaving, under random pulsed jamming.}
    \label{fig:EbJ05dB_23_RandJam}
\end{figure}

\begin{figure}[t]
    \centering
    \resizebox{0.80\columnwidth}{!}{\definecolor{mycolor1}{rgb}{0.31373,0.78431,0.47059}
\definecolor{mycolor2}{rgb}{1.00000,0.00000,1.00000}

\begin{tikzpicture}

\begin{axis}[
width=4.521in,
height=3.566in,
at={(0.758in,0.481in)},
scale only axis,
xmin=2.5,
xmax=6.5,
xlabel style={font=\color{white!15!black}},
xlabel={$E_b/N_0$ [dB]},
ymode=log,
ymin=1e-8,
ymax=1,
yminorticks=true,
ylabel style={font=\color{white!15!black}},
ylabel={BER},
axis background/.style={fill=white},
xmajorgrids,
ymajorgrids,
yminorgrids,
legend style={at={(0.01,0.01)}, anchor=south west, legend cell align=left, align=left, draw=white!15!black, text opacity=1}
]

\addplot [color=red, line width=0.6pt, mark=x, mark options={red}, forget plot]
  table[row sep=crcr]{
2.50 5.8558E-02\\
3.00 4.6594E-02\\
3.50 3.0299E-02\\
4.00 1.0096E-02\\
4.50 1.4667E-03\\
5.00 8.0645E-05\\
5.50 2.6186E-06\\
6.00 1.4932E-07\\
6.5 3.7419E-08\\
};

\addplot [color=red, dashed, line width=0.6pt, mark=x, mark options={red}, forget plot]
  table[row sep=crcr]{
2.50 6.3970E-02\\
3.00 5.5874E-02\\
3.50 4.8587E-02\\
4.00 4.1584E-02\\
4.50 3.6447E-02\\
5.00 3.3929E-02\\
5.50 2.7761E-02\\
6.00 2.6458E-02\\
6.50 2.4925E-02\\
7.00 2.4997E-02\\
7.50 2.1981E-02\\
8.00 2.2964E-02\\
8.50 1.9926E-02\\
};

\addplot [color=orange, line width=0.6pt, mark=x, mark options={orange}, forget plot]
  table[row sep=crcr]{
2.50 5.7001E-02\\
3.00 4.5574E-02\\
3.50 2.9190E-02\\
4.00 4.6526E-03\\
4.50 2.4235E-05\\
4.75 7.8656E-07\\
};

\addplot [color=orange, dashed, line width=0.6pt, mark=x, mark options={orange}, forget plot]
  table[row sep=crcr]{
2.50 6.4197E-02\\
3.00 5.5105E-02\\
3.50 4.7553E-02\\
4.00 4.1905E-02\\
4.50 3.7035E-02\\
5.00 3.1946E-02\\
5.50 3.0529E-02\\
6.00 2.6746E-02\\
6.50 2.5911E-02\\
};

\addplot [color=green, line width=0.6pt, mark=x, mark options={green}, forget plot]
  table[row sep=crcr]{
2.50 6.5284E-02\\
3.00 5.5575E-02\\
3.50 4.4732E-02\\
4.00 1.6250E-02\\
4.50 1.3475E-03\\
5.00 4.4510E-04\\
5.50 1.9063E-04\\
6.00 1.3168E-04\\
6.50 8.2160E-05\\
};

\addplot [color=green, dashed, line width=0.6pt, mark=x, mark options={green}, forget plot]
  table[row sep=crcr]{
2.50 6.6396E-02\\
3.00 5.7889E-02\\
3.50 5.0151E-02\\
4.00 4.3817E-02\\
4.50 4.0355E-02\\
5.00 4.0031E-02\\
5.50 3.8219E-02\\
6.00 3.5535E-02\\
6.50 3.3850E-02\\
};

\addplot [color=blue, line width=0.6pt, mark=x, mark options={blue}, forget plot]
  table[row sep=crcr]{
2.50 5.8552E-02\\
3.00 4.4092E-02\\
3.50 1.6935E-02\\
4.00 5.2855E-04\\
4.50 2.8564E-05\\
5.00 6.1327E-06\\
5.50 1.4920E-06\\
6.00 1.0275E-06\\
6.50 7.5231E-07\\
};

\addplot [color=blue, dashed, line width=0.6pt, mark=x, mark options={blue}, forget plot]
  table[row sep=crcr]{
2.50 6.4273E-02\\
3.00 5.5276E-02\\
3.50 4.7682E-02\\
4.00 4.1836E-02\\
4.50 3.8725E-02\\
5.00 3.5978E-02\\
5.50 3.3040E-02\\
6.00 2.9923E-02\\
6.50 2.7914E-02\\
7.00 2.5440E-02\\
7.50 2.4047E-02\\
};

\addlegendimage{red, thick} \addlegendentry{AR4JA code, $k = 1024$ \cite{TMst}}
\addlegendimage{orange, thick} \addlegendentry{AR4JA code, $k = 4096$ \cite{TMst}}
\addlegendimage{green, thick}  \addlegendentry{SC-LDPC code \cite{GiulianiSoftCOM2026}}
\addlegendimage{blue, thick}  \addlegendentry{Optimized SC-LDPC code}
\addlegendimage{black, solid}
\addlegendentry{With JSI}
\addlegendimage{black, dashed}
\addlegendentry{No JSI}

\end{axis}

\begin{axis}[
width=5.833in,
height=4.375in,
at={(0in,0in)},
scale only axis,
xmin=0,
xmax=1,
ymin=0,
ymax=1,
axis line style={draw=none},
ticks=none,
axis x line*=bottom,
axis y line*=left
]
\end{axis}
\end{tikzpicture}}
    \caption{\ac{BER} comparison at $R=4/5$, $E_b/J_0=5$~dB, $\rho=0.1$, and no interleaving, under random pulsed jamming.}
    \label{fig:EbJ05dB_45_RandJam}
\end{figure}
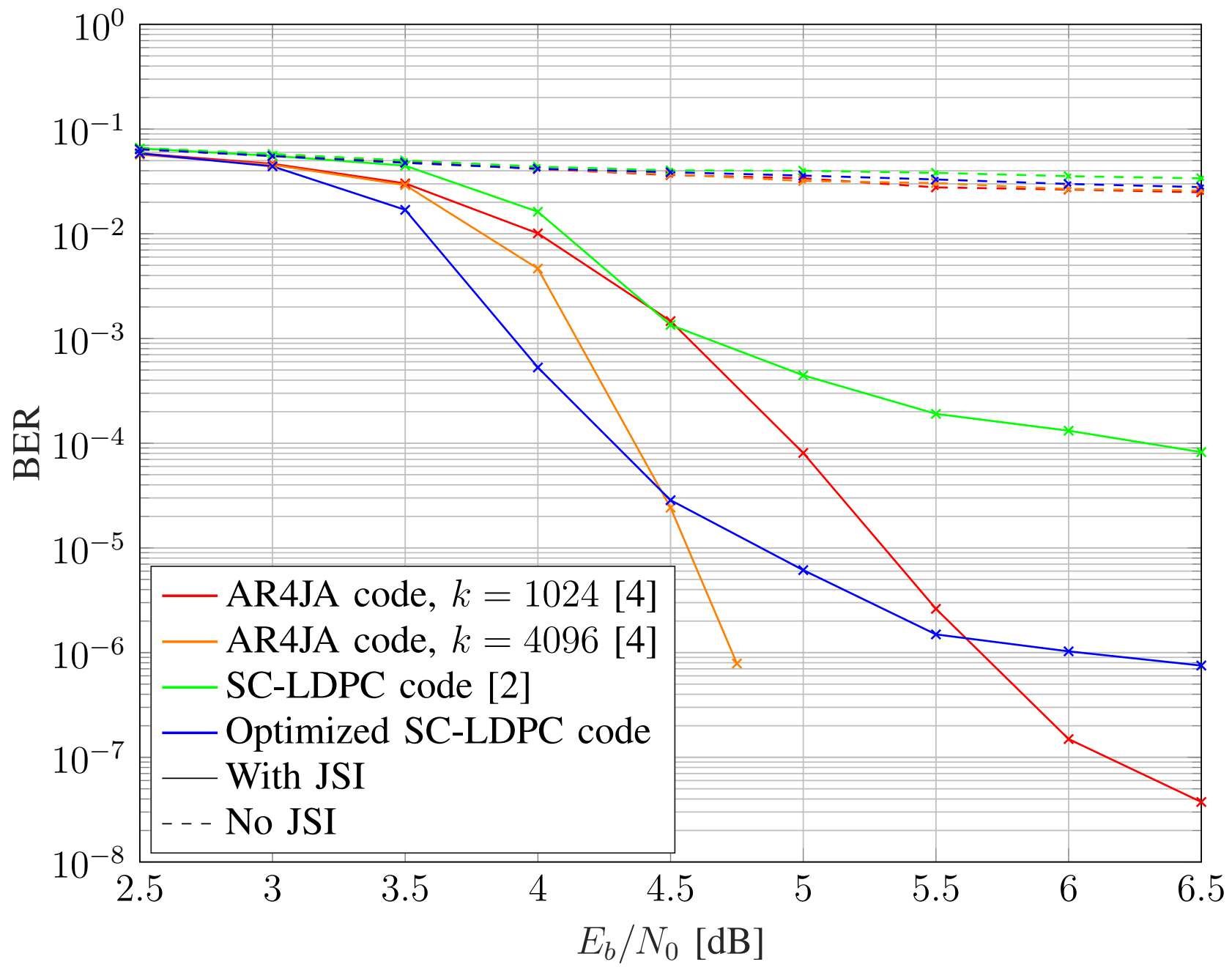

We now consider the ensembles designed for random pulse offsets, which are the main outcome of the design, under random pulsed jamming. Figs.~\ref{fig:EbJ05dB_12_RandJam}, \ref{fig:EbJ05dB_23_RandJam}, and~\ref{fig:EbJ05dB_45_RandJam} report the results for rates $R=1/2$, $R=2/3$, and $R=4/5$, respectively, with the same parameters as in Figs.~\ref{fig:EbJ05dB_12_SingleProfile}--\ref{fig:EbJ05dB_45_SingleProfile}. In these figures, the optimized code is the one designed for random pulse offsets in Section~\ref{sec:designed}-\ref{subsec:selected_ensembles}. No single threshold applies to this case, since the jammed fractions change from frame to frame; the design objectives of these ensembles over the random validation profiles are reported in Table~\ref{tab:design_validation}.

For $R=1/2$, with ideal \ac{JSI}, the optimized code reaches a \ac{BER} of $10^{-4}$ before all the benchmarks. Its margin is only a few hundredths of a dB over the \ac{AR4JA} code with $k=4096$ and about $0.14$~dB over the code of~\cite{GiulianiSoftCOM2026}, whereas the \ac{AR4JA} code with $k=1024$ is about $0.6$~dB behind; the steeper waterfall of the optimized code increases its gain over the code of~\cite{GiulianiSoftCOM2026} at lower error rates. Without \ac{JSI}, the advantage persists and is larger: the optimized code reaches $10^{-4}$ about $0.2$~dB before the \ac{AR4JA} code with $k=4096$ and about $0.4$~dB before the code of~\cite{GiulianiSoftCOM2026}, whereas the \ac{AR4JA} code with $k=1024$ is far behind.

For $R=2/3$, with ideal \ac{JSI}, the optimized code reaches $10^{-4}$ about $0.2$~dB before the \ac{AR4JA} code with $k=4096$ and slightly before the code of~\cite{GiulianiSoftCOM2026}, whereas the \ac{AR4JA} code with $k=1024$ is more than $1$~dB behind. As in the rate-$1/2$ case, the gain over the code of~\cite{GiulianiSoftCOM2026} grows at lower error rates, to about $1$~dB at $10^{-6}$. Without \ac{JSI}, none of the codes reaches the waterfall region in the plotted range.

For $R=4/5$, with ideal \ac{JSI}, the optimized code reaches $10^{-4}$ slightly before the \ac{AR4JA} code with $k=4096$ and well before the \ac{AR4JA} code with $k=1024$ and the code of~\cite{GiulianiSoftCOM2026}. Its \ac{BER} then decreases slowly, so that both \ac{AR4JA} codes eventually become better: the code with $k=4096$ shortly below $10^{-4}$, and the code with $k=1024$ below about $10^{-6}$, both with a clear advantage at the lowest simulated error rates. This is consistent with a distance-limited floor: two degree-two variable-node types of the ensemble form cycles along the coupled chain, which typically produces low-weight codewords. Again, without \ac{JSI}, none of the codes reaches the waterfall region in the plotted range.

\subsection{Effect of the jammer duty cycle}
\label{subsec:rho_results}

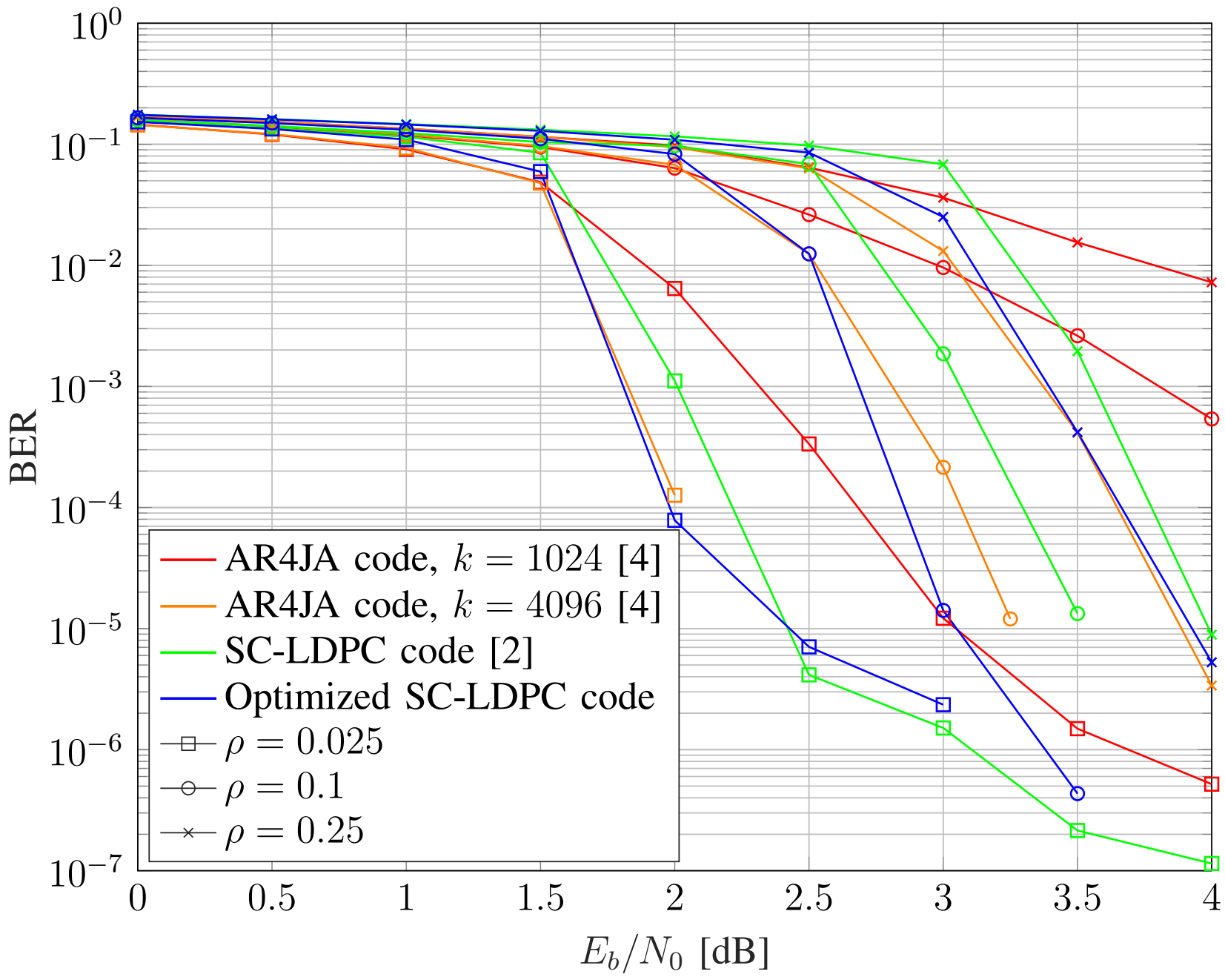
\begin{figure}[t]
    \centering
    \resizebox{0.80\columnwidth}{!}{\definecolor{mycolor1}{rgb}{0.31373,0.78431,0.47059}
\definecolor{mycolor2}{rgb}{1.00000,0.00000,1.00000}
\begin{tikzpicture}

\begin{axis}[
width=4.521in,
height=3.566in,
at={(0.758in,0.481in)},
scale only axis,
xmin=0,
xmax=4,
xlabel style={font=\color{white!15!black}},
xlabel={$E_b/N_0$ [dB]},
ymode=log,
ymin=1e-7,
ymax=1,
yminorticks=true,
ylabel style={font=\color{white!15!black}},
ylabel={BER},
axis background/.style={fill=white},
xmajorgrids,
ymajorgrids,
yminorgrids,
legend style={at={(0.01,0.01)}, anchor=south west, legend cell align=left, align=left, draw=white!15!black}
]

\addplot [color=red, line width=0.6pt, mark=o, mark options={red}, forget plot]
  table[row sep=crcr]{%
0.00 1.6001E-01\\
0.50 1.4051E-01\\
1.00 1.1841E-01\\
1.50 9.4984E-02\\
2.00 6.3560E-02\\
2.50 2.6238E-02\\
3.00 9.5949E-03\\
3.50 2.6207E-03\\
4.00 5.3855E-04\\
4.50 1.4878E-04\\
5.00 4.8933E-05\\
5.50 2.1671E-05\\
6.00 9.2693E-06\\
};

\addplot [color=red, line width=0.6pt, mark=x, mark options={red}, forget plot]
  table[row sep=crcr]{%
 0.00 1.7046E-01\\
 0.50 1.5346E-01\\
 1.00 1.3462E-01\\
 1.50 1.1522E-01\\
 2.00 9.7645E-02\\
 2.50 6.4155E-02\\
 3.00 3.6250E-02\\
 3.50 1.5464E-02\\
 4.00 7.2443E-03\\
 4.50 2.3539E-03\\
 5.00 8.5107E-04\\
 5.50 2.8906E-04\\
 6.00 8.5740E-05\\
};

\addplot [color=red, line width=0.6pt, mark=square, mark options={red}, forget plot]
  table[row sep=crcr]{%
 0.00 1.4551E-01\\
 0.50 1.2053E-01\\
 1.00 9.0852E-02\\
 1.50 4.8590E-02\\
 2.00 6.4381E-03\\
 2.50 3.3483E-04\\
 3.00 1.2176E-05\\
 3.50 1.4900E-06\\
 4.00 5.1870E-07\\
};

\addplot [color=orange, line width=0.6pt, mark=x, mark options={orange}, forget plot]
  table[row sep=crcr]{%
 0.00 1.7096E-01\\
 0.50 1.5458E-01\\
 1.00 1.3441E-01\\
 1.50 1.1519E-01\\
 2.00 9.4637E-02\\
 2.50 6.3172E-02\\
 3.00 1.3132E-02\\
 3.50 4.1418E-04\\
 4.00 3.3816E-06\\
 };

\addplot [color=orange, line width=0.6pt, mark=square, mark options={orange}, forget plot]
  table[row sep=crcr]{%
0.00 1.4447E-01\\
0.50 1.2165E-01\\
1.00 9.3193E-02\\
1.50 4.7801E-02\\
2.00 1.2624E-04\\
};

\addplot [color=orange, line width=0.6pt, mark=o, mark options={orange}, forget plot]
  table[row sep=crcr]{%
0.00 1.6058E-01\\
0.50 1.4128E-01\\
1.00 1.2031E-01\\
1.50 9.6643E-02\\
2.00 6.7707E-02\\
2.50 1.2400E-02\\
3.00 2.1462E-04\\
3.25 1.2016E-05\\
};

\addplot [color=green, line width=0.6pt, mark=o, mark options={green}, forget plot]
  table[row sep=crcr]{%
0.0 1.5842E-01\\
0.5 1.4167E-01\\
1.0 1.2383E-01\\
1.5 1.0421E-01\\
2.0 9.6975E-02\\
2.5 6.8704E-02\\
3.0 1.8591E-03\\
3.5 1.3296E-05\\
};

\addplot [color=green, line width=0.6pt, mark=x, mark options={green}, forget plot]
  table[row sep=crcr]{%
0.00 1.7569E-01\\
0.50 1.6140E-01\\
1.00 1.4704E-01\\
1.50 1.3212E-01\\
2.00 1.1620E-01\\
2.50 9.7623E-02\\
3.00 6.8312E-02\\
3.50 1.9542E-03\\
4.00 8.9006E-06\\
};

\addplot [color=green, line width=0.6pt, mark=square, mark options={green}, forget plot]
  table[row sep=crcr]{%
0.00 1.5405E-01\\
0.50 1.3561E-01\\
1.00 1.1531E-01\\
1.50 8.5143E-02\\
2.00 1.1108E-03\\
2.50 4.1477E-06\\
3.00 1.5102E-06\\
3.50 2.1479E-07\\
4.00 1.1467E-07\\
};

\addplot [color=blue, line width=0.6pt, mark=o, mark options={blue}, forget plot]
  table[row sep=crcr]{%
0.00 1.6555E-01\\
0.50 1.4954E-01\\
1.00 1.3162E-01\\
1.50 1.1111E-01\\
2.00 8.3034E-02\\
2.50 1.2454E-02\\
3.00 1.4126E-05\\
3.50 4.3348E-07\\
};

\addplot [color=blue, line width=0.6pt, mark=x, mark options={blue}, forget plot]
  table[row sep=crcr]{%
0.00 1.7516E-01\\
0.50 1.6076E-01\\
1.00 1.4572E-01\\
1.50 1.2896E-01\\
2.00 1.0902E-01\\
2.50 8.5316E-02\\
3.00 2.5114E-02\\
3.50 4.1894E-04\\
4.00 5.2647E-06\\
};

\addplot [color=blue, line width=0.6pt, mark=square, mark options={blue}, forget plot]
  table[row sep=crcr]{%
0.00 1.5303E-01\\
0.50 1.3383E-01\\
1.00 1.0933E-01\\
1.50 5.9473E-02\\
2.00 7.8362E-05\\
2.50 7.0605E-06\\
3.00 2.3502E-06\\
};

\addlegendimage{red, thick} \addlegendentry{AR4JA code, $k = 1024$ \cite{TMst}}
\addlegendimage{orange, thick} \addlegendentry{AR4JA code, $k = 4096$ \cite{TMst}}
\addlegendimage{green, thick}  \addlegendentry{SC-LDPC code \cite{GiulianiSoftCOM2026}}
\addlegendimage{blue, thick}  \addlegendentry{Optimized SC-LDPC code}
\addlegendimage{mark=square}
\addlegendentry{$\rho=0.025$}
\addlegendimage{mark=o}
\addlegendentry{$\rho=0.1$}
\addlegendimage{mark=x}
\addlegendentry{$\rho=0.25$}

\end{axis}

\begin{axis}[
width=5.833in,
height=4.375in,
at={(0in,0in)},
scale only axis,
xmin=0,
xmax=1,
ymin=0,
ymax=1,
axis line style={draw=none},
ticks=none,
axis x line*=bottom,
axis y line*=left
]
\end{axis}
\end{tikzpicture}}
    \caption{Effect of the duty cycle under random pulsed jamming at $R=1/2$, $E_b/J_0=5~{\rm dB}$, without interleaving or \ac{JSI}.}
    \label{fig:EbJ05dB_rho_comparison}
\end{figure}

We next evaluate whether the conclusions obtained for $\rho=0.1$ are specific to that duty cycle. Fig.~\ref{fig:EbJ05dB_rho_comparison} reports the performance of the rate-$1/2$ codes under random pulsed jamming without \ac{JSI} for different values of $\rho$, while keeping $E_b/J_0=5$ dB and using no symbol interleaving. The $\rho=0.1$ curves are those of Fig.~\ref{fig:EbJ05dB_12_RandJam} without \ac{JSI}.

Fig.~\ref{fig:EbJ05dB_rho_comparison} shows that, for all codes, $\rho=0.025$ gives the lowest and $\rho=0.25$ the highest \ac{BER}, i.e., at this jammer strength concentrating the jammer power on fewer symbols is less harmful than spreading it. This is consistent with the uncoded analysis recalled in Section~\ref{sec:preli}-\ref{subsec:pulsed}~\cite{SimonSS}: at this jammer strength, and with the thermal noise of the considered operating points, it also gives an error probability that increases with $\rho$ over the considered range, and very short pulses become the most harmful choice only when $E_b/J_0$ is large and the thermal noise is negligible. In the waterfall region, the optimized code outperforms the code of~\cite{GiulianiSoftCOM2026} and the \ac{AR4JA} code with $k=1024$ at all three duty cycles, by $0.1$ to $0.4$~dB at a \ac{BER} of $10^{-4}$ with respect to the code of~\cite{GiulianiSoftCOM2026}. Its margin over the \ac{AR4JA} code with $k=4096$, instead, is about $0.2$~dB at the design duty cycle and nearly vanishes for $\rho=0.025$ and $\rho=0.25$, where the two curves are close in the simulated range, with the \ac{AR4JA} code slightly ahead for $\rho=0.25$. Hence, the advantage of the proposed design over the non-optimized \ac{SC-LDPC} code and over the \ac{AR4JA} code with $k=1024$ is not restricted to a single pulse occupancy, whereas its margin over the longer \ac{AR4JA} code is largest at the design duty cycle. For $\rho=0.025$, however, the \ac{BER} of the optimized code decreases slowly below $10^{-5}$, where the code of~\cite{GiulianiSoftCOM2026} becomes better.

\subsection{Effect of symbol interleaving}
\label{subsec:interleaving_results}

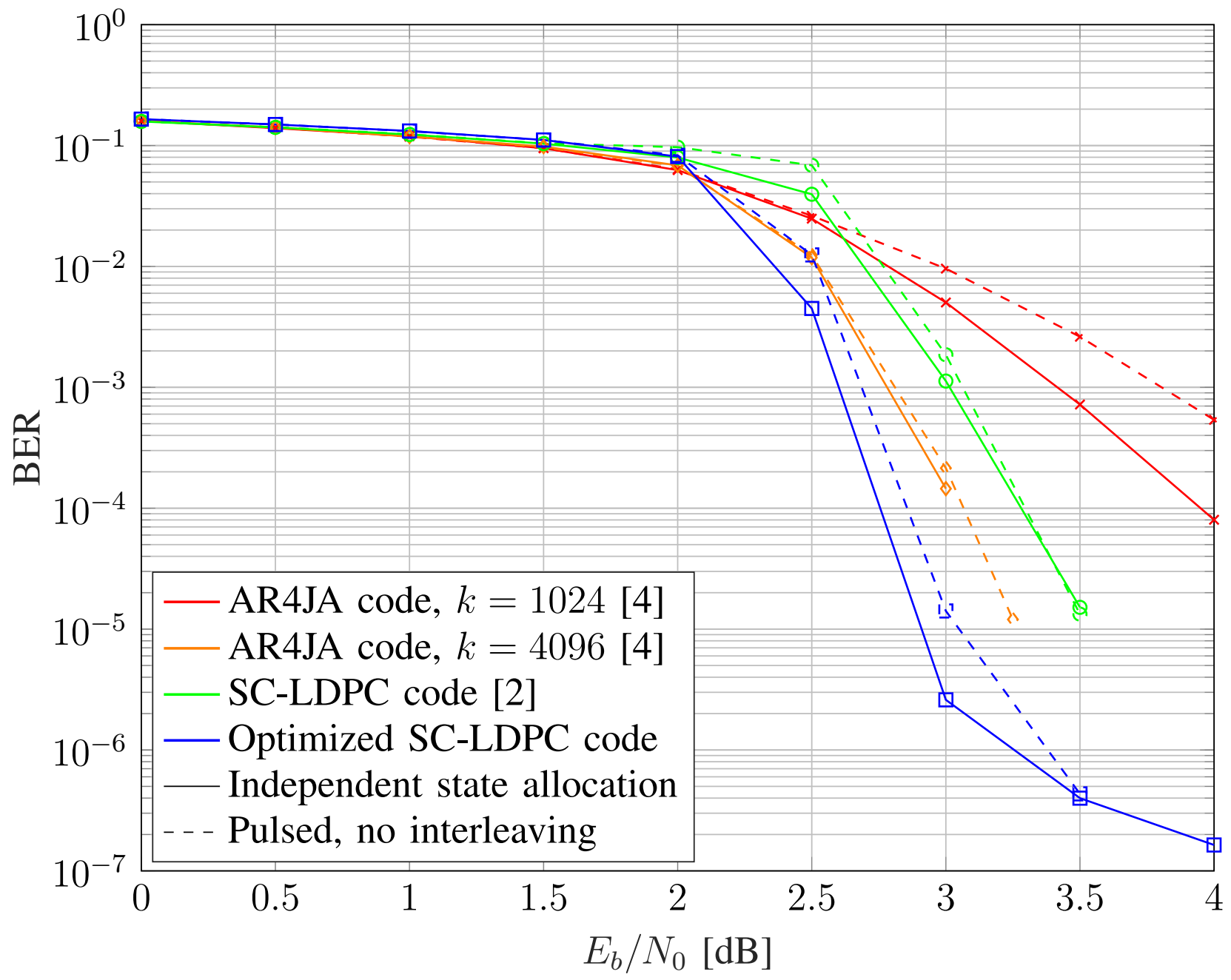
\begin{figure}[t]
    \centering
    \resizebox{0.80\columnwidth}{!}{\definecolor{mycolor1}{rgb}{0.31373,0.78431,0.47059}
\definecolor{mycolor2}{rgb}{1.00000,0.00000,1.00000}

\begin{tikzpicture}

\begin{axis}[%
width=4.521in,
height=3.566in,
at={(0.758in,0.481in)},
scale only axis,
xmin=0,
xmax=4,
xlabel style={font=\color{white!15!black}},
xlabel={$E_b/N_0$ [dB]},
ymode=log,
ymin=1e-7,
ymax=1,
yminorticks=true,
ylabel style={font=\color{white!15!black}},
ylabel={BER},
axis background/.style={fill=white},
xmajorgrids,
ymajorgrids,
yminorgrids,
legend style={at={(0.01,0.01)}, anchor=south west, legend cell align=left, align=left, draw=white!15!black}
]

\addplot [color=red, line width=0.6pt, mark=x, mark options={red}, forget plot]
  table[row sep=crcr]{%
 0.00 1.6007E-01\\
 0.50 1.3923E-01\\
 1.00 1.1936E-01\\
 1.50 9.5043E-02\\
 2.00 6.2617E-02\\
 2.50 2.4838E-02\\
 3.00 5.0336E-03\\
 3.50 7.2070E-04\\
 4.00 8.0231E-05\\
 4.50 1.2474E-05\\
 5.00 2.3574E-06\\
 5.50 4.2673E-07\\
};

\addplot [color=red, dashed, line width=0.6pt, mark=x, mark options={red}, forget plot]
  table[row sep=crcr]{%
0.00 1.6001E-01\\
0.50 1.4051E-01\\
1.00 1.1841E-01\\
1.50 9.4984E-02\\
2.00 6.3560E-02\\
2.50 2.6238E-02\\
3.00 9.5949E-03\\
3.50 2.6207E-03\\
4.00 5.3855E-04\\
4.50 1.4878E-04\\
5.00 4.8933E-05\\
5.50 2.1671E-05\\
6.00 9.2693E-06\\
};

\addplot [color=orange, line width=0.6pt, mark=diamond, mark options={orange}, forget plot]
  table[row sep=crcr]{%
 0.00 1.6015E-01\\
 0.50 1.4108E-01\\
 1.00 1.1871E-01\\
 1.50 9.8060E-02\\
 2.00 6.8539E-02\\
 2.50 1.1830E-02\\
 3.00 1.4515E-04\\
};

\addplot [color=orange, dashed, line width=0.6pt, mark=diamond, mark options={orange}, forget plot]
  table[row sep=crcr]{%
0.00 1.6058E-01\\
0.50 1.4128E-01\\
1.00 1.2031E-01\\
1.50 9.6643E-02\\
2.00 6.7707E-02\\
2.50 1.2400E-02\\
3.00 2.1462E-04\\
3.25 1.2016E-05\\
};

\addplot [color=green, line width=0.6pt, mark=o, mark options={green}, forget plot]
  table[row sep=crcr]{%
0.00 1.5817E-01\\
0.50 1.4133E-01\\
1.00 1.2330E-01\\
1.50 1.0367E-01\\
2.00 7.9640E-02\\
2.50 3.9549E-02\\
3.00 1.1268E-03\\
3.50 1.5138E-05\\
};

\addplot [color=green, dashed, line width=0.6pt, mark=o, mark options={green}, forget plot]
  table[row sep=crcr]{%
0.0 1.5842E-01\\
0.5 1.4167E-01\\
1.0 1.2383E-01\\
1.5 1.0421E-01\\
2.0 9.6975E-02\\
2.5 6.8704E-02\\
3.0 1.8591E-03\\
3.5 1.3296E-05\\
};

\addplot [color=blue, line width=0.6pt, mark=square, mark options={blue}, forget plot]
  table[row sep=crcr]{%
0.00 1.6543E-01\\
0.50 1.4932E-01\\
1.00 1.3199E-01\\
1.50 1.1123E-01\\
2.00 8.1011E-02\\
2.50 4.4910E-03\\
3.00 2.5992E-06\\
3.50 3.9956E-07\\
4.00 1.6378E-07\\
};

\addplot [color=blue, dashed, line width=0.6pt, mark=square, mark options={blue}, forget plot]
  table[row sep=crcr]{%
0.00 1.6555E-01\\
0.50 1.4954E-01\\
1.00 1.3162E-01\\
1.50 1.1111E-01\\
2.00 8.3034E-02\\
2.50 1.2454E-02\\
3.00 1.4126E-05\\
3.50 4.3348E-07\\
};

\addlegendimage{red, thick} \addlegendentry{AR4JA code,  $k=1024$ \cite{TMst}}
\addlegendimage{orange, thick} \addlegendentry{AR4JA code,  $k=4096$ \cite{TMst}}
\addlegendimage{green, thick}  \addlegendentry{SC-LDPC code \cite{GiulianiSoftCOM2026}}
\addlegendimage{blue, thick}  \addlegendentry{Optimized SC-LDPC code}
\addlegendimage{black, solid}
\addlegendentry{Independent state allocation}
\addlegendimage{black, dashed}
\addlegendentry{Pulsed, no interleaving}
\end{axis}

\begin{axis}[
width=5.833in,
height=4.375in,
at={(0in,0in)},
scale only axis,
xmin=0,
xmax=1,
ymin=0,
ymax=1,
axis line style={draw=none},
ticks=none,
axis x line*=bottom,
axis y line*=left
]
\end{axis}

\end{tikzpicture}}
    \caption{\ac{BER} comparison between non-interleaved random pulsed jamming and independent jamming state allocation (ideal interleaving) at $R=1/2$, $E_b/J_0=5~{\rm dB}$, $\rho=0.1$, and without \ac{JSI}.}
    \label{fig:EbJ05dB_Interleaver}
\end{figure}

Finally, we assess the effect of interleaving by comparing non-interleaved random pulsed jamming with an idealized homogeneous state allocation, where each symbol is independently assigned to the jammed state with probability $\rho$ and to the unjammed state with probability $1-\rho$. This allocation reproduces the assumption of the homogeneous two-state \ac{PEXIT} analysis and emulates the mixing effect of an ideal interleaver. Fig.~\ref{fig:EbJ05dB_Interleaver} compares the two models at $R=1/2$, $E_b/J_0=5$~dB, and $\rho=0.1$, without \ac{JSI}; the non-interleaved curves are those of Fig.~\ref{fig:EbJ05dB_12_RandJam} without \ac{JSI}.

For the \ac{AR4JA} code with $k=1024$, the homogeneous allocation gives a clear gain, about $0.4$~dB at a \ac{BER} of $10^{-3}$. For the \ac{AR4JA} code with $k=4096$ and for both \ac{SC-LDPC} codes, instead, the two allocations give close curves, within about $0.15$~dB over the plotted range. Even without interleaving, the optimized code reaches $10^{-3}$ before both \ac{AR4JA} codes with ideal interleaving. Interleaving is therefore beneficial mainly for the short \ac{AR4JA} code, whose codewords span $1.6$ jamming periods and thus intersect from zero to three pulses, possibly truncated, whereas an \ac{AR4JA} codeword with $k=4096$ spans $6.4$ periods and the \ac{SWD} window of the optimized code spans about $8$ periods. For the \ac{SC-LDPC} codes, this behavior also agrees with the analysis of Section~\ref{sec:pexit_noninterleaved}: by Lemma~\ref{lem:nonint_sampling}, the randomized lifting scatters the jammed copies of each variable-node type over the adjacent check nodes, so that only the localization of the bursts along the coupled chain remains. With ideal \ac{JSI}, the corresponding threshold penalty is $0.07$--$0.14$~dB for the designed rate-$1/2$ ensemble (Table~\ref{tab:iteration_budget}), comparable with the interleaving gain of the optimized code observed here without \ac{JSI}. Hence, the \ac{SC-LDPC} codes reach this performance without a symbol interleaver, which avoids the additional buffering that interleaving across spatial positions would introduce in \ac{SWD}.

\section{Conclusion}
\label{sec:concl}

We have analyzed and designed \ac{SC-LDPC} codes for space-telemetry links affected by pulsed jamming. We have extended the \ac{PEXIT} analysis by tracking separately the messages originating from clean and jammed symbols. Without interleaving, we describe each pulse realization by the fraction of jammed symbols seen by each variable-node type, and we have proved that, under random lifting, this description is sufficient for the analysis. The resulting thresholds capture where the bursts fall along the coupled chain, agree with Monte Carlo density evolution within $0.03$~dB, and drive the search over protographs, coupling memories, and edge spreadings.

With ideal \ac{JSI}, the designed codes, decoded through \ac{SWD} and without symbol interleaving, outperform the \ac{AR4JA} codes with $k=1024$ by $0.6$ to more than $1$~dB at a \ac{BER} of $10^{-4}$, at the price of a longer latency. Against the \ac{AR4JA} codes with $k=4096$, whose latency is comparable, the gain is up to about $0.2$~dB at rates $1/2$ and $2/3$, whereas at rate $4/5$ an error floor makes the \ac{AR4JA} codes preferable at low error rates. \ac{SC-LDPC} codes are therefore an attractive option for jammed telemetry links at low and medium rates, when the latency of \ac{SWD} is acceptable.

Future work includes design constraints on the degree-two variable nodes of the coupled protograph aimed at the error floor, a windowed \ac{PEXIT} design objective, and the joint design of the code with practical \ac{JSI} estimation.

\appendix[\texorpdfstring{Proof of Lemma~\ref{lem:nonint_sampling}}{Proof of Lemma 1}]

For each variable-node type $j$, let $\mathcal{I}_j\triangleq\{h\in\{1,\ldots,\kappa\}:j_h=j\}$ and $\kappa_j\triangleq|\mathcal{I}_j|$, so that $\sum_j\kappa_j=\kappa$, and let $\kappa_{j,s}$ be the number of indices $h\in\mathcal{I}_j$ with $s_h=s$. In a lift of size $\widetilde{M}$, type $j$ contains $A_{j,1}=\lfloor\widetilde{M}\widehat{\rho}_j(\mathbf{o})\rfloor$ jammed copies and $A_{j,0}=\widetilde{M}-A_{j,1}$ unjammed copies, so that
\begin{equation}
    \frac{A_{j,s}}{\widetilde{M}}
    =
    p_{j,s}
    +
    O\!\left(
        \frac{1}{\widetilde{M}}
    \right),
    \qquad
    s\in\{0,1\}.
    \label{eq:nonint_type_composition}
\end{equation}
Since the computation neighborhood is tree-like, the $\kappa_j$ copy indices belonging to type $j$ are distinct and are sampled uniformly without replacement from the $\widetilde{M}$ copies of that type. Therefore, the probability of observing the state assignment $\{s_h:h\in\mathcal{I}_j\}$ is
\begin{equation}
    P_j
    =
    \frac{
        \left(A_{j,0}\right)_{\kappa_{j,0}}
        \left(A_{j,1}\right)_{\kappa_{j,1}}
    }{
        \left(\widetilde{M}\right)_{\kappa_j}
    },
    \label{eq:nonint_hypergeometric_probability}
\end{equation}
where $(x)_y\triangleq x(x-1)\cdots(x-y+1)$ denotes the falling factorial. If this assignment has zero probability, the claim is immediate. Otherwise, taking the copies of type $j$ in the order in which they appear in the neighborhood, $P_j$ is the product of $\kappa_j$ conditional probabilities, the $r$-th of which is $q_{j,r}=(A_{j,s_{j,r}}-c_{j,r})/(\widetilde{M}-r+1)$, where $s_{j,r}$ is the required state of the $r$-th copy and $0\leq c_{j,r}\leq r-1<\kappa_j$ is the number of previously sampled copies of type $j$ in the same state. By~\eqref{eq:nonint_type_composition}, $q_{j,r}=p_{j,s_{j,r}}+O(\kappa_j/\widetilde{M})$, uniformly over $r=1,\ldots,\kappa_j$. Since all factors belong to $[0,1]$, the telescoping-product bound $|\prod_r a_r-\prod_r b_r|\leq\sum_r|a_r-b_r|$ gives
\begin{equation}
    \Bigl|
        P_j
        -
        \prod_{h\in\mathcal{I}_j}
        p_{j,s_h}
    \Bigr|
    =
    O\!\left(
        \frac{\kappa_j^2}{\widetilde{M}}
    \right).
    \label{eq:nonint_type_sampling_error}
\end{equation}
By the independent relabeling symmetry of the random-lift ensemble, conditioned on the tree-like type structure, the copy indices associated with each variable-node type form a uniform sample without replacement from that type, and the index assignments factor across distinct types. Consequently, conditioned on $j_1,\ldots,j_\kappa$, the joint probability in~\eqref{eq:nonint_state_sampling} equals $\prod_jP_j$. Applying again the telescoping-product bound, together with~\eqref{eq:nonint_type_sampling_error} and $\sum_j\kappa_j^2\leq(\sum_j\kappa_j)^2=\kappa^2$, yields~\eqref{eq:nonint_state_sampling}. Since the computation depth is fixed, $\kappa$ does not grow with $\widetilde{M}$, and the error term vanishes as $\widetilde{M}\to\infty$. \hfill$\blacksquare$

\section*{Acknowledgment}
The authors used Claude Opus 5.5 (Anthropic) to assist with language editing and organization of the contents of Sections III--VIII, and also to check the consistency between text, tables, and figures. All AI-assisted content was reviewed by the authors, who take full responsibility for the manuscript.

\bibliographystyle{IEEEtran}
\bibliography{bibliography.bib}

\end{document}